\documentclass[a4paper,11pt]{article}
\usepackage{bbold} 
\usepackage{jheppub} 
\usepackage[T2A]{fontenc}
\usepackage{tikz}
\usetikzlibrary{arrows.meta, positioning, quotes}
\usepackage{amsmath}
\usepackage{float}
\usepackage[normalem]{ulem}
\usepackage{lettrine} 
\usepackage{xcolor}
\usepackage{mathrsfs}
\usepackage{xfrac}
\usepackage{upgreek}
\usepackage{booktabs}

\makeatletter
\DeclareRobustCommand{\loplus}{\mathbin{\mathpalette\dog@lsemi{+}}}
\DeclareRobustCommand{\lotimes}{\mathbin{\mathpalette\dog@lsemi{\times}}}
\DeclareRobustCommand{\roplus}{\mathbin{\mathpalette\dog@rsemi{+}}}
\DeclareRobustCommand{\rotimes}{\mathbin{\mathpalette\dog@rsemi{\times}}}

\newcommand{\dog@rsemi}[2]{\dog@semi{#1}{#2}{-90,90}}
\newcommand{\dog@lsemi}[2]{\dog@semi{#1}{#2}{270,90}}
\newcommand{\dog@semi}[3]{%
  \begingroup
  \sbox\z@{$\m@th#1#2$}%
  \setlength{\unitlength}{\dimexpr\ht\z@+\dp\z@\relax}%
  \makebox[\wd\z@]{\raisebox{-\dp\z@}{%
    \begin{picture}(1,1)
    \linethickness{\variable@rule{#1}}
    \roundcap
    \put(0.5,0.5){\makebox(0,0){\raisebox{\dp\z@}{$\m@th#1#2$}}}
    \put(0.5,0.5){\arc[#3]{0.5}}
    \end{picture}%
  }}%
  \endgroup
}
\newcommand{\variable@rule}[1]{%
  \fontdimen8  
  \ifx#1\displaystyle\textfont3\else
    \ifx#1\textstyle\textfont3\else
      \ifx#1\scriptstyle\scriptfont3\else
        \scriptscriptfont3\relax
  \fi\fi\fi
}
\makeatother

\usepackage{nicefrac}
\usepackage{multirow}

\usepackage[most]{tcolorbox}

\usepackage{caption}
\usepackage{subcaption}

\def\Xz{{\cal{X}}}
\def\d{\text{d}}
\def\bt{\tilde{t}}
\def\br{\bar{r}}
\def\bx{\tilde{x}}
\def\by{\tilde{y}}
\def\bz{\tilde{z}}

\newcommand{\nn}{\nonumber}
\usepackage{color}

\title{\boldmath Radiation in Fluid/Gravity and the Flat Limit}

\author[a]{Gabriel Arenas-Henriquez,}
\author[b]{Luca Ciambelli,}
\author[c]{Felipe Diaz,}
\author[d]{Weizhen Jia,}
\author[c]{David Rivera-Betancour}
\emailAdd{arenas2023@tsinghua.edu.cn}
\emailAdd{ciambelli.luca@gmail.com}
\emailAdd{fdiaz@itmp.msu.ru}
\emailAdd{weizhenjia@cuhk.edu.hk}
\emailAdd{driverabetancour@itmp.msu.ru}
\affiliation[a]{Yau Mathematical Sciences Center, Tsinghua University, Beijing, 100084, China}
\affiliation[b]{Perimeter Institute for Theoretical Physics,
31 Caroline St. N., Waterloo ON, Canada, N2L 2Y5}
\affiliation[c]{Institute for Theoretical and Mathematical Physics,
Lomonosov Moscow State University, 119991 Moscow, Russia}
\affiliation[d]{Department of Physics, The Chinese University of Hong Kong, Shatin, New Territories, Hong Kong, China}

\abstract{We explore asymptotically locally anti-de Sitter spacetimes exhibiting gravitational radiative behavior, employing null gauges that allow for a well-defined flat limit. The radiative content in the bulk is captured by the boundary Cotton and stress tensor, which we collect into a radiative vector. We reinterpret this vector holographically in terms of fluid variables in the dual boundary theory. For algebraically special solutions, we uncover a close connection between bulk radiation and dissipative corrections in the boundary stress tensor, demonstrating a direct link between radiation and entropy production in the boundary fluid. This reveals a rich interplay between radiative dynamics in the bulk and out-of-equilibrium conformal physics at the boundary. We then investigate the flat limit of this correspondence in the context of flat-space holography. In this setting, we construct a Carrollian analogue of the radiative vector and introduce Celestial observables, such as energy detectors, which emerge naturally from the bulk's radiative structure. Our analysis shows that bulk radiation sources the Carrollian viscous stress tensor and heat current, which encodes the Bondi news in this framework. We illustrate our results with explicit examples, including Robinson-Trautman spacetimes and accelerating black holes.}

\begin{document}
\maketitle
\flushbottom
\section{Introduction}

In the last three decades, the holographic principle has been a source of tremendous advancements in theoretical physics \cite{tHooft:1993dmi, Susskind:1994vu}. This principle, rigorously formulated in string theory \cite{Maldacena:1997re, Gubser:1998bc, Witten:1998qj}, states that gravity in anti-de Sitter (AdS) spacetimes is dual to a non-gravitating field theory residing on the AdS conformal boundary. In its essence, and from a geometric standpoint, one can appreciate the holographic principle as stemming from the filling-in problem: given some data at the conformal boundary of AdS, can we reconstruct a dual bulk geometry? The answer to this question, given by Fefferman and Graham \cite{Fefferman:1984asd, fefferman2012ambient}, consists of proving that the radial AdS evolution is a Hamiltonian problem \cite{Papadimitriou:2004ap}. That is, given the (conformal class of) boundary metric and its (conformal) extrinsic curvature, one can fill in the bulk perturbatively \cite{Skenderis:1999nb,  deHaro:2000vlm}.\footnote{The bulk reconstruction is generally not convergent, but the geometry near the boundary is faithfully reconstruct up to some region deep enough in the bulk \cite{Skenderis:1999nb, anderson2010structureconformallycompacteinstein, Anderson:2004yi, Anderson:2006ax, Serantes:2022mgl}.}

The holographic principle associates then a physical meaning to these boundary quantities. The boundary metric is understood as the background geometry on which the dual conformal field theory (CFT) lives, whereas the extrinsic curvature is the building block to construct the Brown-York stress tensor \cite{Brown:1992br} which, upon suitable renormalization \cite{Henningson:1998gx, Balasubramanian:1999re}, is understood as the CFT response 
to the boundary geometry \cite{deHaro:2000vlm}. In other words, the boundary metric sources the boundary stress tensor, whose expectation value contains information about the extrinsic curvature in the bulk \cite{Papadimitriou:2004ap}.

Given boundary coordinates $x^i$, calling the boundary metric $g^{(0)}_{ij}$, and the stress tensor built out of the extrinsic curvature $T^{ij}$\footnote{One should refer to this tensor as the expectation value $\langle T^{ij}\rangle$ in the boundary CFT state. We will leave this implicit in the following.}, the bulk Einstein equations in AdS$_4$ impose\footnote{$\nabla^{(0)}_i$ denotes the Levi-Civita connection on the $3d$ boundary, namely the covariant derivative associated with $g^{(0)}_{ij}$ with $\Gamma^{i}_{(0)jk}=\frac12g^{il}_{(0)}\left(\partial_jg^{(0)}_{kl}+\partial_kg^{(0)}_{jl}-\partial_lg^{(0)}_{jk}\right)$.}
\begin{eqnarray}
    \nabla^{(0)}_i T^{ij}=0 \qquad g^{(0)}_{ij}T^{ij}=0.
\end{eqnarray}
These equations express the diffeomorphisms and conformal invariance of the boundary theory, respectively \cite{Imbimbo:1999bj, deHaro:2000vlm, Rooman:2000zi}. They also admit a natural interpretation in terms of the Navier-Stokes equations for a conformal relativistic fluid living at the boundary \cite{Bhattacharyya:2007vjd}. This is the starting point of the fluid/gravity correspondence \cite{Bhattacharyya:2007vjd, Hubeny:2007xt, Son:2007vk, Bhattacharyya:2008xc, Haack:2008cp, Banerjee:2008th, Rangamani:2009xk, Hubeny:2011hd, Bredberg:2011jq, Compere:2011dx, Compere:2012mt, Caldarelli:2013aaa, Pinzani-Fokeeva:2014cka, Grozdanov:2016fkt, Anabalon:2018ydc, Hoult:2021gnb, Ciambelli:2023mvj}, which posits that (Einstein) gravity in the bulk of an asymptotically AdS spacetime is dual to a conformal relativistic fluid in one dimension less. The fluid/gravity correspondence can clearly be understood as a coarse-graining of the microscopic boundary CFT, and thus of AdS/CFT itself, which explains why it is often referred to as a spin-off of the latter. While the fluid/gravity correspondence was originally formulated for AdS black branes, where the boundary derivative expansion is encoded in the boosted brane geometry, it was later generalized to encompass exact bulk solutions such as black holes, providing a boundary stress tensor whose components are geometrically decomposed and capture features of the boundary relativistic fluids. In this extended and more geometric sense, the correspondence acts as a symmetry matching between bulk and boundary, giving access to quantities like energy density, pressure, and heat current without invoking a full hydrodynamic expansion.

By the very nature of the hydrodynamic limit of a field theory, the fluid/gravity correspondence loses information about the microscopic aspects of the boundary theory \cite{Policastro:2002se}. Despite this, it actually allows us to gain much greater geometric control over the duality. Indeed, as already remarked, one can formulate the fluid/gravity correspondence for complicated bulk GR solutions, for which the boundary CFT dual state is hard to construct \cite{Rangamani:2009xk}. This comprises asymptotically-locally AdS spacetimes (AlAdS), in which the boundary metric and stress tensor are complicated, time-dependent quantities, potentially with topological defects reaching the boundary \cite{Bakas:2008gz, Hubeny:2010ry, Caldarelli:2011idw, Leigh:2012jv, Mukhopadhyay:2013gja, Bakas:2014kfa, Petropoulos:2015fba, Gath:2015nxa, Ciambelli:2017wou, Skenderis:2017dnh}. 4-dimensional AlAdS spacetimes are characterized by a non-vanishing conformal tensor at the $3d$ boundary, preventing the boundary metric from being globally conformally flat \cite{Ashtekar:1999jx, Anderson:2004yi}. While the Weyl tensor is identically zero in $3d$, its role as conformal tensor is replaced by the Cotton tensor $\mathscr{C}_{ijk}$, which is the skew-symmetric covariant derivative of the Schouten tensor $\mathscr{S}_{ij}$ \cite{cotton1899varietes}:
\begin{align}\label{Cott_Scho}
    \mathscr{C}_{ijk} \equiv \nabla_i^{(0)} \mathscr{S}_{jk} - \nabla_j^{(0)} \mathscr{S}_{ik}\,, \qquad \mathscr{S}_{ij} \equiv  R^{(0)}_{ij}-\frac14 R^{(0)} g^{(0)}_{ij}\,.
\end{align}
Then, one has
\begin{center}
    Asymptotically-globally AdS spacetime $\quad \Longleftrightarrow \quad \mathscr{C}_{ijk}=0$\,.
\end{center}
The fluid/gravity has been successfully constructed and understood for spacetimes in which $\mathscr{C}_{ijk}\neq 0$ \cite{Caldarelli:2011idw,Mukhopadhyay:2013gja, BernardideFreitas:2014eoi, Petropoulos:2014yaa, Gath:2015nxa, Petropoulos:2015fba}, whereas the AdS/CFT correspondence has so far been elusive in such spacetimes \cite{deHaro:2008gp, Bakas:2008gz}. The primary examples of these topologically non-trivial spacetimes are Taub-NUT spaces (see  \cite{Leigh:2011au, Leigh:2012jv, Kalamakis:2020aaj, Ciambelli:2020qny,Kalamakis:2025daq} for their holographic aspects), Robinson-Trautman spacetimes (see  \cite{Bakas:2014kfa, Ciambelli:2017wou, Skenderis:2017dnh, Adami:2024rkr} for their holographic aspects), and accelerating black holes (see \cite{Anabalon:2018ydc,Arenas-Henriquez:2024ypo,  Roychowdhury:2024oih, Luo:2024cwm} for their holographic and thermodynamic aspects). 

The role of the Cotton tensor, which is so central in the recent developments of the fluid/gravity correspondence in AlAdS spacetimes \cite{Leigh:2007wf, deHaro:2008gp, Bakas:2008gz, Mukhopadhyay:2013gja, Bakas:2014kfa}, has also been appreciated in understanding non-linear gravitational radiation in spacetimes with cosmological constant. While defining gravitational radiation might at first seem an easy task, this turns out to be a formidable challenge in general. Indeed, in a strongly gravitating region of the spacetime, how do we define a non-linear radiative geometric structure on a background? The situation is drastically different if the spacetime has a far-away region that can be approximated by flat space, in which one can clearly isolate radiative degrees of freedom entering such a region. This is the situation arising when observing an isolated system from far away, which is nothing but the original motivation of Bondi-Van der Burg-Metzner and Sachs for studying the asymptotic structure of gravity in asymptotically flat spacetimes \cite{Bondi:1960jsa, Sachs:1961zz,Bondi:1962px, Sachs:1962wk}. And in fact, in such spaces, the notion of gravitational radiation has been understood and fully characterized \cite{Newman:1961qr, Geroch:1977big, Ashtekar:1981bq, Ashtekar:1981hw,  Wald:1999wa, Ashtekar:2014zsa, Flanagan:2015pxa, Fernandez-Alvarez:2019kdd}, resulting in the appreciation of the Bondi mass loss formula as leaking of gravitational radiation, encoded in the news tensor, see e.g.  \cite{Madler:2016xju}. 

This prompts the question: can we define radiation in spacetimes with a non-vanishing cosmological constant? The situation is more complicated, but recently, Fern\'andez-\'Alvarez and Senovilla proposed a definition of radiation for asymptotically de Sitter spacetimes \cite{Fernandez-Alvarez:2019kdd, Fernandez-Alvarez:2020hsv, Fernandez-Alvarez:2021yog, Fernandez-Alvarez:2021zmp, Fernandez-Alvarez:2024bkf}. Applications of these ideas to anti-de Sitter spaces were explored in \cite{Ciambelli:2024kre}, followed by a more recent rigorous analysis \cite{Fernandez-Alvarez:2025qqx}. Remarkably, the main upshot, as we will carefully review in \autoref{Sec:Radiation}, is that radiation at the conformal boundary is controlled by the proportionality between the stress tensor and the Hodge-dualized boundary Cotton tensor.

The important feature of this characterization of radiation is that, in the flat limit, it must reproduce the well-known statement\footnote{This clearly holds in a simplified setup. See e.g. \cite{Fernandez-Alvarez:2019kdd, Ashtekar:2024mme, Ashtekar:2024bpi, Rignon-Bret:2024gcx, Ashtekar:2024xwl, Ashtekar:2024stm} for a more proper analysis and statement.}
\begin{center}No news \quad $\Leftrightarrow$ \quad No radiation.
\end{center}
Nonetheless, performing the flat limit could be a daunting task.
Indeed, the flat limit of AdS has been the subject of intense study in the last decade \cite{Bagchi:2009my,Bagchi:2016bcd,Ciambelli:2018wre,Poole:2018koa, Hijano:2019qmi, Compere:2019bua, Compere:2020lrt,Hijano:2020szl,Komatsu:2020sag,Li:2021snj,Ponomarev:2022qkx,Duary:2022pyv,Campoleoni:2023fug,Bagchi:2023fbj,Bagchi:2023cen,Alday:2024yyj,Kraus:2024gso,Fareghbal:2024lqa, Duary:2024kxl,Marotta:2024sce,Lipstein:2025jfj,Surubaru:2025fmg,Hao:2025btl}. The naive flat limit of the metric written in the Fefferman-Graham gauge is divergent.\footnote{This is equally true for its generalization, the so-called Weyl-Fefferman-Graham gauge \cite{Ciambelli:2019bzz, Jia:2021hgy, Jia:2023gmk, Ciambelli:2023ott, Arenas-Henriquez:2024ypo}.} Nonetheless, it has been appreciated in \cite{Barnich:2012aw} that null gauges --- that is, gauges in which a coordinate is null --- have instead a smooth flat limit from AdS. This has been fully exploited in \cite{Ciambelli:2018wre, Campoleoni:2023fug}, in which it was realized that the fluid/gravity correspondence admits a smooth flat limit, since it is naturally formulated in a null gauge. This provided a road toward the flat space fluid/gravity correspondence. In the latter, the boundary fluid is hosted at null infinity, and is an unusual fluid dubbed Carrollian fluid \cite{deBoer:2017ing, Ciambelli:2018xat, Ciambelli:2018ojf, Campoleoni:2018ltl, deBoer:2021jej, Campoleoni:2022wmf, Petkou:2022bmz, Freidel:2022bai, Redondo-Yuste:2022czg, deBoer:2023fnj, Armas:2023dcz, Bagchi:2023rwd}. 
Yet, the intrinsic Carrollian hydrodynamic and/or geometric interpretation remains unclear. This work aims to advance this understanding by relating bulk radiation and dissipation in asymptotically flat spacetimes to Carrollian tensors, thereby proposing a physically motivated notion of equilibrium and thermal properties for putative dual Carrollian fluids. In this sense, the correspondence we are using (even in AdS) should be more precisely understood as a ``geometro-hydrodynamic'' matching, as we do not perform a Carrollian derivative expansion, a procedure that is yet unknown. One thus expects that these elusive Carrollian fluids provide the macroscopic description of the putative boundary Carrollian field theory \cite{Bagchi:2016bcd, Bagchi:2019xfx, Figueroa-OFarrill:2021sxz, Donnay:2022aba, Donnay:2022wvx, Bagchi:2022emh, Saha:2023hsl, Salzer:2023jqv, Nguyen:2023vfz, Mason:2023mti, Kraus:2024gso, Alday:2024yyj, Ruzziconi:2024zkr, Ciambelli:2025mex, Lipstein:2025jfj, Baulieu:2025itt}. This approach complements the celestial holography program in flat-space holography, which is based on understanding the elusive celestial CFT living on the codimension-2 celestial sphere \cite{strominger2014bms, He:2014laa, Lysov:2014csa, Cachazo:2014fwa,Kapec:2014opa, Strominger:2014pwa, pasterski2016new, Pasterski:2021raf, pasterski2017conformal} (see \cite{Strominger:2017zoo, Raclariu:2021zjz, Pasterski:2021raf} for reviews and further references). Little is known about the quantum aspects of Carrollian field theories,\footnote{Nonetheless, see e.g. \cite{Cotler:2024xhb, Vassilevich:2024vei, Sharma:2025rug, Cotler:2025dau, Bagchi:2025vri} for interesting recent explorations.} and thus the fluid/gravity duality is even more instructive in understanding the macroscopic behavior. Essentially, it indeed amounts to ``put the boundary theory in a box'': whatever the complicated and elusive boundary field theory is microscopically, it must reproduce the fluid/gravity predictions in the macroscopic limit.

Conversely, as already mentioned, gravitational radiation is significantly better understood in asymptotically flat spacetimes than in asymptotically AdS. This asymmetry presents both a challenge and an opportunity: how can one relate the well-understood features of gravitational radiation in flat spacetime to the properties of the dual boundary fluid? And similarly, how are the properties of the boundary fluid in AdS related to the presence or absence of gravitational radiation in the bulk? Precisely because one side of the correspondence is under theoretical control while the other remains more elusive, there is a powerful potential for two-way knowledge transfer. Insights from the gravitating bulk can inform and guide our understanding of boundary dynamics in flat spaces, while the holographic structure of AdS/CFT may offer novel perspectives on radiation in AdS, tied altogether by a smooth and controlled flat limit. This reciprocal interplay is not only conceptually rich but also particularly timely, as it promises to bridge gaps in both frameworks and sharpen our understanding of gravitational dynamics across spacetime asymptotics and its dual boundary interpretation. This can be summarized in the following schema, which makes clear how relating these two concepts can have important, far-reaching repercussions:
\begin{center}
    \begin{tabular}{ |p{4cm}||p{4cm}|p{4cm}|}
 \hline
State of the art & Asymptotically-locally flat spacetimes & Asymptotically-locally AdS spacetimes \\
 \hline\hline
 Gravitational radiation & Well understood & Under construction \\
 \hline
 Boundary dual fluid & Under construction & Well understood \\
 \hline
\end{tabular}
\end{center}

Let us start listing the benefits of understanding the link between radiation and fluids in AdS. As we have discussed, the presence of radiation in AdS is sourcing a non-trivial stress tensor and Cotton tensor at the boundary. One natural expectation is that there exists a direct link between radiation in the bulk and dissipation in the boundary hydrodynamics. This leads to the study of the connection between gravitational radiation and fluid entropy production. Such a connection can be explicitly tested in exact solutions of GR in bulks possessing a non-vanishing and non-proportional boundary Cotton and stress tensor,\footnote{Strictly speaking, we are referring here to the boundary Hodge dual of the Cotton tensor, as defined in \eqref{Cotton_York}.} such as the RT metric. This comprises the C-metric, in which we will show that there is radiation. In turn, this provides a window into the thermodynamic properties of intricate exact solutions, which is a hard endeavor \cite{Appels:2016uha, Appels:2017xoe, Anabalon:2018qfv, Anabalon:2018ydc, Cassani:2021dwa, Kim:2023ncn}. A key feature of the fluid/gravity correspondence is its relationship to the Petrov classification of the bulk exact solution \cite{Petropoulos:2015fba, Gath:2015nxa, Ciambelli:2017wou}. Here, we will leverage this to discuss the simplifications occurring for algebraically-special spaces, which comprise the most relevant examples and applications.

Conversely, establishing a connection between gravitational radiation and Carrollian dual fluids in flat spacetimes promises to yield valuable insights into Carrollian hydrodynamics. Indeed, whether through direct analysis or by taking the flat limit, we identify the Carrollian radiative scalar and vector. These quantities govern the radiative flux reaching null infinity, and we will show that they encapsulate the Carrollian viscous stress tensor at the boundary. This allows us to learn that the Carrollian viscous stress tensor contains information on the time-derivative of the news tensor $\Sigma_{AB}\propto \partial_u N_{AB}$. We can test our flat-space predictions again using the RT and Kerr-NUT spacetimes as examples. Moreover, following \cite{Ciambelli:2024kre}, we can take the flat limit and study energy detector operators in the specific example of the C-metric. 

These are some of the key consequences of establishing a connection between gravitational radiation and boundary fluid dynamics, which is an interplay that lies at the heart of this paper.

The rest of the manuscript is organized as follows: 
\begin{itemize}
    \item In \autoref{Sec:Radiation}, we review the essential aspects of the radiation criteria introduced in \cite{Fernandez-Alvarez:2019kdd, Fernandez-Alvarez:2020hsv, Fernandez-Alvarez:2021yog, Fernandez-Alvarez:2021zmp, Fernandez-Alvarez:2024bkf,Ciambelli:2024kre, Fernandez-Alvarez:2025qqx} and briefly illustrate the geometric nature and holographic implications of the criteria.  

    \item \autoref{Sec:Holo_fluid} is devoted to describing the dual conformal fluid and establishing the connection between dissipative corrections and bulk gravitational radiation. We begin in \autoref{SubSec:BoundaryFluidDynamics} by reviewing the fluid/gravity correspondence and outlining the kinematics and dynamics of the relativistic boundary fluid. In \autoref{SubSec:CovariantNU}, we introduce a covariant Newman-Unti gauge in the bulk, constructed from the boundary geometry and fluid data \cite{Ciambelli:2018wre, Campoleoni:2023fug}. For solutions of algebraically special Petrov type, this gauge allows for a closed-form reconstruction of the full bulk spacetime in general AlAdS backgrounds. Finally, in \autoref{SubSec:GravRadiationAndFluidDynamics}, we express the radiative vector in terms of boundary fluid variables and demonstrate that, for algebraically special spacetimes, bulk gravitational radiation induces dissipative corrections in the boundary fluid. Additionally, we derive an expression for entropy production at the boundary, showing a connection between bulk radiation and out-of-equilibrium boundary physics. As concrete examples, we discuss the AdS-Robinson-Trautman class of solutions and accelerating black holes.

    \item Taking advantage of the adapted null gauges, in \autoref{Sec:Flat_lim} we explore the flat limit of holographic quantities associated with bulk radiation, leading to the construction of the boundary Carrollian fluid and Celestial operators. In \autoref{SubSec:Carroll}, we use the Papapetrou-Randers parametrization of the boundary metric to examine the geometric structure underlying the flat limit. Using a space/time decomposition, we describe the Carrollian manifold that emerges on the boundary and show how the holographic fluid acquires a Carrollian structure in this limit. Then, in \autoref{SubSec:CarrollianPoynting}, we analyze the flat limit of the radiative vector and identify a Carrollian scalar and vector that capture bulk radiation in this regime. These are expressed in terms of the Carrollian viscous stress tensor and heat current. By comparing with the Bondi-Sachs formalism, we show that one of the two Carrollian viscous stress tensor and heat current plays the role of the time and space derivative of the Bondi news tensor, suggesting a Carroll-covariant extension of this quantity. As explicit examples, we consider the Kerr-Taub-NUT and Robinson-Trautman spacetimes. Finally, in \autoref{SubSec:LBMS_Cel_Op}, we briefly review the extension of the Bondi-Sachs framework to arbitrary cosmological constant \cite{Compere:2020lrt}, and demonstrate how radiative spacetimes can be used to construct Celestial operators --- such as energy detector operators --- in the flat limit, using accelerating black holes as explicit examples.

\item Finally, in \autoref{Sec:Conclusions} we conclude with open questions and potential future directions.

\end{itemize}




\section{AdS Radiation and Holography}\label{Sec:Radiation}

In General Relativity, the news tensor ${{N}}_{AB}$ is a crucial quantity to study gravitational waves. It arises within the Bondi-van der Burg-Metzner-Sachs (BMS) formalism \cite{Bondi:1960jsa, Bondi:1962px, Sachs:1962wk} (see \cite{Madler:2016xju} for a pedagogic review), which characterizes asymptotically flat spacetimes at future null infinity ${\cal I}^+$\,. The news tensor encodes the flux of gravitational radiation escaping to infinity. It appears explicitly in the Bondi mass loss formula, which shows that the emission of gravitational waves results in a decrease in the total mass of the system. Moreover, it plays a significant role in classifying asymptotic symmetries via the \textit{peeling} theorem \cite{Sachs:1961zz} and it is intimately related to the gravitational memory effect \cite{Zeldovich:1974gvh, Braginsky:1985vlg, Braginsky:1987kwo, Christodoulou:1991cr, Wiseman:1991ss, Blanchet:1992br, Thorne:1992sdb, Favata:2010zu}. The latter is a subtle DC effect corresponding to a permanent displacement of test masses caused by the passage of gravitational radiation. Remarkably, memory effects are related to Weinberg's soft graviton theorem \cite{Weinberg:1965nx} through a Fourier transform \cite{Strominger:2014pwa}, leading to the Braginsky-Thorne formula for the difference between the initial and final 
transverse traceless components of the metric \cite{Braginsky:1987kwo}. This establishes a deep relation between the infrared behavior of graviton scattering amplitudes and classical memory effects. 

The news tensor describes outgoing gravitational radiation, as can be seen from the Newman-Penrose formalism \cite{Newman:1961qr}, which uses conformal geometry to characterize the spacetime structure. Within this framework, and through Thorne's wave-generation formalism \cite{Thorne:1980ru}, it can be shown that the complex Weyl scalars encode all information about outgoing gravitational waves and can be written in terms of the news tensor. This highlights the remarkable structure and features of asymptotically flat spacetimes, establishing a bridge between different areas of physics and providing valuable insights into the quantum nature of gravity, as reviewed in \cite{Strominger:2017zoo}. 

In the presence of a non-vanishing cosmological constant, the definition of gravitational radiation becomes more subtle, as there is no direct analogue of the news tensor. The absence of null infinity in the standard sense prevents a straightforward generalization of the asymptotic flatness framework, where gravitational radiative degrees of freedom are usually identified.  For $\Lambda > 0$, future infinity becomes spacelike, and therefore the Weyl scalars cannot be directly interpreted as outgoing radiation. Instead, it characterizes gravitational disturbances that freeze out at future infinity. In the asymptotically (locally) anti-de Sitter (AlAdS) case, this situation is even more subtle: with appropriate boundary conditions, the spacetime behaves like a cavity where null rays hit the boundary in finite time and bounce back, preventing energy from escaping. As a result, boundary conditions play a fundamental role in defining and interpreting outgoing radiation, with several implications for the AdS/CFT correspondence \cite{Hubeny:2010ry, Bakas:2014kfa, Ciambelli:2017wou, Liu:2018crr, Kundu:2019ull, Grumiller:2023ahv, McNees:2024iyu} (see \cite{Podolsky:2003gm} for a study of the Newman-Penrose formalism for $\Lambda < 0$). Additionally, imposing Dirichlet boundary conditions, the asymptotic symmetry group of AdS$_4$ is finite-dimensional, in contrast with its asymptotically flat counterpart, and lacks supertranslations, which in Minkowski spacetime relate vacua displaced by the passage of gravitational radiation.  Therefore, defining gravitational radiation when $\Lambda \neq 0$ is a highly non-trivial task.

An interesting criterion for characterizing radiative degrees of freedom for any value of the cosmological constant was proposed in a series of papers by Fern\'andez-\'Alvarez and Senovilla \cite{Fernandez-Alvarez:2019kdd,Fernandez-Alvarez:2020hsv,Fernandez-Alvarez:2021yog, Senovilla:2022pym, Fernandez-Alvarez:2025qqx}.\footnote{See also \cite{Bonga:2023eml} for a description of gravitational radiation with $\Lambda>0$ using the Bondi-Sachs formalism.} This novel proposal provides a criterion based on a geometric construction that utilizes conformal compactification techniques by introducing a non-physical metric $\hat{g} = \hat\Omega^2 g$\,. In this way, the manifold is conformally compactified, and the boundary $\cal B$ is endowed with a metric structure. The method consists of constructing a canonical momentum using a rescaled version of the Bel-Robinson tensor,\footnote{The Bel-Robinson tensor is traceless, totally symmetric, and covariantly conserved. It carries the tidal energy-momentum of the gravitational field and can be related to quasi-local energy-momentum quantities \cite{HorowitzGravEnergy, Senovilla:1999xz}.}
\begin{align}
    \hat{\cal D}_{\mu\nu\rho\sigma} = \hat{w}^\beta{}_{\rho\mu\alpha}\hat{w}^\alpha{}_{\nu\sigma\beta} + \ast\hat{w}^\beta{}_{\rho\mu\alpha}\ast\!\hat{w}^\alpha{}_{\nu\sigma\beta}\,,
\end{align}
where $\hat{w}^\beta{}_{\rho\mu\alpha} \equiv \hat\Omega^{-1}\hat{W}^\beta{}_{\rho\mu\alpha}$ is a rescaled version of the Weyl tensor, defined in terms of the conformally compactified metric $\hat{g}$, and $\ast \hat{w}^\beta{}_{\rho\mu\alpha}$ is its Hodge dual. The rescaling of the Bel-Robinson tensor ensures the correct asymptotic behavior, such that it remains non-vanishing at the boundary. For $\Lambda \geqslant0$~, one can construct a canonical super-momentum vector 
\begin{align}\label{Poynting}
    \hat{ P}^\mu = -\hat{\cal D}^\mu{}_{\alpha\beta\gamma} \hat{n}^\alpha \hat{n}^\beta \hat{n}^\gamma\,,
\end{align}
where $\hat{n}_\mu =\partial_\mu \hat\Omega$ is normal to the boundary, whose normalization at the boundary is given by $\hat{n}_\mu \hat{n}^\mu \overset{\cal B}{=} -\Lambda/3$\,. The crucial feature here is that for $\Lambda \geqslant0$ there is a preferred choice for the vector field $\hat{n}^\mu$, which must be a non-spacelike, future-pointing vector field at the boundary, given by the conformal compactification procedure.

Then, the super-momentum vector can be decomposed along the normal vector $\hat{n}^{\mu}$ as
\begin{equation}\label{supmom_decomp}
    \hat{P}^{\mu}=W \hat{n}^{\mu}+S^{\mu}\,,
\end{equation}
where $W$ is called the super-energy density and $S^{\mu}$ is the super-Poynting vector. In this decomposition, $S^{\mu}$ is transverse to $\hat{n}_{\mu}$, namely $S^{\mu}\hat{n}_{\mu}=0$, and 
\begin{align}
    W&=\frac{1}{N^2}\hat{P}^{\mu}\hat{n}_{\mu}\,,\label{W_dens}
    \\
    S^{\mu}&=\Delta^{\mu}_{\hphantom{\mu}\nu}\hat{P}^{\nu}\,,\label{Sup_Mom}
\end{align}
with
\begin{equation}\label{ProjectorD}
   N^2=\hat{n}^{\mu}\hat{n}_{\mu}\qquad \Delta^{\mu}_{\hphantom{\mu}\nu}=\delta^{\mu}_{\nu}-\frac{1}{N^2}\hat{n}^{\mu}\hat{n}_{\nu}\,,
\end{equation}
where $\Delta^{\mu}_{\hphantom{\mu}\nu}$ represents the projector onto the space transverse to $\hat{n}^{\mu}$.

In the absence of matter,\footnote{In \cite{Fernandez-Alvarez:2019kdd,Fernandez-Alvarez:2021yog}, it is proven that \eqref{DivergencePoynting} still holds in the presence of matter if and only if $T_{\mu\nu}$ decays as ${\cal O}(\hat\Omega^3)$ when approaching the conformal boundary.} the super-momentum vector is divergence-free at the boundary\footnote{$\hat\nabla_\mu$ denotes the covariant derivative with respect to the unphysical metric $\hat g$\,.}
\begin{align}\label{DivergencePoynting}
    \hat\nabla_\mu \hat{P}^\mu \overset{\cal B}{=} 0\,.
\end{align}
A non-trivial asymptotic value of the super-Poynting vector, denoted by $\hat{\mathscr{P}}^i$ and dubbed the boundary super-Poynting vector, indicates the presence of gravitational radiation at the boundary as it measures the flux of tidal energy. For asymptotically flat spacetimes described in Bondi coordinates, $\hat{\mathscr{P}}^i$ vanishes precisely when the news tensor is trivial, thereby reproducing the standard criterion for the absence of radiation. For the asymptotically locally de Sitter case, Fernandez-Alvarez and Senovilla showed that the asymptotic value of the super-Poynting vector $S^\mu$ indeed provides a robust criterion for gravitational radiation by analyzing various known solutions and boundary submanifolds. This demonstrates that this new approach consistently recovers the known flat space results and can be viewed as a natural extension for $\Lambda \neq 0$\,. 

In the case of asymptotically locally AdS spacetimes ($\Lambda<0$), the situation is more subtle, as there is no preferred choice of non-spacelike vector at the boundary. Indeed, the normal $\hat{n}^\mu$ to the boundary is now spacelike. However, as long as one considers any vector $\hat{v}^\alpha$ such that $\hat{v}^\alpha \hat{v}_\alpha=-1$ and $\hat{v}^\alpha  \hat{n}_{\alpha}=0$, a similar description to the one above can be performed, replacing $\hat{n}^\alpha$ with $\hat{v}^\alpha$, leading to the super-momentum
\begin{align}
    \hat{ P}^\mu = -\hat{\cal D}^\mu{}_{\alpha\beta\gamma} \hat{v}^\alpha \hat{v}^\beta \hat{v}^\gamma\,,
\end{align}
and super-Poynting $S^{\mu}(\hat{v})$ defined via
\begin{equation}\label{susu}
    \hat{P}^{\mu}=W \hat{v}^{\mu}+S^{\mu}(\hat{v})\,,
\end{equation}
which is now a function of $\hat{v}^\alpha$. Then, the condition for the total absence of radiation (incoming and outgoing) at the boundary of asymptotically AdS spaces has been found in \cite{Fernandez-Alvarez:2025qqx} to be
\begin{equation}
    \hat{P}^{\mu} \hat{n}_\mu=0\,,
\end{equation}
which is equivalent to $S^{\mu}(\hat{v})\hat{n}_\mu=0$. This can be rewritten covariantly, and independently of the choice of $\hat{v}$, in terms of the boundary data. Defining the Cotton-York tensor as \begin{equation}\label{Cotton_York}
\mathscr{C}_{ij}=\frac12\eta_i^{\hphantom{i}kl}\mathscr{C}_{klj}\,,
\end{equation}
where $\eta_{ijk}\equiv\sqrt{-g_{(0)}}\epsilon_{ijk}$\,, one finds that absence of radiation is obtained requiring 
\begin{equation}\label{lin}
\alpha \mathscr{C}_{ij}=\beta T_{ij}\,,
\end{equation}
for any $\alpha$ and $\beta$ functions at the boundary. Here, $T_{ij}$ is the holographic stress tensor.\footnote{For the holographic stress tensor we use the convention of \cite{Ciambelli:2018wre,Campoleoni:2023fug} which differs from the standard one \cite{Balasubramanian:1999re} by a factor of $-\ell^{-1}$, i.e., $T_{ij}=-\frac{1}{\ell}T^{\text{std}}_{ij}$.}

The contracted version of this condition was studied in \cite{Ciambelli:2024kre}, and later understood in \cite{Fernandez-Alvarez:2025qqx} to be only a necessary condition for the absence of radiation. Let us introduce the boundary vector field\footnote{This is related to \cite{Ciambelli:2024kre} through $\hat{\mathscr{P}}^i=\frac{1}{4\ell^4}\hat{\mathscr{P}}^i_{\text{CPT}}$\,.}
\begin{align}
    \hat{\mathscr{P}}^i
    =\frac{16\pi G}{\ell^2} \mathscr{C}^i{}_{jk}T^{jk}\,.\label{Super_P1}
\end{align}
This vector was called the super-Poynting vector in \cite{Ciambelli:2024kre}. This is, however, a misnomer, because it is not equal to the boundary value of the vector $S^\mu(\hat{v})$ introduced in \eqref{susu}, as pointed out in \cite{Fernandez-Alvarez:2025qqx}. Nonetheless, this vector is an extremely useful quantity in AdS spacetimes. Indeed, it gives a necessary criterion for the absence of radiation, and it packages together the Cotton and stress tensor into a rank-1 object. As such, we will call it from now on the ``radiative vector'', and study its role in the fluid/gravity correspondence. The reason why we confine our attention to the radiative vector is that it allows us to work with a simple and explicit expression for radiation in terms of the boundary data. Albeit it gives a necessary condition for the absence of radiation, one can turn the story upside down: if the radiative vector is not zero, then we are certain that the bulk gravitational solution is radiating. Moreover, in the examples below, the necessary condition for the absence of radiation turns out to also be sufficient. In conclusion, one notes that this radiative vector is actually a ``radial super-Poynting''. The radial evolution from the boundary in AdS spaces is a Hamiltonian problem \cite{Anderson:2004yi, fefferman2012ambient}, and this vector plays a crucial role as a book-keeping device for the boundary conformal structure and Brown-York stress tensor.

The expression \eqref{Super_P1} has a clear and immediate implication: radiative AdS spacetimes necessarily possess a non-conformally-flat boundary geometry, since a non-zero boundary radiative vector requires a non-vanishing Cotton tensor $\mathscr{C}_{ijk}$. As a result, dual theories to such bulk configurations do not contain the maximal number of conformal isometries,\footnote{Indeed, the Cotton tensor plays in $3d$ the role that the Weyl tensor plays in higher dimensions: it is zero if and only if the number of conformal isometries is maximal.} in contrast to flat space holography, where the presence of gravitational waves does not affect the asymptotic symmetry group: supertranslations preserve the underlying Carrollian structure. However, it is important to stress that a non-vanishing Cotton tensor is not a sufficient condition for radiation. 

As we just reviewed, in \cite{Fernandez-Alvarez:2025qqx}, for the AdS case, the super-Poynting vector is constructed by contracting the Bel-Robinson tensor with timelike vectors (orthogonal to the normal vector). This prescription provides a robust criterion for the absence of radiation: Beyond the vanishing of \eqref{Super_P1}, it additionally requires that the holographic stress tensor and the Cotton tensor be linearly dependent, as stated in \eqref{lin}. This property is commonly regarded as energy-momentum/Cotton tensor duality  \cite{Leigh:2007wf,deHaro:2007fg,Bakas:2008gz, Mansi:2008br, Mansi:2008bs, deHaro:2008gp, Bakas:2009pbm,Miskovic:2009bm, Mukhopadhyay:2013gja,  Ciambelli:2020qny,Corral:2024lva,Kalamakis:2025daq}, and corresponds to a generic feature of self-dual gravitational solutions, i.e., spacetimes in which the electric part of the Weyl tensor equals (up to a sign) its magnetic counterpart. 
An example of this is the AdS version \cite{Hawking:1998ct} of the Taub-NUT solution \cite{Taub:1950ez}, which consists of a 1-parameter extension of the Schwarzschild-AdS black hole. The new parameter can be interpreted as gravitational magnetic charge, and the Euclidean version of this spacetime has been understood as a gravitational instanton \cite{Eguchi:1978xp}. Although the holographic interpretation of the solution is rather complicated due to the presence of the Misner string \cite{Misner:1963fr, Bonnor:1969ala}, it is known that the dual theory corresponds to a perfect fluid state whose velocity is a geodesic congruence with no shear nor expansion of the boundary metric but with non-trivial constant vorticity due to the NUT parameter. See \cite{Caldarelli:2011idw, Leigh:2012jv,Leigh:2011au, Ciambelli:2020qny, Kalamakis:2020aaj, Cano:2021qzp, Kalamakis:2025daq} for further remarks on the holographic interpretation of the solution.

In contrast, the Robinson-Trautman (RT) family of solutions provides examples of genuinely radiative, asymptotically locally AdS spacetimes. These geometries are characterized by an expanding, twist-free, and shear-free congruence of null geodesics \cite{Robinson:1962zz}. Their AdS extensions have been studied in the context of holography \cite{Bakas:2014kfa,BernardideFreitas:2014eoi,Ciambelli:2017wou, Skenderis:2017dnh, Fareghbal:2018xii},
and they possess a non-vanishing radiative vector when the Gaussian curvature is constant and the $P$ function (see \eqref{RobinsonTrautman}) is time-independent, in agreement with \cite{Skenderis:2017dnh}.

 A particularly illustrative case within this class of solutions is the AdS C-metric, which describes a pair of uniformly accelerating black holes connected by a cosmic string. As we will see shortly, this solution has a non-trivial boundary radiative vector with linearly independent stress and Cotton tensors, such that the spacetime radiates.

\section{Holographic Fluid Perspective}\label{Sec:Holo_fluid}

The fluid/gravity duality provides a powerful holographic framework for the study of non-equilibrium phenomena. It relates long-wavelength solutions of Einstein's equations with a negative cosmological constant to the dynamics of a non-linear conformal fluid at the timelike boundary. This has been successfully constructed in the pioneering work \cite{Bhattacharyya:2007vjd,Baier:2007ix,Bhattacharyya:2008xc}, where conformal fluid configurations in the Landau-Lifshitz frame are used to reconstruct asymptotically AdS spacetimes order by order in the derivative expansion (see also \cite{Rangamani:2009xk,Hubeny:2011hd} for comprehensive reviews). 

In what follows, we review the main features of conformal fluid dynamics and its relation to AlAdS spacetimes through the fluid/gravity correspondence. Within this framework, we will reinterpret gravitational radiation in terms of holographic fluid variables.

\subsection{Boundary Fluid Dynamics}\label{SubSec:BoundaryFluidDynamics}

We consider a three-dimensional relativistic fluid living at the conformal boundary of a four-dimensional AlAdS spacetime, with negative cosmological constant $\Lambda=\nicefrac{-3}{\ell^2}$, where $\ell$ is the AdS radius. The boundary is timelike and endowed with a metric $g^{(0)}_{ij}$\,, where lowercase Latin indices label the boundary coordinates. The remaining boundary data is encoded in the energy-momentum tensor $T^{ij}$ of the holographic fluid configuration, which is covariantly conserved
\begin{equation}\label{DT}
    \nabla^{(0)}_iT^{ij}=0\,,
\end{equation}
and the timelike congruence $u^{i}$, interpreted as the fluid velocity field.\footnote{For a non-perfect fluid, $u^i$ has a gauge freedom: each choice of the velocity field represents a hydrodynamic frame \cite{Kovtun:2019hdm,Petkou:2022bmz, Campoleoni:2022wmf}.} The fluid energy-momentum tensor is decomposed along the velocity field as\footnote{This decomposition is valid for any rank-2 symmetric tensor. We will use this decomposition when discussing the properties of the boundary Cotton-York tensor.}
\begin{equation}\label{fluid_stress_tensor}
    T_{ij}=\left(\varepsilon+p\right)\ell^2u_iu_j+ pg^{(0)}_{ij}+\tau_{ij}+\ell^2u_iq_j+\ell^2u_jq_i\,,
\end{equation}
where $\varepsilon$ is the local energy density, $p$ is the local pressure, $q_i$ is the heat current and $\tau_{ij}$ is the viscous stress tensor. These hydrodynamic variables are defined through the energy-momentum tensor as
\begin{align}
    \varepsilon&=\ell^2 T_{ij}u^iu^j\,,\label{energy_dens}\\
    q_i&=-\varepsilon u_i-T_{ij}u^j\,,\label{heat_curr}\\
    \tau_{ij}&=P^{k}_{i}P^{l}_{j}T_{kl}-pP_{ij}\,,\label{viscous_stress}
\end{align}
where the velocity field $u^i$ is normalized with respect to the AdS radius of the bulk solution as
\begin{equation}\label{norm_u}
    g^{(0)}_{ij}u^iu^j=-\frac{1}{\ell^2}\,,
\end{equation}
and 
\begin{equation}\label{projector_to_h}
    P_{ij}=g^{(0)}_{ij} +\ell^2u_{i}u_{j}
\end{equation}
is the projector onto the space transverse to $u^i$\,. In this decomposition, both the heat current and the viscous stress tensor are transverse to $u^i$\,; that is, they satisfy $q_iu^i=0=\tau_{ij}u^j$\,.

In the hydrodynamic regime, the dissipative terms, such as the heat current and viscous stress tensor, are determined by constitutive relations. These terms consist of an infinite derivative expansion of local temperature $T$ and velocity field $u^i$\,.\footnote{It could also include derivatives of some local chemical potential in the presence of currents associated with matter fields.} For the present discussion, we keep $q_i$ and $\tau_{ij}$ as arbitrary functions, namely, our ``abstract'' boundary fluid may include both hydrodynamic and non-hydrodynamic modes.

Conformal invariance requires the energy-momentum tensor to be traceless, which implies the following conditions:
\begin{equation}\label{eq_state}
    \varepsilon=2p\,\quad\text{and}\quad \tau^{i}_{\hphantom{i}i}=0\,,
\end{equation}
where the first relation corresponds to the well-known equation of state for a conformal fluid in three dimensions. 

The irreducible decomposition of the velocity field $u^i$ provides the kinematic properties of the fluid, which are encoded in
\begin{equation}\label{Du_decomp}
\nabla^{(0)}_iu_{j}=\sigma_{ij}+\omega_{ij}+\frac12\Theta P_{ij}-\ell^2u_ia_{j}\,,
\end{equation}
where the quantities 
\begin{align}
\Theta={}&\nabla^{(0)}_i  u^i \,,\label{def20}
\\ a_i ={}&u^j \nabla^{(0)}_j u_i , \label{def21}
\\ \omega_{ij}={}& \nabla^{(0)}_{[i} u_{j]} + \ell^2u_{[i}a_{j] }\,\label{def23}
\\ \sigma_{ij}={}&\nabla^{(0)}_{(i } u_{j )} + \ell^2u_{(i } a_{j)} -\frac{1}{2} \Theta\,P_{ij}\,,
\label{def24}
\end{align}
represent, respectively, the expansion, acceleration, vorticity, and shear of the fluid.

\paragraph{Boundary Weyl covariance:}

Relativistic fluids defined at the conformal boundary of an AlAdS solution are known to be invariant under Weyl transformations \cite{Loganayagam:2008is,Rangamani:2009xk}. The latter acts as a rescaling of the boundary metric and fluid velocity as
\begin{equation}\label{Weyl-scaling}
    g^{(0)}_{ij}\to\mathscr{B}^{-2}g_{ij}^{(0)}\,,\quad\text{and}\quad u^i\to\mathscr{B}u^i\,,
\end{equation}
with $\mathscr{B}(x)$ an arbitrary function depending on the coordinates of the background geometry. Tensors that rescale with a power $w$ of $\mathscr{B}$ are said to be Weyl-covariant tensors of weight $w$.\footnote{For example, the boundary metric $g^{(0)}_{ij}$ is Weyl covariant of weight $w=-2$ and the velocity field $u^i$ is a Weyl-covariant vector of weight $w=1$.}

For the boundary fluid configuration at hand, one can introduce a Weyl-covariant derivative $\mathscr{D}_i$ that maintains the Weyl covariance of tensors. This is implemented with the inclusion of a Weyl connection $A_i$ which follows the transformation rule $A\to A-\text{d}\ln{\mathscr{B}}$. For conformal relativistic fluids, the Weyl connection\footnote{The expression for the Weyl connection used in this work follows from requiring $\mathscr{D}_iu^i=0=u^j\mathscr{D}_j u_i$\,. See \cite{Loganayagam:2008is} for a more detailed derivation.} is defined as 
\begin{equation}\label{Weyl_conn}
    A_i=\ell^2\left(a_i-\frac{1}{2}\Theta u_i\right)\,.
\end{equation}

The Weyl-covariant derivative acts on scalars $\Phi$ and vectors $V^{j}$ of weight $w$ as 
\begin{align}
    \mathscr{D}_i\Phi&=\partial_i\Phi+wA_i\Phi\,,\label{WD_scalar}
    \\
    \mathscr{D}_iV^j&=\nabla^{(0)}_iV^j+\left(w-1\right)A_iV^j+A^jV_i-\delta^{j}_iA_kV^k\,.\label{WD_vec}
\end{align}
By construction, $\mathscr{D}_{i}$ is metric compatible, i.e., $\mathscr{D}_ig^{(0)}_{jk}=0\,.$ Further properties and definitions of Weyl-covariant geometric quantities can be found in \autoref{App:Weyl}.

\paragraph{Boundary Cotton tensor:} 
 Another important ingredient for the boundary analysis of AlAdS spacetimes is the Cotton tensor of the boundary geometry. The Cotton tensor defined in \eqref{Cott_Scho} measures the deviation of the boundary metric from conformal flatness, and it can be dualized to a rank-2 symmetric tensor known as the Cotton-York tensor, defined as in \eqref{Cotton_York}, that is,
\begin{equation}
\mathscr{C}_{ij}=\frac12\eta_i^{\hphantom{i}kl}\mathscr{C}_{klj}\,,
\end{equation}
where we recall that $\eta_{ijk}\equiv\sqrt{-g_{(0)}}\epsilon_{ijk}$\,.

The Cotton-York tensor admits a decomposition along the timelike congruence $u^i$\,, in the same way as the boundary energy-momentum tensor, as
\begin{equation}\label{Cotton_decom}
     \mathscr{C}_{ij}=\frac{c}{2\ell}\left(g^{(0)}_{ij}+3\ell^2 u_iu_j\right)-\ell c_{ij}+\ell u_ic_j+\ell u_jc_i\,.
\end{equation}
Here, $c$, $c_i$, and $c_{ij}$ correspond to the Cotton density, Cotton current, and Cotton stress tensor, respectively, and are defined as
\begin{align}
    c&=\ell^3 \mathscr{C}_{ij}u^iu^j\,,\label{Cott_dens}\\
    c_i&=-cu_i-\ell \mathscr{C}_{ij}u^j\,,\label{Cott_curr}\\
    c_{ij}&=-\frac{1}{\ell}P^{k}{}_{i}P^{l}{}_{j}\mathscr{C}_{kl}+\frac{c}{2\ell^2}P_{ij}\,.\label{Cott_stress}
\end{align}

Just as the boundary energy-momentum tensor, the Cotton-York tensor $\mathscr{C}_{ij}$ is traceless and covariantly conserved, namely
\begin{equation}\label{Tr_DivC}
    g^{ij}_{(0)}\mathscr{C}_{ij}=0 \quad\text{and}\quad\nabla^{(0)}_i\mathscr{C}^{ij}=0\,,
\end{equation}
with the difference that the above equations are off-shell geometric identities. From a gravitational perspective, the boundary Cotton tensor has been used in the literature to define magnetic charges associated with the isometries of the bulk, just as the holographic stress tensor can be used for the definition of gravitational mass and angular momentum \cite{deHaro:2007fg}.

\subsection{Covariant Newman-Unti Gauge}\label{SubSec:CovariantNU}

The hydrodynamic variables of the boundary conformal fluid, together with the boundary geometry, form a complete set of data required to reconstruct the dual four-dimensional AlAdS solutions of the vacuum Einstein equations
\begin{equation}\label{Einstein}
    \mathcal{E}_{\mu\nu}=R_{\mu\nu}-\frac12g_{\mu\nu}R-\frac{3}{\ell^2}g_{\mu\nu}=0\,,
\end{equation}
where the indices $\mu,\nu$ labels the bulk coordinates $x^\mu=\{r,x^i\}$\,, $g_{\mu\nu}$ is the metric of the bulk manifold, $R_{\mu\nu}$ is the Ricci tensor, and $R=g^{\mu\nu}R_{\mu\nu}$ is the Ricci scalar.

The reconstruction procedure is typically performed order by order in an asymptotic expansion using Eddington-Finkelstein-like coordinates, as shown in \cite{Rangamani:2009xk} for the case of AlAdS spacetimes with the boundary fluid described in the Landau-Lifshitz frame. This approach was later extended in \cite{Ciambelli:2018wre, Campoleoni:2023fug} to more general asymptotics and arbitrary hydrodynamic frames via the covariant Newman-Unti gauge. Here, we will follow the latter approach based on the covariant Newman-Unti gauge, whose advantage relies on explicit covariance with respect to the boundary structure and the possibility to take a smooth flat limit $\ell\to\infty$ of the AlAdS solution.

Following \cite{Ciambelli:2018wre, Campoleoni:2023fug}, for AlAdS bulk spacetimes in four dimensions, the covariant Newman-Unti gauge is dictated by the gauge conditions\footnote{The gauge conditions for the standard Newman-Unti gauge are given by $g_{rr}=0$, $g_{ru}=-1$ and $g_{rA}=0$, where $u$ is the retarded time while $A$ and $B$ label the angular coordinates. The covariant Newman-Unti gauge is a relaxation of the last two conditions.}
\begin{equation}\label{cNU_condi}
    g_{rr}=0 \quad\text{and}\quad g_{ri}=\ell^2u_{i}\,,
\end{equation}
and the line element reads
\begin{equation}\label{general_cNU}
    {\rm d}s^2_{\rm bulk}=2\ell^2u_{i}\text{d}x^i\text{d}r+\sum_{n\geqslant0}r^{2-n}g^{(n)}_{ij}\text{d}x^i\text{d}x^j\,,
\end{equation}
where $r$ is the holographic radial coordinate and the conformal boundary is located at $r=\infty$. In the asymptotic expansion, $g^{(0)}_{ij}$ is the metric of the conformal boundary and $g^{(3)}_{ij}$ encodes the information of the holographic energy-momentum tensor $T_{ij}$. For any $g^{(n)}_{ij}$ with $n\neq0$ and $n\neq3$, their components are determined in terms of the set of boundary data $\{g^{(0)}_{ij},u_i,\varepsilon,q_i,\tau_{ij}\}$ by using the Einstein equations. Up to ${\cal O}(1/r)$, partially solving the Einstein equations fixes
\begin{align}
    g^{(1)}_{ij}&=2\ell^2u_{(i}A_{j)}+\mathcal{C}_{ij}\,,\label{g1}
    \\
    g^{(2)}_{ij}&=\ell^4S_{ij}\,,\label{g2}
\end{align}
where $\mathcal{C}_{ij}$ is the shear of the null geodesic congruence $l^{\mu}\partial_\mu=\partial_r$. By using the Einstein equations, $\mathcal{C}_{ij}$ is identified with the shear $\sigma_{ij}$ of the boundary velocity $u^{i}$ through the following relation:\footnote{The shear $\mathcal{C}_{ij}$ is different from the one typically defined in the Bondi-Sachs coordinates. This is due to the different nature of the radial coordinate. In both gauges $\partial_r$ is tangent to a null geodesic congruence and affinely parameterized in the Newman-Unti gauge, while in the Bondi coordinates the radial coordinate is the luminosity distance (see \eqref{BondidetCond}).}
\begin{equation}\label{NU_shear}
    \frac{1}{\ell^2}\mathcal{C}_{ij}=-2\sigma_{ij}\,.
\end{equation}
In \eqref{g2}, $S_{ij}$ is a Weyl-covariant symmetric tensor of weight-0 defined as
\begin{equation}\label{SW}
    S_{ij}=2u_{(i}\mathscr{D}_k \left(\sigma^{\hphantom{j)}k}_{j)}+\omega^{\hphantom{j)}k}_{j)}\right)-\frac{1}{2}\mathscr{R} u_{i}u_{j}
+\left( \sigma_{ik}+ \omega_{ik}\right)
\left(\sigma_j^{\hphantom{j}k}+\omega_j^{\hphantom{j}k} 
\right)\,,
\end{equation}
with $\mathscr{R}$ the Weyl-covariant Ricci scalar defined in \eqref{Weyl_rel_Rscalar}.

The components of $g^{(3)}_{ij}$ are parameterized as
\begin{equation}\label{g3}
    g^{(3)}_{ij}=8\pi G\ell^2\left(\ell^2\varepsilon u_{i}u_{j}+\frac43\ell^2\Delta q_{(i}u_{j)}+\frac23\Delta\tau_{ij}\right)\,,
\end{equation}
with 
\begin{equation}
    \Delta q_{i}=q_i-\frac{1}{8\pi G}\ast\!c_i\,,\quad\text{and}\quad\Delta\tau_{ij}=\tau_{ij}+\frac{\ell^2}{8\pi G}\ast\!c_{ij}\,,
\end{equation}
where the dual operator $\ast$ (for the space transverse to $u^{i}$) acts, respectively, on vectors and rank-2 tensors as $\ast V_i=\eta^{j}_{\hphantom{j}i}V_j$ and $\ast B_{ij}=\eta^{k}_{ \hphantom{k}i}B_{kj}$\,, with $\eta_{ij}=-\ell\eta_{kij}u^{k}$. The above parameterization is such that the remaining Einstein equations are exactly the components of the divergence of $T^{ij}$, that is, the Einstein field equations are mapped to the boundary fluid equations \eqref{DT} at order $\nicefrac{1}{r^2}$. For more details on the derivation of the above, see \cite{Campoleoni:2023fug}.

\paragraph{Algebraically special Petrov type:} 

The full reconstruction of general AlAdS spacetimes in the covariant Newman-Unti gauge involves an infinite series expansion in inverse powers of the radial coordinate.
However, in certain cases, this expansion can be resummed into a closed form for the bulk metric when the boundary geometry is suitably constrained by tuning the boundary data so that the metric \eqref{general_cNU}  reduces to the line element for a subclass of solutions. The latter procedure can be achieved by appropriately engineering the properties of the Weyl tensor of the bulk manifold through the boundary Cotton-York tensor. Indeed, it was shown in \cite{Gath:2015nxa, Petropoulos:2015fba} that the necessary conditions that one must impose in the boundary configuration to obtain algebraically special Petrov type bulk solutions\footnote{These are solutions that admit at least one multiple principal null direction. In the Petrov classification, these solutions correspond to type D, II, III, or N.} are
\begin{equation}\label{resumm_condi1}
    \sigma_{ij}=0\,,\quad q_{i}=\frac{1}{8\pi G}\ast\!c_i\,,\quad\text{and}\quad\tau_{ij}=-\frac{\ell^2}{8\pi G}\ast\!c_{ij}\,.
\end{equation}

The first condition requires the boundary timelike congruence $u^i$ to be shear-free. This translates in the bulk as a shear-free condition for the null geodesic congruence $\partial_r$. The remaining two conditions impose precise relations between the dissipative sector of the holographic fluid and the geometric data encoded in the Cotton tensor: the heat current $q_i$ is set by the dual of the Cotton current \eqref{Cott_curr}, and the transverse-traceless viscous stress tensor $\tau_{ij}$ is proportional to the dual of the Cotton stress \eqref{Cott_stress}. 

As a result, once the conditions \eqref{resumm_condi1} are satisfied, the line element \eqref{general_cNU} is resummed into an exact bulk metric:
\begin{equation}\label{Ress_NU}
    {\rm d}s^2_{\text{res. bulk}}=2\ell^2u_i\text{d}x^i\text{d}r+\left(r^2g^{(0)}_{ij}+2r\ell^2u_{(i}A_{j)}+\ell^4S_{ij}+\frac{\ell^4\left(8\pi G\varepsilon r+c\gamma\right)}{r^2+\gamma^2}u_{i}u_{j}\right)\text{d}x^i\text{d}x^j\,,
\end{equation}
with 
\begin{equation}\label{gamma_1}
    \gamma^2=\frac{\ell^4}{2}\omega_{ij}\omega^{ij}\,.
\end{equation}
This provides a powerful boundary criterion for reconstructing exact bulk configurations within the Petrov algebraically special class of solutions.

\subsection{Gravitational Radiation from Fluid Dynamics}\label{SubSec:GravRadiationAndFluidDynamics}

We now have the necessary tools to analyze radiative phenomena in AlAdS spacetimes through the properties of the holographic conformal fluid by using the boundary radiative vector \eqref{Super_P1}. In what follows, we adopt an alternative expression for this vector, writing it in terms of the Cotton-York tensor and the boundary energy-momentum tensor as
\begin{equation}\label{super_P2}
    \frac{1}{16\pi G}\hat{\mathscr{P}}^{i}=\frac{1}{\ell^{2}} \mathscr{C}_{j}^{\hphantom{j}k}T_{kl}\eta^{jli}\,.
\end{equation}
Using the decompositions \eqref{fluid_stress_tensor} and \eqref{Cotton_decom} of the energy-momentum tensor $T_{ij}$ and the Cotton tensor $\mathscr{C}_{ij}$ along the congruence $u^i$, we obtain
\begin{equation}\label{SP_fluid1}
    \frac{1}{8\pi G}\hat{\mathscr{P}}^i=-\frac{3}{\ell^2}\left(\varepsilon\ast\!c^i- c\ast\! q^i\right)-\frac{2}{\ell^2}\left(\ell^2\ast\!c^{ij}q_j+\ast\tau^{ij}c_j\right)-\frac{2}{\ell}\eta^{ijk}\left(c_{j}^{\hphantom{j}l}\tau_{lk}+c_jq_k\right)\,.
\end{equation}

The above formula shows that for a general AlAdS spacetime expressed in the covariant Newman-Unti gauge \eqref{general_cNU}, gravitational radiation can be fully characterized in terms of the dual conformal fluid dynamics and boundary geometry. Moreover, as stated in \cite{Ciambelli:2024kre}, a non-vanishing $\hat{\mathscr{P}}^i$ requires the boundary to be non-conformally flat. This, in turn, implies that the bulk solution must be an AlAdS spacetime. However, not every AlAdS spacetime has a non-vanishing boundary radiative vector. A clear counterexample is provided by a bulk spacetime dual to a perfect conformal fluid living on a non-conformally flat background. In this case, the boundary data satisfies $q_i=0$, $\tau_{ij}=0$ and $\mathscr{C}_{ij}\neq0$. For this configuration, the boundary radiative vector reads 
\begin{equation}\label{P_perf}
    \frac{1}{8\pi G}\hat{\mathscr{P}}^i=-\frac{3}{\ell^2}\varepsilon\ast\!c^i\,.
\end{equation}
From \eqref{P_perf} we see that the minimum necessary condition for the solution to radiate is a non-vanishing boundary Cotton current (with non-zero energy density). In other words, if the non-conformally flat background where the perfect fluid is hosted has a vanishing $c^i$, there will be no gravitational radiation in the bulk. 

An example of the latter is the Taub-NUT-AdS spacetime. The holographic dual of this solution is a perfect fluid defined on a non-conformally flat background in which the only non-vanishing piece of the boundary Cotton-York tensor is determined by the Cotton density $c$ while the Cotton current $c_i$ and Cotton stress $c_{ij}$ vanish. Therefore, the boundary radiative vector for the Taub-NUT-AdS spacetime vanishes, implying the absence of gravitational radiation,\footnote{This is because the stronger condition \eqref{lin} is also satisfied.} a fact that is already known in the literature.

\paragraph{Algebraically special case:}

We now turn our attention to algebraically special solutions.  As we saw previously, the conditions \eqref{resumm_condi1} that are imposed on the boundary dynamics reduce the solution space of the bulk to the algebraically special class. For this class of solutions, the boundary radiative vector becomes
\begin{equation}\label{P_aesp}
    \frac{1}{8\pi G}\hat{\mathscr{P}}^i=-\frac{3}{\ell^2}\left(8\pi G\varepsilon q^i-c\ast\!q^i\right)+\frac{16\pi G}{\ell^2}\left(2\tau^{ij}q_j+u^i\left(\tau^{kl}\tau_{kl}-\ell^2q^kq_k\right)\right)\,.
\end{equation}
From the above formula, one can see that a non-vanishing $\hat{\mathscr{P}}^i$ depends fully on having a non-vanishing heat current $q^i$ and a viscous stress tensor $\tau^{ij}$. This implies that for an algebraically special type of solution to radiate, the holographic fluid on the conformal boundary must be non-perfect, leading to dissipative processes that are mapped to the bulk spacetime as gravitational radiation. Again, a concrete example of this phenomenon can be seen from the RT solution, whose dual fluid possesses a non-vanishing heat current and viscous stress tensor, and therefore radiates. Later in this work, we will see that the above observation also applies to AlAdS accelerating black holes as they fall in the algebraically special class of solutions and whose holographic dual fluid description contains dissipative corrections \cite{Hubeny:2009kz}, which indicates that accelerating black holes radiate.

\paragraph{Boundary Entropy Production:} Another important object for the description of out-of-equilibrium phenomena in conformal fluid dynamics is the entropy current. In the absence of chemical potential (and therefore without extra charged current), the entropy current at first order is canonically expressed as \cite{ISRAEL1981204,Kovtun:2012rj}
\begin{equation}\label{Entropy_curr1}
    s^i=\frac{1}{T}\left(pu^i-T^{ij}u_j\right)\,,
\end{equation}
where $T$ is the local equilibrium temperature, which is related to the local energy density through the standard Stefan's law as
\begin{equation}\label{Stefan}
    \varepsilon=\sigma T^3,
\end{equation}
with the Stefan-like constant $\sigma=\nicefrac{8\pi^2}{27G}$\,.\footnote{See \cite{Bhattacharyya:2008mz} for more details of the fluid/gravity duality dictionary.}
By using \eqref{heat_curr} and \eqref{eq_state}, the above can be recast as
\begin{equation}\label{Entropy_curr2}
    s^i = \frac{1}{T}\left(\frac{3}{2}\varepsilon u^i + q^i \right)\,.
\end{equation}
One can then define the entropy density $s$ measured by an observer with velocity $u^i$ as $s=s_iu^i$\,.

The entropy current \eqref{Entropy_curr2} is invariant under changes of the hydrodynamic frame in the absence of matter fields. Furthermore, in accordance with the second law of thermodynamics, its divergence satisfies the \textit{maximum entropy principle} 
\begin{equation}\label{entropy_production}
    \nabla^{(0)}_is^i\geqslant0\,,
\end{equation}
which quantifies the local entropy production due to dissipative processes.

To extract the information about entropy production \eqref{entropy_production}, we can use the results in \cite{Ciambelli:2024kre}. Consider the algebraically special class of solutions given by \eqref{Ress_NU}. In particular, the radiative vector \eqref{Super_P1} can be derived as the boundary value of the vector obtained by contracting the rescaled Bel-Robinson tensor directly with the spacelike normal vector $\hat{n}^\alpha$\footnote{This should not be confused with the super-Poynting vector, which is obtained by contracting with a non-spacelike vector $u^\alpha$.}
\begin{equation}
    \hat{F}^\mu=-\hat{\cal D}^\mu{}_{\alpha\beta\gamma} \hat{n}^\alpha \hat{n}^\beta \hat{n}^\gamma\,.
\end{equation}
This vector, at the boundary, is automatically covariantly conserved for vacuum Einstein equations. Its decomposition along and transverse to $\hat{n}^\alpha$ gives
\begin{equation}
  \hat{F}^\mu=Y \hat{n}^\mu+{\mathscr{Q}}^\mu,
\end{equation}
such that the boundary value of ${\mathscr{Q}}^\mu$ is precisely $\hat{\mathscr{P}}^i$, see \cite{Ciambelli:2024kre}.

In the NU gauge, we can use this fact and the asymptotic expansion of this vector to gather
\begin{equation}\label{boundary_divP}
    \hat\nabla_\mu \hat{F}^\mu \overset{\cal B}{=} 0\to\nabla^{(0)}_i\left(\hat{\mathscr{P}}^i+Y^{(0)}u^i\right)= \hat{F}^r_{(1)}\,,
\end{equation}
where we have expanded 
\begin{align}
    \hat{F}^r ={}& r^2\hat{F}_{(0)}^r + r\hat{F}^r_{(1)}+\dots \,, \\ \nonumber Y ={}& Y^{(0)} + r^{-1} Y^{(1)}+\dots\,,
\end{align}
and for the leading and subleading terms of this vector's radial component, we find
\begin{align}
    \hat{F}^r_{(0)}&=-\frac{1}{\ell^2}Y^{(0)}
    \\
    \hat{F}^{r}_{(1)} &= -\frac{16\pi G}{\ell^2}T^{i}{}_j{\mathscr{E}}^j_{(1)}{}_i - \frac{2}{\ell}\mathscr{C}^i{}_j \mathscr{M}^j_{(1)}{}_{i}\,,
\end{align}
where 
\begin{align}
    {\mathscr{E}}^\mu{}_{\nu} :={}&-\hat{W}^\mu{}_{\alpha\beta\nu}\hat{n}^\alpha \hat{n}^\beta \,= r^{-1}{\mathscr{E}}^\mu_{(0)}{}_\nu +   r^{-2}{\mathscr{E}}^\mu_{(1)}{}_\nu+ \dots\,,\\
    {\mathscr{M}}^\mu{}_{\nu} :={}&-\ast \hat{W}^\mu{}_{\alpha\beta\nu}\hat{n}^\alpha \hat{n}^\beta = r^{-1}{\mathscr{M}}^\mu_{(0)}{}_\nu + r^{-2}{\mathscr{M}}^\mu_{(1)}{}_\nu+ \dots\,,
\end{align}
are the electric and magnetic parts of the Weyl tensor, respectively. For the leading orders of the boundary components, we find
\begin{align}
    \mathscr{E}^i_{(0)}{}_j ={}& -\frac{8\pi G}{\ell^2}T^i{}_j\,,
   \qquad \mathscr{M}^i_{(0)}{}_j = -\frac{1}{\ell}\mathscr{C}^i{}_j\,. 
\end{align}
Furthermore, 
\begin{align}
    Y^{(0)} ={}& -\frac{64\pi^2 G^2}{\ell^2}T^{ij}T_{ij} - \mathscr{C}^{ij}\mathscr{C}_{ij}\,, \\ \nonumber Y^{(1)} ={}& -2\ell^2 Y^{(0)}A^i u_i - \ell^2 \hat{F}_{(1)}^r\,,
\end{align}
whose components for the algebraically special case reduce to
\begin{equation}\label{W0AlgSpe}
    Y^{(0)} = -\frac{64\pi^2 G^2}{\ell^2}\left(\frac32\varepsilon^2 + 2\tau_{ij}\tau^{ij} -4\ell^2 q_iq^i\right) -\frac{3c^2}{2\ell^2} \,.
\end{equation}

The full expressions for the subleading orders of the electric ${\mathscr{E}}^i_{(1)}{}_j$ and magnetic ${\mathscr{M}}^i_{(1)}{}_j$ components of the Weyl tensor expanded in terms of fluid variables is cumbersome and of a high algebraic complexity. This prevents us from identifying a simplified or physically insightful structure, and thus, we do not present them explicitly. Nonetheless, by expanding the remaining terms, we can isolate the divergence of the entropy current and show that $\hat{F}^r_{(1)}$ is not the sole source, as additional contributions arise from dissipative terms.

By substituting the value for the algebraically special family of solutions of the radiative vector \eqref{P_aesp} and \eqref{W0AlgSpe} into the boundary value of the divergence of the vector \eqref{boundary_divP}, we obtain
\begin{equation}\label{Sproduction}
   \boxed{ \nabla_i^{(0)}s^i =  \frac{1}{\varepsilon T}\nabla^{(0)}_i {\vartheta^i}-s^i\nabla^{(0)}_i \log(\varepsilon T) - \frac{\ell^2}{192\pi^2 G^2 }\frac{1}{\varepsilon T}\hat{F}^r_{(1)}\,,}
\end{equation}
where 
\begin{align}
    {\vartheta^i} \equiv \frac43 \tau^{ij}q_j + \frac{2}{3}\ell^2q^j q_j u^i + \varepsilon^2 u^i - \frac{c^2}{16\pi G} u^i - \frac{c}{8\pi G}\ast q^i\,.
\end{align}

This relation demonstrates that the boundary radiative vector encodes information about non-equilibrium physics at the conformal boundary.  In particular, it provides a geometric probe of energy transfer processes in the dual field theory. We will now test this criterion by applying it to exact radiative solutions. We conclude by noticing 2 aspects of our result. First, we have studied the first order expression of the entropy current in the derivative expansion. In the future, it would be interesting to study the second order expression as found in, e.g., \cite{Rezzolla:2013dea}, and its relationship to bulk radiation. Second, we imported here the results of \cite{Ciambelli:2024kre}. While this gave us a correct characterization of entropy production, it would be interesting to relate the right-hand side of \eqref{Sproduction} to the super-momentum vector defined in \eqref{susu}. We leave this for future work.

\subsubsection*{Example I: AdS Accelerating Black Holes}

The AdS C-metric is an exact vacuum solution to Einstein’s equations, representing a pair of uniformly accelerating black holes in AdS spacetime. It generalizes the AdS-Schwarzschild solution through an additional acceleration parameter, with the motion driven by an external force mediated by a cosmic string, which is a codimension-2 topological defect. In Boyer-Lindquist coordinates, the metric takes the form
\begin{equation}\label{Cmetric2}
    {\rm d}s^2 = \frac{1}{\omega^2(\bar{r},x)}\left[-f(\bar{r})\d t^2 + \frac{\d\bar{r}^2}{f(\bar{r})} + \bar{r}^2\left(\frac{\d x^2}{\Sigma(x)} + \Sigma(x) \frac{\d\phi^2}{{\cal K}^2}\right)\right]\,,
\end{equation}
with 
\begin{align}
    f(\bar{r}) ={}& \frac{\bar{r}^2}{\ell^2}+\left(1-A^2\bar{r}^2\right)\left(1-\frac{2m}{\bar{r}}\right)\,, \label{Cmetric_functions1}
    \\ 
    \omega(\bar{r},x)={}&1+A \bar{r}x\,,\qquad  \Sigma(x)= \left(1-x^2\right)\left(1+2Am x\right)\,\label{Cmetric_functions2}\,.
\end{align}
Here the conformal boundary is located at $\bar{r}_{\rm cb} = -\nicefrac{1}{Ax}$ and the angular coordinates are defined with the azimuthal angle $\phi \in [0,2\pi]$ and the polar angle $\theta = \arccos x$ such that $x \in [-1,1]$\,, and the physical parameter of the spacetime are: $m$ the mass parameter, $A$ the acceleration parameter, and $\cal K$ the parameter that governs the conical deficit induced by the acceleration which is given by the cosmic string tension. In the weak field limit ($m\to 0$), the solution reduces to the spacetime of a Rindler-accelerated observer, recovering the expected behavior for uniform acceleration in the absence of curvature singularities. 

In this spacetime, there are in total three Killing horizons, two of which are associated with a black hole’s trapped surface, and the third one is a non-compact Killing horizon that causally disconnects the two black holes. Nonetheless, for the specific parameter range where the cosmological constant satisfies $A\ell < 1$\,, the spacetime contains only a single black hole, as the Rindler horizon merges with conformal infinity, which effectively pushes the second black hole beyond its reach. This regime is commonly referred to as the \textit{slowly accelerating phase}. 

The solution is used to study the effects of acceleration on black holes, and it helps understand how cosmic strings could influence black hole dynamics. Notably, accelerating black holes with $\Lambda \geqslant 0$ have been shown to be radiative spacetimes \cite{Fernandez-Alvarez:2024bkf}. As we will demonstrate, this radiative behavior persists even in the AdS case, making these solutions particularly valuable for exploring holographic implications of radiation.\footnote{See \cite{Anabalon:2018ydc, Arenas-Henriquez:2022www, Arenas-Henriquez:2023hur, Arenas-Henriquez:2023vqp, Tian:2023ine, Roychowdhury:2024oih, Arenas-Henriquez:2024ypo, Li:2025rzl} for some holographic aspects of accelerating black holes.} Further details on their physical and geometric properties are discussed in \autoref{App:WFG}, which contains a thorough overview of accelerated black holes in AdS.

We now analyze the holographic conformal fluid using the previously established framework. To facilitate this analysis, we first express the line element in the covariant Newman-Unti gauge through the coordinate transformation
\begin{equation}\label{Trans1}
    \text{d}t=\text{d}u+\frac{1}{f(\bar{r})}\text{d}\bar{r}\,.
\end{equation}
By using \eqref{Trans1}, the line element \eqref{Cmetric2} is recast as
\begin{equation}\label{NU_1}
    {\rm d}s^2=-\frac{2}{\omega^2}\text{d}u\text{d}\bar{r}-\frac{f}{\omega^2}\text{d}u^2+\frac{\bar{r}^2}{\omega^2}\left(\frac{1}{\Sigma}\text{d} x^2+\frac{\Sigma}{\mathcal{K}^2}\text{d}\phi^2\right)\,,
\end{equation}
Additionally, to simplify the analysis at the conformal boundary, it is convenient to perform the following redefinition for the radial coordinate
\begin{equation}\label{radial_trans}
    \bar{r}=\frac{r}{1-Arx}\,,
\end{equation}
such that the conformal boundary defined by $\omega(r_{\text{cb}})=0$ is located at $r=\infty$\,. Then, the C-metric with a negative cosmological constant can be expressed in the covariant Newman-Unti gauge as
\begin{equation}\label{NU_2}
    \text{d}s^2=-2\text{d}u\text{d}r-2r^2A\text{d}u\text{d}x-\frac{f(r,x)}{\omega^2(r,x)}\text{d}u^2+r^2\left(\frac{\text{d}x^2}{\Sigma(x)}+\frac{\Sigma(x)}{\mathcal{K}^2}\text{d}\phi^2\right)\,,
\end{equation}
where now
\begin{align}\label{metric_funct2}
    f(r,x)=\omega^2\frac{r^2}{\ell^{2}}+\left(1-A^2r^2\omega^2\right)\left(1-\frac{2M}{\omega r}\right)\,,
\qquad   \omega(r,x)=\frac{1}{1-Arx}\,.
\end{align}

We can now identify the boundary quantities. The timelike conformal boundary metric appears at ${\cal O}\left(r^2\right)$ in the asymptotic expansion around $r=\infty$. That is
\begin{align}
    g^{(0)}_{ij}\text{d}x^i\text{d}x^j&=\lim_{r\to\infty}r^{-2}g_{\mu\nu}\text{d}x^{\mu}\text{d}x^{\nu}\nonumber
    \\
    &=-\frac{1}{\ell^2}\left(1-A^2\ell^2\Sigma\right)\text{d}u^2-2A\text{d}u\text{d}x+\frac{1}{\Sigma}\text{d}x^2+\frac{\Sigma}{\mathcal{K}^2}\text{d}\phi^2\,.\label{g0}
\end{align}
The boundary timelike congruence $u^i$ is encoded in the $g_{ri}$ component of the bulk metric, which reads
\begin{equation}\label{u_vec}
u^i\partial_i=\partial_u+A\Sigma\partial_x\,,
\end{equation}
and its associated 1-form is given by
\begin{equation}\label{u_bdry}
    u_i\text{d}x^i=-\frac{1}{\ell^2}\text{d}u\,,
\end{equation}
with normalization $g^{(0)}_{ij}u^ju^j=-1/\ell^2$.

The kinematics of the fluid are given by \eqref{def20}, \eqref{def21}, \eqref{def23} and \eqref{def24}. For the configuration at hand, the only non-vanishing quantity is the expansion, which reads
\begin{equation}\label{expansion}
    \Theta=2A\left(Am-3Amx^2-x\right)\,.
\end{equation}

Having the boundary metric, the Cotton-York tensor is computed using \eqref{Cott_Scho} and \eqref{Cotton_York}. For the boundary of the C-metric, we get
\begin{equation}\label{Cotton_comp}
    \mathscr{C}_{ij}=\frac{3Am\Sigma}{\mathcal{K}}\begin{pmatrix}
        0 & 0 & \frac{\left(1-A^2\ell^2\Sigma\right)}{\ell}\\
        0 & 0 & A\ell\\
        \frac{\left(1-A^2\ell^2\Sigma\right)}{\ell} & A\ell & 0
    \end{pmatrix}\,.
\end{equation}
In the decomposition \eqref{Cotton_decom}, the Cotton density $c$ vanishes, the Cotton current is
\begin{equation}\label{Cotton_vec}
    c_{i}=-\frac{3Am\Sigma}{\mathcal{K}}\delta^{\phi}_i\,,
\end{equation}
while the Cotton stress tensor is given by
\begin{equation}\label{Cotton_sress}
    c_{ij}=\frac{3A^2m\Sigma}{\mathcal{K}}\begin{pmatrix}
        0 & 0 & A\Sigma\\
        0 & 0 & -1\\
        A\Sigma & -1 & 0
    \end{pmatrix}\,.
\end{equation}

The C-metric belongs to the algebraically special type of solutions. As discussed previously, this implies that the non-perfect pieces $q^i$ and $\tau^{ij}$ of the boundary fluid stress tensor are fixed in terms of the Cotton current and Cotton stress tensor through the relation \eqref{resumm_condi1}. For the components of the heat current, we find
\begin{equation}\label{q_comp}
    q_{u}=\frac{3A^2m\Sigma}{8\pi G}\,,\quad q_{x}=-\frac{3Am}{8\pi G}\,,\quad\text{and}\quad q_{\phi}=0\,,
\end{equation}
while for the viscous stress tensor $\tau^{ij}$ we get
\begin{equation}\label{tau_comp}
    \tau_{ij}=\frac{3\ell^2A^2m}{8\pi G}\begin{pmatrix}
        A^2\Sigma^2 & -A\Sigma & 0\\
        -A\Sigma & 1 & 0\\
         0 & 0 & -\frac{\Sigma^2}{\mathcal{K}}
    \end{pmatrix}\,.
\end{equation}
The energy density $\varepsilon$ is extracted from the term at $\mathcal{O}(1/r)$ in the bulk metric expansion. By matching with \eqref{Ress_NU}, we have
\begin{equation}
\varepsilon = \frac{m}{4\pi G}\,.
\end{equation}
The constant value of $\varepsilon$ implies that the local-equilibrium fluid temperature remains constant as well. With the above, the boundary stress tensor of the fluid is constructed using \eqref{fluid_stress_tensor} and takes the following form:
\begin{align}
    T^i{}_j = \frac{m}{8\pi G}\begin{pmatrix}
        3A^2\ell^2\Sigma-2 & -3A\ell^2 & 0 \\ 0 & 1 & 0 \\ 0 & 0 & 1-3A^2\ell^2\Sigma
    \end{pmatrix}\,.
\end{align}
It can be easily checked that for the dual of the C-metric, the boundary stress tensor is traceless and conserved.

We can now compute the boundary radiative vector associated with the dual holographic fluid to the C-metric by using \eqref{P_aesp}. We get
\begin{align}\label{P_C_metric}
    \hat{\mathscr{P}}^{u}&=-18m^2A^2\Sigma\left(1-2A^2\ell^2\Sigma\right)\,,
    \\
    \hat{\mathscr{P}}^x&=\frac{18m^2A\Sigma}{\ell^2}\left(1-2A^2\ell^2\Sigma\right)\left(1-A^2\ell^2\Sigma\right)\,,
    \\
    \hat{\mathscr{P}}^{\phi}&=0\,.
\end{align}
Furthermore, we find that the holographic stress tensor and the boundary Cotton tensor are not linearly dependent. This shows that the AdS C-metric describes a radiative black hole solution for any non-zero acceleration parameter $A$ and any value of the cosmological constant using the radiation criteria of \cite{Ciambelli:2024kre, Fernandez-Alvarez:2025qqx}. Moreover, one can easily see that in the limit $A\to 0$ the solution ceases to radiate, thereby relating acceleration and radiation.

Finally, the entropy current \eqref{Entropy_curr2} for the case at hand becomes a constant
\begin{align}
    s^i = \frac{3m}{8\pi G T}\delta^i_u\,,
\end{align}
such that it is covariantly conserved and saturates \eqref{entropy_production}. 
As a consistency check, we can now evaluate \eqref{Sproduction} on the accelerating black hole background, finding
\begin{align}
    \nabla_i^{(0)} s^i = 0\,,
\end{align}
showing that there is no boundary entropy production at first order, despite the solution describing a radiative spacetime. 
We will revisit this point in the next example, where we show that the same property holds for the entire class of RT solutions.

\subsubsection*{Example II: AdS-Robinson-Trautman Spacetimes}\label{AdSRTexample}

RT spacetimes are a class of non-stationary exact solutions of Einstein’s vacuum equations, describing radiative spacetimes with applications in modeling slowly decaying black holes emitting spherical gravitational waves.\footnote{See \cite{Adami:2024mtu} for a generalization to any compact wavefront.} These solutions are characterized by a shear-free, twist-free, and expanding null geodesic congruence, and they belong to the algebraically special class of spacetimes (Petrov type II). See \cite{Stephani:2003tm} for a detailed review on this topic.

The AdS extension of RT solutions has provided key insights into the holographic duality between gravitational radiation in the bulk and non-equilibrium dynamics in the boundary CFT \cite{Bakas:2014kfa, BernardideFreitas:2014eoi, Skenderis:2017dnh, Ciambelli:2017wou, Fareghbal:2018xii, Ciambelli:2024kre}. The line element of the RT-AdS metric in Newman-Unti coordinates takes the form
\begin{align}\label{RT_AdS_lineelement}
    \d s^2 = -2\d u\d r - F(r,u,x^A)\d u^2 + 2r^2 e^{\Phi(u,x^A)}\d\zeta\d\bar\zeta\,, 
\end{align}
with
\begin{align}\label{FAdS_RT}
    F(r,u,x^A) = r\dot\Phi - \Delta \Phi - \frac{2m}{r} +\frac{r^2}{\ell^2}\,,
\end{align}
where $x^A=\{\zeta,\bar{\zeta}\}$ are the holomorphic coordinates of the constant $u,r$ hypersurface, and $\Phi(u,x^A)$ is an arbitrary function related to the wavefront, referred to as the RT field. Here, we have defined $\dot\Phi \equiv \partial_u\Phi$ and the two-dimensional Laplacian $\Delta \equiv e^{-\Phi}\partial_\zeta\partial_{\bar\zeta}$. After fixing a reparametrization ambiguity, $m$ is a constant that can be related to the ADM energy at late times. The Einstein equations constrain $\Phi$ to satisfy a parabolic fourth-order nonlinear differential equation on the transverse plane:
\begin{align}\label{RT_equation}
    \Delta\Delta\Phi + 3m\dot\Phi = 0\,.
\end{align}
This is known as the RT equation, which is independent of the value of the cosmological constant. 

Let us now apply the formalism developed in the previous section to find the radiative vector and check if there is boundary entropy production. We find that the only non-trivial kinematic fluid quantity is again the expansion, given by
\begin{align}
    \Theta = \dot\Phi\,.
\end{align}

The boundary metric is given by 
\begin{align}
    g^{(0)}_{ij}\d x^i \d x^j = -\frac{1}{\ell^2}\d u^2 + 2e^{\Phi}\d\zeta\d\bar\zeta\,, 
\end{align}
with Cotton-York tensor
\begin{align}\label{RT_Cotton}
    \mathscr{C}_{ij} = \frac{i}{2\ell}\begin{pmatrix}
        0 & \Delta\partial_\zeta\Phi -\partial_\zeta\Delta\Phi &  \partial_{\bar\zeta}\Delta\Phi-\Delta\partial_{\bar\zeta}\Phi  \\ \Delta\partial_\zeta\Phi -\partial_\zeta\Delta\Phi & \ell^2 \left(\partial_{\zeta}^2\dot\Phi - \partial_\zeta\dot\Phi \partial_\zeta\Phi\right) & 0 \\  \partial_{\bar\zeta}\Delta\Phi-\Delta\partial_{\bar\zeta}\Phi & 0 & \ell^2\left(\partial_{\bar\zeta}\Phi\partial_{\bar\zeta}\dot\Phi - \partial^2_{\bar\zeta}\dot\Phi\right)
    \end{pmatrix}\,.
\end{align}
In the Cotton decomposition \eqref{Cotton_decom}, the Cotton density $c$ vanishes and we find
\begin{align}
    c_u ={}&0 \,, \\c_\zeta ={}& \frac{i}{2}\left(\partial_\zeta\Phi \Delta\Phi - \Delta\partial_{\zeta}\Phi \right)\,, \\ c_{\bar\zeta} ={}& -\frac{i}{2}\left(\partial_{\bar\zeta}\Phi \Delta\Phi - \Delta\partial_{\bar\zeta}\Phi \right)\,,
\end{align}
for the Cotton current, and
\begin{align}
    c_{ij} = \frac{i}{2}\text{diag}\left[0,\partial_{\zeta}\Phi\partial_{\zeta}\dot\Phi - \partial^2_\zeta \Phi, \partial_{\bar\zeta}\Phi\partial_{\bar\zeta}\dot\Phi - \partial^2_{\bar\zeta} \Phi\right]\,,
\end{align}
for the Cotton stress tensor. Now, as the solution belongs to the algebraically special type of solutions, the existence of the nontrivial Cotton current and stress tensor implies that the boundary fluid suffers from dissipation, and there is a nontrivial heat current
\begin{align}
    q^u ={}& 0\,,\\ q^\zeta ={}& \frac{e^{-\Phi}}{16\pi G}\left(\Delta\partial_{\bar\zeta}\Phi - \partial_{\bar\zeta}\Phi\Delta\Phi\right)\,, \\ q^{\bar\zeta} ={}& \frac{e^{-\Phi}}{16\pi G}\left(\Delta\partial_{\zeta}\Phi - \partial_{\zeta}\Phi\Delta\Phi\right)\,,
\end{align}
and a non-trivial viscous stress tensor
\begin{align}
    \tau_{ij} = \frac{\ell^2}{16\pi G}\text{diag}\left[0,\partial_\zeta\Phi\partial_\zeta\partial_{\bar\zeta}\Phi - \partial_\zeta^2\dot\Phi ,  \partial_{\bar\zeta}\Phi \partial_{\bar\zeta}\dot\Phi - \partial_{\bar\zeta}\dot\Phi \right]\,.
\end{align}
The energy density can be read off from the metric and is given by
\begin{align}
    \varepsilon = \frac{m}{4\pi G}\,.
\end{align}

Finally, the holographic stress tensor of the dual fluid is given by
\begin{align}\label{RT_Stress}
    16\pi G T_{uu} ={}& \frac{4m}{\ell^2}\,, \\ 16\pi G T_{uA} ={}& \partial_A\Phi\Delta\Phi-\Delta\partial_A\Phi  \\ 16\pi G T_{\zeta\zeta} ={}& \ell^2\left(\partial_\zeta\Phi\partial_\zeta\dot\Phi - \partial_\zeta^2\dot\Phi\right) \,,\\ 16\pi G T_{\bar\zeta\bar\zeta} ={}& \ell^2\left(\partial_{\bar\zeta}\Phi \partial_{\bar\zeta}\dot\Phi - \partial_{\bar\zeta}^2\dot\Phi \right)\,, \\16\pi G T_{\zeta\bar\zeta} ={}& 2m e^\Phi\,.
\end{align}

We can now compute the boundary radiative vector by using \eqref{P_aesp}. We get
\begin{align}
    \hat{\mathscr{P}}^u ={}& e^{-2\Phi}\left[\ell^2 \left(\partial_{\bar\zeta}\Phi\partial_{\bar\zeta}\dot\Phi - \partial_{\bar\zeta}^2\dot\Phi\right)\left(\partial_\zeta\Phi\partial_\zeta\dot\Phi - \partial_\zeta^2\dot\Phi\right)- \left(\partial_{\bar\zeta}\Phi\Delta\Phi - \Delta\partial_{\bar\zeta}\Phi\right)\left(\partial_{\zeta}\Phi\Delta\Phi - \Delta\partial_{\zeta}\Phi\right)\right]\,, \\ \hat{\mathscr{P}}^{\zeta}={}& e^{-2\Phi}\left[\frac{3m}{\ell^2}e^\Phi\left(\partial_{\bar\zeta}\Phi\Delta\Phi - \Delta\partial_{\bar\zeta}\Phi\right) - \left(\Delta\partial_\zeta\Phi - \partial_\zeta\Phi\Delta\Phi\right)\left(\partial_\zeta\Phi\partial_{\bar\zeta}\dot\Phi - \partial^2_{\bar\zeta}\dot\Phi\right)\right]\,,   \\ \hat{\mathscr{P}}^{\bar\zeta}={}& e^{-2\Phi}\left[\frac{3m}{\ell^2}e^\Phi\left(\partial_{\zeta}\Phi\Delta\Phi - \Delta\partial_{\zeta}\Phi\right) - \left(\Delta\partial_{\bar\zeta}\Phi - \partial_{\bar\zeta}\Phi\Delta\Phi\right)\left(\partial_{\bar\zeta}\Phi\partial_{\zeta}\dot\Phi - \partial^2_{\zeta}\dot\Phi\right)\right]\,.
\end{align}
This indicates that the RT-AdS solutions radiate provided that any component of the boundary radiative vector is non-vanishing and/or that its Cotton tensor \eqref{RT_Cotton} is linearly independent of its holographic stress tensor \eqref{RT_Stress}; that is, one can have a vanishing radiative vector but \eqref{lin} may still not hold, leading to radiation.

Finally, the first order entropy current for the RT family reads
\begin{align}
    s^u ={}& \frac{3m}{8\pi G T}\,\\ s^\zeta ={}& \frac{e^{-\Phi}}{16\pi G T}\left(\Delta\partial_{\bar\zeta}\Phi - \partial_{\bar\zeta}\Phi\Delta\Phi\right)\,, \\ s^{\bar\zeta} ={}& \frac{e^{-\Phi}}{16\pi G T}\left(\Delta\partial_{\zeta}\Phi - \partial_{\zeta}\Phi\Delta\Phi\right)\,,
\end{align}
and taking its divergence, one finds
\begin{align}
    \nabla_i^{(0)}s^i = \frac{1}{8\pi GT }\left(\Delta\Delta\Phi + 3m\dot\Phi\right)\,,
\end{align}
which vanishes provided that the RT field satisfies the RT equation \eqref{RT_equation}. As a consistency check, if we use our relation for boundary entropy production \eqref{Sproduction}, one gets
\begin{align}
    \nabla^{(0)}_i s^i = 0\,.
\end{align}

The lack of first order boundary entropy production for the dual fluid of the RT-AdS class is already discussed in \cite{Ciambelli:2017wou}. The existence of radiation in the bulk suggests that the boundary is not at equilibrium, but this does not imply that there is entropy production in the dual fluid if the kinematical out-of-equilibrium quantities are chosen to coincide with their thermodynamic local-equilibrium counterparts.\footnote{As demonstrated in \cite{Becattini:2023ouz}, this property persists even in the presence of spin currents, with the result being exact as well. The entropy production vanishes when the conserved current remains unchanged from its local equilibrium value. This is a condition whose definition depends on the choice of hydrodynamic frame.} As shown in \cite{Ciambelli:2017wou}, one can understand this as a thermodynamic Moutier isothermal path with the irresistibility process being of a geometric nature. 

Note that, on the other hand, the task of defining an entropy current for a relativistic fluid is subtle, and its interplay with fluid frame covariance is delicate. Here, we confined our attention to the first-order entropy current \eqref{Entropy_curr1}. This differs from the analysis in \cite{Bakas:2014kfa,Skenderis:2017dnh}, in which the dual fluid of RT-AdS solutions exhibits boundary entropy production. While the analysis in these papers is performed in the Landau-Lifshitz frame, one would need to go to higher orders in the derivative expansion to compare the entropy currents. Another subtlety in comparing these results is that  \cite{Skenderis:2017dnh} considers linear perturbations to the Schwarzschild-AdS black hole that are required to satisfy the RT equation. These metric perturbations induce a gradient expansion in the boundary fluid, with the boundary tensors computed only perturbatively in this expansion. This contrasts with our exact results, derived in a generic hydrodynamic frame. We currently do not have an example of a bulk vacuum exact solution whose dual fluid has entropy production at first order, but we are investigating various concrete possibilities. One should note that the first order hydrodynamic expansion is generally parabolic and thus not causal, unless we are in special hydrodynamic frames, such as BDNK --- see \cite{Ciambelli:2023mvj} for a holographic analysis. Therefore, as already stated, an avenue worth-pursuing would be to push the analysis to higher orders in the hydrodynamic expansion.

The main takeaway message of this section is the understanding of gravitational radiation in AlAdS spaces as sourcing dissipation in the dual boundary fluid, and its subtle link with the fluid entropy current. Our findings are supported by non-trivial checks in radiative exact bulk solutions.


\section{Flat Limits}\label{Sec:Flat_lim}
Asymptotically flat spacetimes can be obtained from asymptotically AdS geometries by taking the limit $\ell\to\infty$ in a suitable coordinate system. This procedure has been mostly implemented in two gauges. First, in the covariant extension of the Newman-Unti gauge \cite{Ciambelli:2018wre, Campoleoni:2023fug}. Second, in the generalization of Bondi-Sachs coordinates to non-zero cosmological constant \cite{Compere:2019bua}, known as the $\Lambda$-{BMS} gauge. In the flat limit, these two gauges yield different intrinsic characterizations of asymptotically flat spacetimes and their boundary structures, which may play complementary roles in the formulation of flat holography.
The flat limit of the $\Lambda$-BMS gauge makes the asymptotic symmetries and radiative data at null infinity manifest, providing a natural framework to explore aspects of celestial holography. This includes the derivation of soft theorems, the analysis of memory effects, and the characterization of celestial operators defined on the sphere at null infinity.
In contrast, the covariant Newman-Unti gauge for Ricci-flat spacetimes exhibits manifest covariance under the Carrollian geometry induced at null infinity. This makes it especially well-suited for the Carrollian approach to flat holography, where the boundary geometry and the dynamics of the Carrollian field theory are naturally expressed in hydrodynamic language. Despite their differing formulations, these two approaches are deeply connected, as it has been shown \cite{Donnay:2022wvx, Donnay:2022aba,Fiorucci:2025twa} that the flux-balance laws of BMS symmetries can be derived from sourced Ward identities of a conformal Carrollian field theory after taking an appropriate integral transform.

In what follows, we first use the covariant Newman-Unti gauge and the machinery developed in \cite{Ciambelli:2018wre, Campoleoni:2023fug} to interpret gravitational radiation in terms of Carrollian fluid dynamics through a Carrollian decomposition of \eqref{P_aesp} for the algebraically special subclass. Later, we employ the $\Lambda$-BMS gauge to extract celestial operators from the smooth $\ell \to \infty$ limit of the radiative AlAdS spacetimes. We show an explicit example using the AdS C-metric by mapping the solution in the Weyl-Fefferman-Graham (WFG) gauge (see \autoref{App:WFG}) to the Bondi gauge in an asymptotic expansion.

\subsection{Ricci-Flat Spacetimes and Boundary Carrollian Fluids}\label{SubSec:Carroll}

As shown in \cite{Ciambelli:2018wre,Ciambelli:2020eba, Ciambelli:2020ftk,Campoleoni:2023fug}, one can obtain Ricci-flat spacetimes by considering the flat limit of \eqref{general_cNU} for the general case or the flat limit of \eqref{Ress_NU} for the algebraically special class of solutions, which will be the case we consider in this section. To that end, it is convenient to parametrize the boundary geometry \textit{\`a la} Papapetrou-Randers
\begin{equation}\label{PR_gauge}
g^{(0)}_{ij}\text{d}x^i\text{d}x^j=-\frac{1}{\ell^2}\left(\Omega\text{d}u-b_{A}\text{d}x^A\right)^2+a_{AB}\text{d}x^A\text{d}x^{B}\,,
\end{equation}
where the functions $\Omega$\,, $b_A$\,, and $a_{AB}$ depend on the coordinates $\{u,x^A\}$ with $x^A$ labeling the transverse directions (e.g., the angles of the sphere) and $u$ the retarded time. These functions are also chosen to be independent of the AdS radius.

The above metric has the advantage of being covariant under Carrollian diffeomorphisms given by the transformations $u'=u'(u,\textbf{x})$ and $\textbf{x}'=\textbf{x}'(\textbf{x})$. Under Carrollian diffeomorphisms, $\Omega$ transforms as a scalar, $b_{A}$ transforms as a connection, and $a_{AB}$ transforms as a rank-two tensor corresponding to the metric of the transverse section.

For simplicity, we also consider the boundary velocity $u^i$ to be at rest, which in this parameterization reads as
\begin{equation}\label{u_PR}
    u^i\partial_i=\frac{1}{\Omega}\partial_u\,,\quad\text{and}\quad u_i\text{d}x^i=-\frac{1}{\ell^2}\left(\Omega\text{d}u-b_{A}\text{d}x^A\right)\,,
\end{equation}
with norm given by
\begin{equation}\label{normu}
    g^{(0)}_{ij}u^iu^j=-\frac{1}{\ell^2}\,.
\end{equation}
One can see that at $\mathcal{O}(r^2)$ in the asymptotic expansion of \eqref{Ress_NU}, the flat limit of $g^{(0)}_{ij}$ leads to a degenerate metric
\begin{align}\label{degen_metric}
    \text{d}s^2_{\text{Carr}}=\lim_{\ell\to\infty}g^{(0)}_{ij}\text{d}x^i\text{d}x^j=0\cdot\text{d}u^2+a_{AB}\text{d}x^A\text{d}x^B\,,
\end{align}
whose kernel is generated by the vector field
\begin{align}\label{Carr_vec}
    \upupsilon =\lim_{\ell\to\infty}u^{i}\partial_i=\frac{1}{\Omega}\partial_u\,,
\end{align}
which is known in the literature as the \textit{Carroll vector} or \textit{field of observers}. This Carroll vector has a dual 1-form given by
\begin{align}\label{clock_form}
    \upmu =\lim_{\ell\to\infty}\ell^2u_{i}\text{d}x^i =-\Omega\text{d}u+b_A\text{d}x^A\,,
\end{align}
such that $\upupsilon(\upmu)=-1$, which is typically known as the \textit{clock-form}, where $b_A$ serves as an \textit{Ehresmann connection} \cite{Ciambelli:2019lap}. 
Therefore, a flat limit in the bulk acts as an ultralocal limit in the boundary geometry, leading to a Carroll structure consisting of a $2+1$-dimensional manifold $\mathcal{M}=\mathbb{R}\times\mathcal{S}$, endowed with a degenerate metric \eqref{degen_metric} and a Carroll vector \eqref{Carr_vec}. 

A strong Carroll structure is also accompanied by an ambiguous connection that is metric-compatible and Carroll vector-compatible and respects the time/space splitting used throughout this work. Here, we follow the time/space splitting formalism of \cite{Ciambelli:2018wre,Ciambelli:2019lap, Campoleoni:2023fug} to define the spatial and temporal Carroll-covariant derivatives $\{\tilde{\nabla}_A,\frac{1}{\Omega}\tilde{D}_u\}$, as well as their Weyl-Carroll covariant versions $\{\tilde{\mathscr{D}}_A,\frac{1}{\Omega}\tilde{\mathscr{D}}_u\}$, which are defined in \eqref{Carr_covD_space}, \eqref{Carr_covD_time}, \eqref{WC_covD_spatial} and \eqref{WC_covD_time}, respectively. The latter acts on Carrollian scalars $\Phi$ and vectors\footnote{In the time/space splitting formalism, vectors are defined with respect to the base space $\mathcal{S}$ and their indices are raised and lowered with the two-dimensional metric $a_{AB}$.} $V^A$ while maintaining their transformation properties under Carrollian diffeomorphism and Weyl rescaling. For more details on the Carroll structure considered here and its corresponding geometric quantities defined in the time/space splitting formalism, see \autoref{App:Carroll_boundary}. Here, since we are interested in null infinity, we keep the time/space split and restrict our attention to Carrollian diffeomorphisms. Comparing with the diffeomorphism-covariant Carrollian analysis of \cite{Freidel:2022bai, Freidel:2022vjq, Ciambelli:2023mir, Ciambelli:2023mvj, Ciambelli:2025mex}, which is performed in abstract index notation, one must require that the adapted horizontal frames (defined in the frame bundle) are actually coordinates on the Carrollian manifold, breaking therefore diffeomorphisms down to Carrollian diffeomorphisms. Then, the Carrollian connection discussed in these works and thoroughly analyzed in \cite{Ciambelli:2023xqk} becomes the connection used in this manuscript once projected to the horizontal sub-bundle and converted into frame bundle indices. \

Returning to the flat limit of \eqref{Ress_NU}, the rest of the components of the bulk metric are found by taking the flat limit at different orders in the asymptotic expansion. For the $(u,r)$ component, one finds
\begin{equation}
    \lim_{\ell\to\infty}2g_{ur}\text{d}u\text{d}r=2\upmu\text{d}r\,.
\end{equation}
for all orders in $r$. At $\mathcal{O}(r)$ in the expansion of $g^{(n)}_{ij}$ one has
\begin{equation}\label{Carr_g1}
    \lim_{\ell\to\infty}g^{(1)}_{ij}\text{d}x^i\text{d}x^j=-\theta\upmu^2+2\varphi_A\upmu\text{d}x^A\,,
\end{equation}
where $\theta$ is the Carrollian expansion \eqref{Carr_expan}, and $\varphi_A$ is the Carrollian acceleration \eqref{vort_acc}. For the next order in the expansion, one gets
\begin{equation}\label{Carr_g2}
    \lim_{\ell\to\infty}g^{(2)}_{ij}\text{d}x^i\text{d}x^j=-\frac{1}{2}\tilde{\mathscr{R}}\upmu^2-2\ast\!\tilde{\mathscr{D}}_A\ast\!\varpi\upmu\text{d}x^A+\ast\!\varpi^2a_{AB}\text{d}x^A\text{d}x^B\,,
 \end{equation}
where $\tilde{\mathscr{R}}$ is the Weyl-Ricci-Carroll scalar \eqref{WeylRicc}, and $\ast\varpi$ is the dual of the Carrollian vorticity $\varpi_{AB}$ defined in \eqref{vort_acc} and \eqref{dual_Carr_vort}. Here, the 2-dimensional Hodge dual $\ast$ is defined in \eqref{2dHodge_dual}.

The rest of the components $g^{(n)}_{ij}$ for $n\geqslant3$ involve the energy density $\varepsilon$ and the Cotton density $c$, which are pieces of $T^{ij}$ and $\mathscr{C}^{ij}$. Thus, we need to know the behavior of the components of $T^{ij}$ and $\mathscr{C}^{ij}$ with respect to the AdS radius $\ell$. 

The expansion of the Cotton pieces is exact, and for the configuration at hand, is given by
\begin{align}
     c&=c_{(0)}+\frac{1}{\ell^2}c_{(1)}\,,\label{c_exp}
     \\
      c^A&=\chi^A+\frac{1}{\ell^2}\psi^A\,,\label{ci_exp}
     \\
     c^{AB}&=X^{AB}+\frac{1}{\ell^2}\Psi^{AB}\,,\label{cij_exp}
\end{align}
with each piece given in \eqref{c1}, \eqref{c0}, \eqref{psi-Carroll}, \eqref{chi-Carroll}, \eqref{Psi-Carroll} and \eqref{X-Carroll}.\footnote{In general, more terms appear in the expansion of the Cotton pieces, but those are built in terms of the Carrollian shear, which vanishes for the configuration dual to all Ricci-flat spacetimes. For more details, see \cite{Campoleoni:2023fug}.}

For the algebraically special case, the expansion of the boundary energy-momentum tensor $T^{ij}$ in powers of $\ell$ has been obtained in \cite{Ciambelli:2018wre,Campoleoni:2023fug}. Its components are given by
\begin{align}
\varepsilon&=\varepsilon_{(0)}\,,\label{e_expanded}
    \\
    q^{A}&=Q^A+\frac{1}{\ell^2}\pi^A\,,\label{q_expanded}
    \\
    \tau^{AB}&=-\ell^2\Sigma^{AB}-\Xi^{AB}\,,\label{tau_expanded}
\end{align}
where $\varepsilon_{(0)}$ is the energy density of the dual Carrollian fluid, $Q^A$ and $\pi^A$ are two Carrollian heat currents (Carroll energy flux and momentum, respectively), and $\Sigma^{AB}$ and $\Xi^{AB}$ are the Carroll viscous stress tensors. The dissipative terms are fixed by some pieces of the Carrollian Cotton tensors as\footnote{The functions $Q^A$, $\Sigma^{AB}$ and $\Xi^{AB}$ are determined by imposing finiteness conditions in the flat limit of \eqref{general_cNU}, while $\pi^A$ is fixed as part of the constraints required to restrict the bulk solution space to the algebraically special class. For a detailed derivation, see \cite{Campoleoni:2023fug}.}
\begin{equation}\label{Carroll_heat}
    Q^A=\frac{1}{8\pi G}\ast\!\chi^A\,,\qquad\pi^A=\frac{1}{8\pi G}\ast\!\psi^A\,,
\end{equation}
\begin{equation}\label{Carroll_viscousstress}
    \Sigma^{AB}=\frac{1}{8\pi G}\ast\!X^{AB}\,,\qquad\Xi^{AB}=\frac{1}{8\pi G}\ast\!\Psi^{AB}\,,
\end{equation}
while the Carroll energy density $\varepsilon_{(0)}$ remains arbitrary.

Using the above, the flat limit of $g^{(3)}_{ij}$ and $g^{(4)}_{ij}$ gives
\begin{align}\label{Carr_g3}
    \lim_{\ell\to\infty}g^{(3)}_{ij}\text{d}x^i\text{d}x^j&=8\pi G\varepsilon_{(0)}\upmu^{2}\,,\qquad\lim_{\ell\to\infty}g^{(4)}_{ij}\text{d}x^i\text{d}x^j=\ast\varpi c_{(0)}\,,
\end{align}
while the rest of the components $g^{(n)}_{ij}$ that appear deeper in the asymptotic expansion are expressed in terms of $\varepsilon_{(0)}$, $c_{(0)}$ and different powers of $\ast\varpi$ (see \cite{Mittal:2022ywl} for the subleading terms).

The line element for a Ricci-flat spacetime in this gauge can be resummed in an exact form for the case at hand (algebraically special case). The latter reads
\begin{align}\label{flatRess_cNU}
    \text{d}s^2&=2\upmu\text{d}r+\left(r^2+\ast\varpi^2\right)a_{AB}\text{d}x^A\text{d}x^B-r\left(\theta\upmu-2\varphi_A\text{d}x^A\right)\upmu\nonumber
    \\
    &-\left(\frac{\upmu}{2}\tilde{\mathscr{R}}+2\ast\!\tilde{\mathscr{D}}_A\ast\varpi\text{d}x^A\right)\upmu+\frac{8\pi G\varepsilon_{(0)}r+\ast\varpi c_{(0)}}{r^2+\ast\varpi^2}\upmu^2\,,
\end{align}
where the complete solution space for algebraically special spacetimes is given by the Carrollian geometry of null infinity $\{\Omega,b_{A},a_{AB}\}$ and the energy density $\varepsilon_{(0)}$ of the holographic Carrollian fluid. 

The vacuum Einstein equations evaluated on \eqref{flatRess_cNU} are translated to Carrollian fluid equations that constrain the time and space evolution of $\varepsilon_{(0)}$ as \cite{Ciambelli:2018xat,Petkou:2022bmz, Ciambelli:2018wre,Mittal:2022ywl,Campoleoni:2023fug}
\begin{align}
    \frac{1}{\Omega}\tilde{\mathscr{D}}_u\varepsilon_{(0)}+\frac{1}{8\pi G}\tilde{\mathscr{D}}_A\ast\!\chi^A&=0\,,\label{Carroll_uevo}
    \\
    \tilde{\mathscr{D}}_A\varepsilon_{(0)}-\frac{1}{8\pi G}\ast\!\tilde{\mathscr{D}}_{A}c_{(0)}&=0\,.\label{Carroll_Aevo}
\end{align}

\subsection{Holographic Carrollian Radiative Vector}\label{SubSec:CarrollianPoynting}

The idea now is to extract information about gravitational radiation in asymptotically flat spacetimes from its conformal boundary by building a Carroll analogue of the boundary radiative vector involving the dynamics of the dual Carrollian fluid, just as done for the AdS case in relation to its holographic relativistic fluid. Here we focus on the case of algebraically special Ricci-flat spacetimes given by \eqref{flatRess_cNU}, whose dual describes general, potentially non-perfect, Carrollian fluids.

We start with the boundary radiative vector of algebraically special AlAdS spacetimes given by \eqref{P_aesp} and consider its expansion in powers of the AdS radius. We take advantage of the Papapetrou-Randers parameterization of the boundary metric \eqref{PR_gauge} where, for a given vector $V^{i}$ with indices $i, j=\{u,A\}$, the $V_{u}$ component transforms as a Carrollian scalar and $V^A$ transforms as a Carrollian vector. 

In the above decomposition, from the temporal part, we find

\begin{equation}\label{Carr_exp_Pu}
    \frac{1}{\Omega}\hat{\mathscr{P}}_u=\hat{\varrho}+\frac{1}{\ell^2}\hat{\varrho}_{(1)}+\frac{1}{\ell^4}\hat{\varrho}_{(2)}+\frac{1}{\ell^6}\hat{\varrho}_{(3)}\,,
\end{equation}
where the leading term gives the Carrollian scalar,
\begin{equation}\label{Carr_Pu0}
\boxed{
    \hat{\varrho}=-128\pi^2 G^2\Sigma^{AB}\Sigma_{AB}\,.}
\end{equation}
For the rest of the subleading pieces in the expansion of $\hat{\mathscr{P}}_u$, we obtain
\begin{align}
    \hat{\varrho}_{(1)}&=128\pi^2 G^2\left(Q^AQ_A-2\Sigma_{AB}\Xi^{AB}\right)\,,
    \\
    \hat{\varrho}_{(2)}&=128\pi^2 G^2\left(2Q^A\pi_A-\Xi^{AB}\Xi_{AB}\right)\,,
    \\
    \hat{\varrho}_{(3)}&=128\pi^2 G^2\pi^A\pi_A\,.
\end{align}
In the same way, we expand $\hat{\mathscr{P}}^A$ in terms of the AdS radius. We find
\begin{equation}\label{Carr_exp_PA}
    \hat{\mathscr{P}}^A=\hat{\Upsilon}^A+\frac{1}{\ell^2}\hat{\Upsilon}^A_{(1)}+\frac{1}{\ell^4}\hat{\Upsilon}^A_{(2)}+\frac{1}{\ell^6}\hat{\Upsilon}^A_{(3)}\,,
\end{equation}
where the leading term in the expansion gives
\begin{equation}\label{Carr_PA0}
\boxed{
    \hat{\Upsilon}^A=-256\pi^2 G^2\Sigma^{AB}Q_{B}\,.}
\end{equation}
The rest of the terms $\Upsilon^A_{(n)}$ are found as
\begin{align}
    \hat{\Upsilon}^A_{(1)}&=-24\pi G\left(8\pi G\varepsilon_{(0)} Q^A-c_{(0)}\ast\!Q^A\right)-256\pi^2 G^2\left(\Sigma^{AB}\pi_B+\Xi^{AB}Q_{B}\right)\,,
    \\
     \hat{\Upsilon}^A_{(2)}&=-24\pi G\left(8\pi G\varepsilon_{(0)} \pi^A-c_{(0)}\ast\!\pi^A-c_{(1)}\ast\!Q^A\right)-256\pi^2 G^2\Xi^{AB}\pi_B\,,
     \\
     \hat{\Upsilon}^A_{(3)}&=-24\pi Gc_{(1)}\ast\!\pi^A\,.
\end{align}
From the above expressions, we see that a flat limit of the boundary radiative vector dual to asymptotically AdS spacetimes leads to a pair of finite contributions expressed in terms of Carrollian boundary data $\{\hat{\varrho},\hat{\Upsilon}^A\}$. We call $\hat{\varrho}$ and $\hat{\Upsilon}^A$ the \textit{Carroll-radiative scalar} and \textit{Carroll-radiative vector}, respectively. The latter gives a criterion for the presence of gravitational radiation in Ricci-flat spacetimes based on the Carrollian boundary fluid dynamics. Indeed, in order for the bulk solution to radiate, the holographic dual must be a non-perfect Carrollian fluid with a non-vanishing viscous stress tensor $\Sigma^{AB}$ and a heat current $Q^A$. Moreover, the minimum requirement for a non-vanishing $\hat{\varrho}$ and $\hat{\Upsilon}^A$ is to have $\Sigma^{AB}\neq0$. The rest of the dissipative terms, including the second heat current $\pi^{A}$ and the viscous stress tensor $\Xi^{AB}$, are not related to radiation phenomena in the bulk. This is an important turning point in understanding the physics behind these Carrollian data at the boundary.

Note that, as shown in \cite{Ciambelli:2024kre}, the flat limit of the radiative vector lands exactly on the super-Poynting vector for flat space. Here, compared to that analysis in Bondi-Sachs coordinates, it can be seen that the viscous stress tensor $\Sigma^{AB}$ and the heat current $Q^A$ capture the information of the news tensor $N_{AB}$ (defined in \autoref{SubSec:LBMS_Cel_Op}) through its time and spatial derivatives. Indeed, for an asymptotically locally AdS spacetime in Bondi-Sachs coordinates \eqref{LBMSgauge} where we fixed $q^{(0)}_{AB}$ to be the metric of the unit sphere and $\beta_0=U^0_{A}=0$, the components of the boundary super-Poynting vector are given in \eqref{P_expBondi}, which are expressed in terms of a power expansion of the cosmological constant. Therefore, in the flat limit, the relation between  \eqref{Carr_Pu0}, \eqref{Carr_PA0}, and the expression \eqref{P_expBondi} below, suggests the following identification:
\begin{equation}\label{relation_PNU_PBondi}
    \Sigma_{AB}\sim \partial_u N_{AB}\,,\qquad Q_A\sim D^{(0)}_B N^{B}_{\hphantom{B}A}\,.
\end{equation}
This gives an interesting interpretation of the radiative properties of the spacetime captured by the news tensor $N_{AB}$ in terms of Carrollian boundary data. It also makes manifest that a vanishing news tensor implies the absence of radiation.

To clarify the role of the dissipative terms of the holographic Carrollian fluid with respect to gravitational radiation in the bulk, we will now consider two examples.

\subsubsection*{Example I: Kerr-Taub-NUT Spacetimes}

Consider the case for a type D stationary Ricci-flat spacetime with rotation and gravitomagnetic monopole. This corresponds to the Kerr-Taub-NUT black hole solution whose line element in the covariant Newman-Unti gauge \eqref{flatRess_cNU} reads
\begin{align}\label{Kerr_TN}
    \text{d}s^2={}&-2\left(\text{d}u-b_{\phi}\text{d}\phi\right)(\text{d}r-a\sin^2\theta\text{d}\phi)+\left(r^2+\left(n-a\cos\theta\right)^2\right)(\text{d}\theta^2+\sin^2\theta\text{d}\phi^2)\nonumber
    \\
    &+\left(\frac{2mr+2n\left(n-a\cos\theta\right)}{r^2+\left(n-a\cos\theta\right)^2}-1\right)\left(\text{d}u-b_{\phi}\text{d}\phi\right)^2\,,
\end{align}
where $m$ is the mass parameter, $a$ is the rotational parameter, $n$ is the NUT charge, also known as the magnetic mass parameter, and 
\begin{equation}\label{b_phi}
    b_{\phi}=2n\left(1-\cos\theta\right)-a\sin^2\theta\,.
\end{equation}

The Carroll structure at null infinity is given by the degenerate metric
\begin{equation}\label{boundary_KTN}
    \text{d}s^2_{Carr}=0\cdot\text{d}u^2+\text{d}\theta^2+\sin^2\theta\text{d}\phi^2\,,
\end{equation}
with the Carroll vector
\begin{equation}\label{Carr_Vec_KTN}
    \upupsilon=\partial_u
\end{equation}
and dual clock form 
\begin{equation}\label{clock_KTN}
    \upmu=-\text{d}u+b_{\phi}\text{d}\phi\,.
\end{equation}
The holographic dual of this solution corresponds to a Carrollian fluid with 
\begin{equation}\label{fluidvar_KTN}
    \varepsilon_{(0)}=\frac{m}{4\pi G}\,,\qquad\pi_{A}=\frac{3}{4\pi G}a\sin\theta\left(a\cos\theta-n\right)\delta{^\phi}_A\,.
\end{equation}
The fact that we have a dissipative contribution due to a non-vanishing heat current $\pi^A$ indicates that, contrary to its AdS counterpart where the dual to Kerr-Taub-NUT-AdS is a perfect fluid, here the Carrollian fluid dual to a Kerr-Taub-NUT spacetime is non-perfect, albeit, admittedly, the interpretation of $\pi^A$ as heat current in Carrollian fluids still requires a more rigorous derivation. And in fact, this analysis brings new perspectives on the physics associated with this current, and might even suggest that $\pi^A$ should be treated as a geometric tensor, not necessarily associated with heat or dissipation. Indeed, this vector is of geometric nature, since it is related to the non-conformal flatness of null infinity due to \eqref{Carroll_heat}. 

Gravitational radiation is described by studying the Carroll-radiative scalar \eqref{Carr_Pu0} and vector \eqref{Carr_PA0}. Due to the absence of a heat current $Q^A$ and a viscous stress tensor $\Sigma^{AB}$, we have
\begin{equation}
    \hat{\varrho}=0\qquad \hat{\Upsilon}^A=0,
\end{equation}
and thus this is a non-radiative spacetime, consistent with the known fact that, indeed, Kerr-Taub-NUT black holes do not radiate.

\subsubsection*{Example II: Robinson-Trautman Spacetimes}

Let us now consider RT spacetimes, studied in \autoref{AdSRTexample} for the AdS case, for vanishing cosmological constant. The metric function \eqref{FAdS_RT} now reads
\begin{equation}\label{F_RT}
    F(r,u,x^A)=r\dot{\Phi}-\Delta\Phi-\frac{2m}{r}\,.
\end{equation}
At null infinity, the Carrollian boundary structure is given by the degenerate metric
\begin{equation}\label{degen_RT}
\text{d}s^2=0\cdot\text{d}u^2+2e^{\Phi}\text{d}\zeta\text{d}\bar{\zeta}
\end{equation}
with the Carroll vector and clock form expressed as
\begin{equation}\label{Carr_vec_clock}
    \upupsilon=\partial_u\quad\text{and}\quad\upmu=-\text{d}u\,.
\end{equation}

The above Carroll structure has non-vanishing Carrollian-Cotton tensors given by 
\begin{align}
    \chi_A\text{d}x^A&=-\frac{i}{2}\left(\partial_{\zeta}\Phi\Delta\Phi-\Delta\partial_\zeta\Phi\right)\text{d}\zeta+\frac{i}{2}\left(\partial_{\bar{\zeta}}\Phi\Delta\Phi-\Delta\partial_{\bar{\zeta}}\Phi\right)\text{d}\bar{\zeta}\,,\label{chi_RT}
    \\
    X_{AB}\text{d}x^A\text{d}x^B&=\frac{i}{2}\left(\partial^2_{\zeta}\dot{\Phi}-\partial_{\zeta}\Phi\partial_{\zeta}\dot{\Phi}\right)\text{d}\zeta^2-\frac{i}{2}\left(\partial^2_{\bar{\zeta}}\dot{\Phi}-\partial_{\bar{\zeta}}\Phi\partial_{\bar{\zeta}}\dot{\Phi}\right)\text{d}\bar{\zeta}^2\,,\label{X_RT}
\end{align}
while $c_{(0)}=c_{(1)}=0$ and $\psi_A=\Xi_{AB}=0$. This implies that the boundary is not conformally flat,\footnote{See \cite{Miskovic:2023zfz} for the analysis on the Carroll-Cotton tensors and the notion of conformal flatness in Carroll structures.} and thus the bulk is asymptotically locally flat.

The holographic Carrollian fluid is given by the energy density
\begin{equation}
    \varepsilon_{(0)}=\frac{m}{4\pi G}\,,
\end{equation}
and the heat current/energy flux $Q_A$ and viscous stress tensor $\Sigma_{AB}$ are fully determined by the Carroll-Cotton descendants through \eqref{Carroll_heat} and \eqref{Carroll_viscousstress}. The latter reads
\begin{align}
    Q_{A}\text{d}x^A&=-\frac{1}{16\pi G}\left(\left(\partial_{\zeta}\Phi\Delta\Phi-\Delta\partial_\zeta\Phi\right)\text{d}\zeta+\left(\partial_{\bar{\zeta}}\Phi\Delta\Phi-\Delta\partial_{\bar{\zeta}}\Phi\right)\text{d}\bar{\zeta}\right)\,,\label{Q_RT}
    \\
    \Sigma_{AB}\text{d}x^A\text{d}x^B&=\frac{1}{16\pi G}\left(\left(\partial^2_{\zeta}\dot{\Phi}-\partial_{\zeta}\Phi\partial_{\zeta}\dot{\Phi}\right)\text{d}\zeta^2+\left(\partial^2_{\bar{\zeta}}\dot{\Phi}-\partial_{\bar{\zeta}}\Phi\partial_{\bar{\zeta}}\dot{\Phi}\right)\text{d}\bar{\zeta}^2\right)\,.\label{Sig_RT}
\end{align}

In the above configuration, the Carroll-radiative scalar and vector are non-vanishing due to the presence of dissipation given by \eqref{Q_RT} and \eqref{Sig_RT}. Using \eqref{Carr_Pu0} and \eqref{Carr_PA0} we find
\begin{align}
    \hat{\varrho}={}&-e^{-2\Phi}\left(\partial^2_{\zeta}\dot{\Phi}-\partial_{\zeta}\Phi\partial_{\zeta}\dot{\Phi}\right)\left(\partial^2_{\bar{\zeta}}\dot{\Phi}-\partial_{\bar{\zeta}}\Phi\partial_{\bar{\zeta}}\dot{\Phi}\right)\,,\label{Pu0_RT}
    \\
    \hat{\Upsilon}_{A}\text{d}x^A={}&e^{-\Phi}\left(\partial^2_{\zeta}\dot{\Phi}-\partial_{\zeta}\Phi\partial_{\zeta}\dot{\Phi}\right)\left(\partial_{\bar{\zeta}}\Phi\Delta\Phi-\Delta\partial_{\bar{\zeta}}\Phi\right)\text{d}\zeta\nonumber
    \\
    &+e^{-\Phi}\left(\partial^2_{\bar{\zeta}}\dot{\Phi}-\partial_{\bar{\zeta}}\Phi\partial_{\bar{\zeta}}\dot{\Phi}\right)\left(\partial_{\zeta}\Phi\Delta\Phi-\Delta\partial_\zeta\Phi\right)\text{d}\bar{\zeta}\,,\label{PA0_RT}
\end{align}
This indicates that the presence of gravitational radiation in the bulk solution is driven by the dissipative phenomena of its Carrollian dual, captured by the heat current $Q_{A}$ and the viscous stress tensor $\Sigma_{AB}$, or equivalently, the Carroll-Cotton tensors $\chi_A$ and $X_{AB}$. This further supports our thesis that $Q_{A}$ and $\Sigma_{AB}$ are the Carrollian dissipative tensors, whereas their subleading descendants in the flat limit might have only a geometric origin.

\subsubsection*{Type I Spacetimes?}

All the analysis of this section has been done under the assumption that the dual Ricci-flat spacetime belongs to the algebraically special subclass. For a more general analysis of the Carroll analogue of the boundary radiative vector and its relation to the boundary Carrollian fluid, one should perform a flat limit of the expression \eqref{SP_fluid1} and then derive the associated Carroll-radiative scalar and vector. We will not pursue such an analysis in full generality here, but instead, we consider the flat limit of some type I AlAdS spacetime dual to a holographic perfect fluid. 
For that case, we already showed that a non-vanishing boundary radiative vector can exist as long as the Cotton current $c_{i}$ is non-vanishing, as seen in \eqref{P_perf}. Taking an expansion on \eqref{P_perf} in powers of $\ell$, we find
\begin{equation}\label{exp_Pperf}
    \hat{\mathscr{P}}^A=-\frac{192\pi^2 G^2}{\ell^2}\varepsilon_{(0)}\ast\!\chi^A+\mathcal{O}\left(\ell^{-4}\right)\,.
\end{equation}
 Therefore, a flat limit of the above configuration implies that the Carroll-radiative scalar and vector vanish, which means that there is no gravitational radiation in the dual Ricci-flat spacetime. 
 
 This demonstrates a key difference from the AlAdS case: the minimal condition for gravitational radiation to exist in the bulk is the presence of dissipative effects in the holographic Carrollian fluid. For type I solutions, dissipation does not depend on boundary conformal flatness, as one can readily verify that the boundary radiative vector \eqref{P_perf} vanishes in the flat limit. Furthermore, the boundary conformal flatness is irrelevant here, since the asymptotic symmetry group includes supertranslations that relate vacua differing due to the passage of gravitational waves. Therefore, in conclusion, the relationship between dissipative and gravitational radiation in flat spacetime allows us to understand more in depth the physical role played by the various Carrollian data at the boundary.

\subsection{Celestial Operators}\label{SubSec:LBMS_Cel_Op}

For asymptotically flat spacetimes, Bondi coordinates provide a suitable framework for approaching null infinity through an asymptotic expansion and analyzing the flux of energy due to gravitational radiation. Although this is specific to the vanishing cosmological constant, an adapted version, known as the $\Lambda$-BMS gauge,\footnote{See \cite{Barnich:2012aw} for an early extension for three-dimensional AdS gravity.} was introduced for any value of the cosmological constant in \cite{Poole:2018koa,Compere:2019bua}. This adapted gauge is described by
\begin{align}\label{LBMSgauge}
    {\rm d}s^2 = e^{\beta}\frac{V}{r} \d u^2 - 2e^{\beta}\d u\d r + g_{AB}\left(\d x^A - U^A \d u\right)\left(\d x^B - U^B \d u\right)\,,
\end{align}
where $\beta$, $V$, $U^{A}$, and $g_{AB}$ are, in principle, arbitrary functions of the coordinates. In this gauge, null surfaces are foliated by $u$, with null geodesics generated affinely parametrized by $r$\,. 
The conformal infinity is now located at $r=\infty$, which, upon taking the flat limit $\Lambda \to 0$, turns into the null infinity of the Bondi gauge. The transverse metric $g_{AB}$ satisfies an extra gauge condition
\begin{align}\label{BondidetCond}
    \partial_r\,  {\rm det}\left(\frac{g_{AB}}{r^2}\right) = 0\,, 
\end{align}
which is equivalent to requiring the determinant to be $\mathrm{det}(g_{AB})= r^{4} \chi(u,x^{A})$, where $\chi(u,x^{A})$ remains unfixed. Imposing the minimal fall-off condition
\begin{align}
    g_{AB} = r^2 q_{AB} + r {C}_{AB} + {D}_{AB} + \frac{1}{r} {E}_{AB} + \frac{1}{r^2}{F}_{AB}+\dots\,,
\end{align}
 allows the Einstein equations to be solved asymptotically \cite{Compere:2019bua,Compere:2020lrt}. 
Then, the gauge condition over the determinant \eqref{BondidetCond} implies that $\det q = \det\overset{\circ}{q}$\,, with $\overset{\circ}{q}_{AB}$ the metric of the unit-round sphere, and that
\begin{align}
    g^{AB}\partial_r g_{AB}=\frac{4}{r} \Rightarrow q^{AB}{C}_{AB} = 0\,,
\end{align}
such that ${C}_{AB}$ is symmetric and traceless with respect to $q_{AB}$.

The equations of motion impose that the solution space for Einstein gravity is given by\footnote{$D_A$ is the covariant derivative with respect to $q_{AB}$\,.}
\begin{align}
   \beta ={}& \beta_0(u,x^A) - \frac{1}{32r^2}{C}^{AB}{C}_{AB} + \dots\\ U^A ={}& U^A_0(u,x^B) + \frac{2}{r}e^{2\beta_0}\partial^A \beta_0 - \frac{1}{r^2}e^{2\beta_0}\left(\frac12 D_B {C}^{AB} + {C}^{AB}\partial_B \beta_0\right)+\dots \\ \frac{V}{r} ={}& \frac{\Lambda}{3}e^{2\beta_0}r^2 - r\left(D_A U_0^A+\hat{l}\right)\\ {}&- e^{2\beta_0}\left[\frac12 R[q] + \frac{\Lambda}{16}{C}_{AB}{C}^{AB} + 2D_A\partial^A\beta_0 + 4\partial_A\beta_0\partial^A\beta_0\right]+\frac{2m}{r} + \dots\,, \nonumber 
\end{align}
where all functions depend on the boundary coordinates $(u,x^A)$, $m$ in the flat limit corresponds to the Bondi mass, and 
\begin{align}
    \hat l \equiv \partial_u \log \sqrt{q} = \frac12 q^{AB}\partial_u q_{AB}\,.
\end{align}
Furthermore, the radial Hamiltonian momentum constraint imposes that
\begin{align}\label{CABdiffeq}
    \frac{\Lambda}{3}{C}_{AB} = e^{-2\beta_0}\left[\left(\partial_u  - D_C U^C_0- \hat{l}\right)q_{AB} + D_A U_B^0 + D_B U_A^0\right]\,,
\end{align}
 which shows that in the flat limit, the symmetric tensor ${C}_{AB}$ becomes free data, and the $u$-dependence of the transverse metric becomes constrained. For details on the construction and constraints of the subleading coefficients, see \cite{Compere:2019bua}. 

The boundary conformal structure is given by
\begin{align}
    \d s^2_{(0)} = \left(\frac{\Lambda}{3}e^{4\beta_0} + U_0^A U_A^0\right)\d u^2 - 2U_A^0 \d u\d x^A + q_{AB}\d x^A \d x^B\,,
\end{align}
which can always be brought to the following using the boundary gauge freedom:
\begin{align}
    U_0^A = 0\,, \qquad \beta_0 = 0\,,\qquad \sqrt{q} = \sqrt{\overset{\circ}{q}}\,,
\end{align}
such that it reduces to a null metric in the flat limit. 
The transformation from this gauge to the FG gauge allows the identification of 
 (A)dS analogues of the standard Bondi quantities in flat spacetime \cite{Barnich:2012aw,Compere:2019bua,Compere:2020lrt}.  Using this dictionary, one finds that the holographic stress tensor is
\begin{align}
    T_{ij} = \frac{1}{8\pi G}\begin{pmatrix}
2 m^{(\Lambda)} & & \,& & N_B^{(\Lambda)} \\
 N_B^{(\Lambda)} & & & &-\frac{3}{\Lambda}q_{AB}m^{(\Lambda)} - J_{AB}
\end{pmatrix}\,,
\end{align}
where $m_{\rm B}^{(\Lambda)}$ is the (A)dS version of the Bondi mass which transforms under gauge symmetries to preserve the Bondi gauge and corresponds to the holographic energy density \cite{Compere:2020lrt}, $N_A^{(\Lambda)}$ is the (A)dS version of the angular momentum aspect, and $J_{AB}$ is a symmetric traceless tensor. These quantities can be expressed explicitly in terms of the metric expansion as
\begin{align}
    m_{\rm B}^{(\Lambda)} ={}& m_{\rm B} + \frac{1}{16}\left(\partial_u + \hat{l}\right){C}_{AB}{C}^{AB}\,,\\
    N_A^{(\Lambda)} ={}& N_A -\frac{3}{2\Lambda}D^B \left({N}_{AB}-\frac{1}{2}\hat{l} {C}_{AB}\right) -\frac34\partial_A\left(\frac{1}{\Lambda}R[q]-\frac38 {C}^{CD}{C}_{CD}\right)\,,\\
        \nonumber J_{AB} ={}& - \frac{3}{\Lambda^2}\left[\partial_u\left({N}_{AB} - \frac12\hat{l} {C}_{AB}\right) + \left(D_{A}D_B -\frac12 q_{AB} D_C D^C\right)\hat{l}\right] \\ \nonumber {}&- \frac{1}{\Lambda}\Bigg[D_{(A}D^C {C}_{B)C} - \frac12q_{AB}D^CD^D {C}_{CD} - \frac12R[q]{C}_{AB}\Bigg]\nonumber \\ {}& +\frac{5}{16}{C}_{AB}{C}_{CD}{C}^{CD} - {E}_{AB} \,,
\end{align}
where $N_A$ is the angular momentum aspect in the flat limit, and ${N}_{AB} \equiv \partial_u {C}_{AB}$ is the corresponding Bondi news tensor in the flat limit.

To find the radiative vector \eqref{Super_P1} in this gauge, we need to obtain the Cotton tensor. However, this is quite a difficult task since it depends on three derivatives of the metric. In order to do so, let us consider an expansion around flat space  of the following form \cite{Ciambelli:2024kre}:
\begin{align}\label{Lexpansion}
    q_{AB} ={} q^{(0)}_{AB} + \sum_{n\geqslant1} \Lambda^{n}q^{(n)}_{AB}\,, \qquad  U_A^0 ={} u_A^{(0)} + \sum_{n\geqslant1} \Lambda^{n}u^{(n)}_{A}\,,
\end{align}
where now Einstein equations impose that the leading order coefficients must satisfy\footnote{$D^{(0)}_A$ is the covariant derivative with respect to $q_{AB}^{(0)}$\,.}
\begin{align}
    \partial_uq^{(0)}_{AB} - \frac12 q^{(0)}_{AB}q^{CD}_{(0)}\partial_u q^{(1)}_{CD} ={} 0\,, \qquad  D^{(0)}_{(A}u_{B)}^{(0)} - \frac12 q^{(0)}_{AB}D^{(0)}_C u^c_{(0)}={}0\,,
\end{align}
such that
\begin{align}
    {C}_{AB} = 3e^{-2\beta_0}\Bigg[\partial_u q^{(1)}_{AB}- \frac12q^{(0)}_{AB}\left(q_{(0)}^{CD}\partial_u q_{CD}^{(1)}+q_{(1)}^{CD}\partial_u q_{CD}^{(0)}\right)-\frac12q^{(1)}_{AB}q_{(0)}^{CD}\partial_u q_{CD}^{(0)}\Bigg]+{\cal O}(\Lambda)\,.
\end{align}

The boundary gauge freedom allows us to assume that $q^{(0)}_{AB}$ is the metric of the unit round two-sphere by considering $U_A^0= 0 $ and  $\beta_0=0$. Then, the Cotton-York tensor is
\begin{align}
    \mathscr{C}_{ij} = \begin{pmatrix}
-\sqrt{\tfrac{|\Lambda|^3}{3}} \eta^{AB}\big(D^{(0)}_{A} D^D_{(0)}\partial_u q^{(1)}_{BD} + \frac12 {N}_B{}^C\partial_u q^{(1)}_{AC}\big) + {\cal O}\big(\Lambda^{\frac52}\big) && -\sqrt{\tfrac{|\Lambda|}{3}}\eta^{CD}D^{(0)}_C {N}_{DB} +{\cal O}\big(\Lambda^{\frac32}\big) \\
 -\sqrt{\tfrac{|\Lambda|}{3}}\eta^{CD}D^{(0)}_C {N}_{DB} +{\cal O}\big(\Lambda^{\frac32}\big) && -\sqrt{\tfrac{3}{|\Lambda|}}\eta_{AC}\partial_u {N}^C{}_B + {\cal O}\big(\Lambda^{\frac12}\big)
\end{pmatrix}\,,
\end{align}
with $\eta_{AB} \equiv \sqrt{q^{(0)}}\epsilon_{AB}$ the transverse volume 2-form. Therefore, the boundary radiative vector in the AlAdS case  becomes 
\begin{align}
   \hat{\mathscr{P}}^u ={}& \tfrac{3}{2|\Lambda|} \partial_u {N}^{AB}\partial_u {N}_{AB} + \tfrac13 D^{(0)}_BN^{BC}D^{(0)}_A N^A{}_C +  {\cal O}\left(\Lambda\right) \,, \nonumber \\\hat{\mathscr{P}}^A ={}& \tfrac{1}{2}\partial_u {N}^{AB}D^{(0)}_C {N}^C{}_B + {\cal O}\left(\Lambda\right) \,,\label{P_expBondi} 
\end{align}
where the leading order terms in the expansion vanish in the absence of radiation. In this way, it is clear that the radiative vector obtained in the $\Lambda$-BMS gauge contains information about the Bondi news tensor, which appears at the leading order divergence in the $\Lambda$ expansion. Moreover, we can exactly obtain the news tensor using this expansion by analyzing  the Schouten tensor
\begin{align}
    \lim_{\Lambda \to 0} \mathscr{S}_{ij} =\begin{pmatrix}
        0 &&& 0 \\ 0 &&&  -\frac{1}{2} {N}_{AB}
    \end{pmatrix}\,,
\end{align}
 and obtain the derivative of the news tensor through the leading divergences of the stress tensor
\begin{align}
\lim_{\Lambda\to0}\frac{|\Lambda|}{3}T_{ij} =\begin{pmatrix}
        0 &&& 0 \\ 0 &&&  \frac{1}{16\pi G}\partial_u {N}_{AB}
    \end{pmatrix} 
\end{align}
 and the Cotton tensor
\begin{align}
\lim_{\Lambda\to0}\left(\frac{|\Lambda|}{3}\right)^{\frac12}\mathscr{C}_{ij} =\begin{pmatrix}
        0 &&& 0 \\ 0 &&&  - \eta_{AC}\partial_u {N}^C{}_B
    \end{pmatrix}\,.
\end{align}

As discussed in \autoref{SubSec:Carroll}, in the flat limit, conformal infinity becomes the null infinity of asymptotically flat spacetimes described by a Carrollian CFT. Remarkably, it was shown in \cite{pipolo2024celestial, hijano2019flat, Hijano:2020szl} that, using the shadow transformation, one can construct Celestial operators and soft theorems by taking the flat limit of holographic operators obtained in the AdS/CFT correspondence. This shows a remarkable connection between holographic proposals.  Celestial holography \cite{strominger2014bms,Strominger:2017zoo} (see  \cite{costello2022celestial, costello2023top} for a top-down approach and \cite{Pasterski:2021raf} for a review) is a program based on a codimension-2 hologram aimed at describing gravitational scattering processes in an asymptotically flat spacetime. It can be connected to Carrollian holography \cite{Donnay:2022aba,Donnay:2022wvx} by considering flat spacetime with leaky boundary conditions. These different approaches exhibit non-trivial relations through a codimensional ladder of dualities involving various limits of the moduli parameters of the theory. 

As we have seen, for radiative spacetimes, we can extract the information of the news tensor and its derivative. As it is shown in \cite{Ciambelli:2024kre}, now that the news tensor is found through the flat limit, it is possible to construct Celestial operators by using dimensional reduction \cite{strominger2014bms, kapec20172d}, particularly
\begin{align}
    P_z = \frac{1}{4G}D^z_{(0)}\int \d u\, {N}_{zz}\,,\qquad T_{zz} = \frac{i}{8\pi G}\int \d^2w\, \frac{1}{z-w}D^2_w \bar{D}_{(0)}^{\bar{w}}\int du\,u {N}_{\omega\omega}\,,
\end{align}
are the supertranslation and superrotation parameters, respectively, as holomorphic currents that physically encode memory effects \cite{Strominger:2014pwa, pasterski2016new}. When inserted into the $S$-matrix elements, the supertranslation parameters take the form of a Kac-Moody current, while the superrotation gives the two-dimensional stress tensor Ward identity. Furthermore, one can realize the $w_{1+\infty}$ symmetry algebra of Celestial CFT  \cite{guevara2021holographic, strominger2021w, himwich2022celestial} by changing basis \cite{donnay2022goldilocks}
\begin{align}
    {\cal O}_{\Delta}^\pm (z,\bar{z}) =  \frac{\Gamma(\Delta-2)}{4\pi i^{\Delta+2} }\int \d u \,u^{2-\Delta}\partial_u {N}_{zz}\,.
\end{align}
This gives a consistent, continuous spectrum of celestial weights \cite{de2003holographic, pasterski2017conformal}, and by analyzing collinear limits, the operator product expansion (OPE) of these modes closes under a radial bracket, which forms a representation of the $w_{1+\infty}$ algebra \cite{adamo2022celestial}.

Another interesting celestial operator that can be constructed by analyzing the flat limit of radiative AlAdS spacetimes is the energy cell operator \cite{Hofman:2008ar,Gonzo:2020xza} (see \cite{Herrmann:2024yai} for a nice review and an extension to quantum gravity). Consider a calorimeter cell in an idealized collider experiment, placed asymptotically far away from the collision at some angular position $\hat{\mathbf{n}}$ on the celestial sphere $S^2$. The energy operator is defined in terms of null integrals of stress tensor components
\begin{align}
    {\cal E}(\hat{\mathbf{n}}) \propto \lim_{r\to \infty}\int_{0}^\infty \d\tilde{t}\, r^2\left(\hat{\mathbf{n}}^i T_{0i}\right)\,,
\end{align}
where the energy momentum tensor is understood to be normal-ordered, and $\hat{t}$ represents the working time of the detector. These operators are also referred to as averaged null energy condition (ANEC) operators, as they satisfy the ANEC \cite{faulkner2016modular,hartman2017averaged}, ensuring that the null-time average of these operators is positive in any QFT. Upon considering $n$-point correlation functions for the energy operator, it allows us to understand the amount of energy radiated at different points of $S^2$.  These have been holographically described in \cite{Hofman:2008ar}. For the graviton\footnote{Following \cite{Herrmann:2024yai}, let us consider a metric perturbation around Minkowski background $g_{\mu\nu} = \eta_{\mu\nu} + \sqrt{32\pi G}h_{\mu\nu}$\,, and define the asymptotic graviton field
\begin{align}
    h_{\mu\nu}(\alpha,z) = \lim_{\epsilon \to \infty} \epsilon h_{\mu\nu}(x+\epsilon z)\,,
\end{align}
where $\alpha = 2x\dot z$ is the retarded detector time and $z^\mu$ is the future pointing null-vector. Then we have
\begin{align}
{\cal E}_h(z) =2\int_{-\infty}^\infty d\alpha :(\partial_\alpha h_{\mu\nu}(\alpha,z))(\partial_\alpha h^{\mu\nu}(\alpha,z)):\,,   
\end{align}
which is in agreement with the effective stress tensor for gravitational waves \cite{maggiore2008gravitational}.
}, in Bondi coordinates, the energy operator can be written in terms of the shear tensor as \cite{Herrmann:2024yai,Gonzo:2020xza}
\begin{align}\label{EShear}
    {\cal E}(\hat{\mathbf{n}}) = \frac{1}{16\pi G}\int_{-\infty}^\infty \d u\, \partial_u {C}_{zz} \partial_u {C}^{zz}\,,
\end{align}
where $ \hat{\mathbf{n}}$ can be expressed in stereographic coordinates as
\begin{align}
    \hat{\mathbf{n}} = \left(\frac{z+\bar{z}}{1+z \bar{z}},-\frac{i(z-\bar{z})}{1+z\bar{z}}, \frac{1-z\bar{z}}{1+z\bar{z}}\right)\,.
\end{align}

The algebra and Carrollian aspects of these operators have been studied in detail for different theories in \cite{Liu:2022mne, Liu:2023qtr, Liu:2023gwa, Li:2023xrr,Liu:2023jnc,Guo:2024qzv}. A generalization of these operators is the light-ray operators  \cite{casini2017modular,Kravchuk:2018htv, belin2019einstein, kolouglu2020shocks, kolouglu2021light, chang2022transverse} whose algebra has been studied from different approaches in \cite{Cordova:2018ygx, Belin:2020lsr}, revealing a BMS subalgebra. 
As can be seen from \eqref{EShear}, it is possible to construct such operators by extracting the shear tensor from the flat limit of \eqref{LBMSgauge} when the radiation is non-vanishing, revealing an interesting connection between energy correlators and gravitational radiation. It would be interesting to study the uplift of this discussion to AdS, and how gravitational radiation there intertwines with these operators.

\subsubsection*{Example: Accelerating Black Holes}
As an example, let us consider the rapidly accelerating black hole discussed earlier. Firstly, we need to write the solution in the $\Lambda$-BMS gauge \eqref{LBMSgauge}. In order to do so, we first recast the metric in the WFG gauge, as detailed in \autoref{App:WFG}. For simplicity, we consider a trivial boundary Weyl connection, i.e., we take the line element \eqref{WFG} with $\alpha(\bx^i) = 0$\,. Furthermore, we will use a radial coordinate $\frac{1}{\tilde{r}} = z$ instead of the Poincar\'e coordinate, so that the conformal boundary is located at $\tilde{r} = \infty$\,. Then, we propose the following diffeomorphism:
\begin{align}\label{CoordTransformation}
    \bt ={}& \sum_{n\geqslant0}U_{(n)}(u,x^A)r^{-n}\,, \\  \tilde{r} ={}& R_{(\rm{L})}r+\sum_{n\geqslant0}R_{(n)}(u,x^A)r^{-n}\,, \\ \bar{x}^A ={}& \sum_{n\geqslant0}X^A_{(n)}(u,x^A)r^{-n}\,,
\end{align}
where $X^A_{(n)} = (X_{(n)},Y_{(n)})$, and all coefficients $U_{(n)}, R_{(n)}, X^A_{(n)}$ can be determined by solving the Bondi gauge conditions $g_{rr} = 0 = g_{rA}$ and \eqref{BondidetCond} order by order in the asymptotic expansion. The gauge condition for $g_{rA}$ imposes that
\begin{align}
    Y_{(0)}(u,x^A) ={} y + {\cal Y}(u,x)\,, \qquad Y_{(n\geqslant1)}(u,x^A) ={} 0\,,
\end{align}
and
\begin{align}
    X_{(0)}(u,x^A) ={}&  {\cal X}(u)\,, \\  X_{(1)}(u,x^A) ={}& {\left(1+A\ell^2\left(\dot{\cal X} - A\Upsilon\right)\right)}\Upsilon{\dot{\cal X}}^{-1}\,, \\ X_{(2)}(u,x^A) ={}& \left(A\ell\Upsilon\left(6A\ell + \tfrac{5-6A^2\ell^2\Upsilon}{\sqrt{\Upsilon\left(A^2\ell^2\Upsilon-1\right)}}\right)-2\right){\dot\Upsilon}\ell^2{\dot\Xz}^{-1}\,,
\end{align}
where 
\begin{align}
    \Upsilon \equiv \left(1+2Am {\cal X}\right)\left({\cal X}^2 - 1\right)\,.
\end{align}
Here, ${\cal X}$ satisfies the differential equation
\begin{align}
    {\ell^2\dot{\cal X}^2+ \Upsilon} = \Upsilon A^2\ell^2\left(1+{\cal X}^2-2 A m {\cal X}^2\left(1-{\cal X}\right)\right) \,.
\end{align}
The higher order terms impose
\begin{align}
    U_{(0)}(u,x^A) ={}& u + {\cal U} + \frac{\ell x}{\sqrt{\Upsilon\left(A^2\ell^2\Upsilon-1\right)}} \,, \\  U_{(1)}(u,x^A) ={}& 0\,,\\  U_{(2)}(u,x^A) ={}& 0\,,
\end{align}
where $\cal U$ is a constant, and
\begin{align}
    R_{(\rm L)} ={} 1\,, \qquad R_{(0)} ={}\frac{\dot\Upsilon\left(2A\ell\left(A\ell\Upsilon-\sqrt{\Upsilon\left(A^2\ell^2\Upsilon-1\right)}\right)-\ell\right)}{4\sqrt{\Upsilon\left(A^2\ell^2\Upsilon-1\right)}}\,.
\end{align}
With these functions, we find that\footnote{In this coordinate system, we have $x=\cos\theta$\,, with $\theta$ the polar angle of the $S^2$\,, such that ${\rm det}\,\overset{\circ}{q}=1$\,. }
\begin{align}
    g_{rr} ={}& {\cal O}\left(r^{-4}\right)\,,
         \\ g_{rA} ={}& {\cal O}\left(r^{-3}\right)\,, \\  {\rm det}\, g_{AB} ={}& {\rm det}\,\overset{\circ}{q} + {\cal O}\left(r^{-3}\right)\,, 
\end{align}
and for the metric tensors, we get $\beta_0=0$ and $\hat{\ell}=0$ together with
\begin{align}
    U^x_{0} ={}& -\frac{\sqrt{\Upsilon\left(A^2\ell^2\Upsilon-1\right)}}{\ell} + \frac12\left(\frac{1}{\Upsilon} + \frac{A^2\ell^2}{A^2\ell^2\Upsilon-1}\right)x\dot\Upsilon\,, \\  U^y_{0} ={}& \left(\frac{\sqrt{\Upsilon\left(A^2\ell^2\Upsilon-1\right)}}{\ell} -\frac12\left(\frac{2}{\Upsilon} + \frac{1}{\Upsilon\left(A^2\ell^2\Upsilon-1\right)}\right)x\dot\Upsilon\right)\partial_x {\cal Y} - \dot{\cal Y}\,. 
\end{align}
For the transverse metric, we have
\begin{align}
 q_{AB}\d x^A\d x^B = \left(\frac{1}{\Upsilon} + \Upsilon\partial_x\left(\cal{Y}\right)^2\right)\d u^2 + 2\Upsilon\partial_x{\cal Y}\d u\d x + \Upsilon \d y^2\,,
\end{align}
while for the Bondi quantities we get
\begin{align}
    m^{(\Lambda)} ={}& m_{\rm B} + \frac{3\dot\Upsilon^2}{32\Upsilon \dot{\cal X}^2\left(\Lambda+3A^2\Upsilon\right)}\,, \\ N_A^{(\Lambda)} ={}& -\frac{3}{2\Lambda}D^B {N}_{AB} + N_A\,, \\ J_{AB} ={}& -\frac{3}{\Lambda}\partial_u {N}_{AB} + {\cal O}\left(\Lambda^{-2}\right)\,.
\end{align} 
And the symmetric shear tensor can be expressed as
\begin{align}
    {C}_{AB} = \frac{\ell \dot\Upsilon}{2\dot\Xz \sqrt{\Upsilon(A^2\ell^2\Upsilon-1)}}\begin{pmatrix}
-\frac{1}{\Upsilon}+\Upsilon \partial_x{\cal Y}^2 &&& \Upsilon \partial_x{\cal Y} \\\Upsilon \partial_x{\cal Y} &&& \Upsilon
    \end{pmatrix}
\end{align}
which satisfies $q^{AB}{C}_{AB} = 0$\,. As a cross-check, we find that the shear tensor satisfies the non-trivial Hamiltonian constraint \eqref{CABdiffeq}.
Indeed, one finds higher-order functions for the coordinate expansion that satisfy the Bondi gauge conditions. However, these expressions are rather lengthy and not necessary for our current purpose, so we shall not present them here.

We can now perform the flat limit. Note that, as previously described, accelerating black holes contain two regimes: rapid and slow acceleration, governed by the product $A \ell$. In the slow acceleration regime $(A \ell \leqslant 1)$\,, the large $\ell$ limit is not trivial. One needs to consider $\ell\to\infty$ while simultaneously taking $A\to0$\,, such that the product remains fixed and less than one. 
In the rapidly accelerating black hole $A \ell >1 $\,, there is no such issue, and the flat limit of the shear tensor gives 
\begin{align}
    {C}_{AB} = \frac{\dot\Upsilon}{2A\dot{\cal X}}\begin{pmatrix}
        \frac{1}{\Upsilon^2}\left[\left(\Upsilon\partial_x{\cal Y}\right)^2-1\right] && \partial_x {\cal Y} \\ \partial_x {\cal Y} && 1
    \end{pmatrix}\,.
\end{align}
Having this, one can reconstruct all the aforementioned celestial operators for the C-metric. The shear becomes free data, fixed by the boundary gauge symmetries in terms of the arbitrary functions $\cal X$ and $\cal Y$\,, and exhibits a time-dependent news tensor. This is in agreement with \cite{bicakgrav,Bicak:1989rb,BicakCosmicStringNews, Podolsky:2009ag, Fernandez-Alvarez:2024bkf}, where the radiative properties of the asymptotically flat C-metric have been investigated using alternative approaches. The arbitrary functions can be restricted by imposing the asymptotic conditions found for the news tensor in \cite{Strominger:2016wns}, which show that finite superrotations relate vacua with the same asymptotic structure, except at singular points of the transformation. These singular points are associated with defects linked to the presence of cosmic strings piercing the null infinity. At early times, the news tensor of the C-metric corresponds to that of a cosmic string with a deficit angle, related to the black holes' conserved quantities. At late times, it transitions to that of pure Minkowski spacetime. Thus, the C-metric describes a black hole pair nucleation process, where the cosmic string snaps into two pieces,\footnote{For more details, see \cite{Griffiths:2002gm, Anabalon:2021wjy}.} generating the acceleration between the black hole pair, pushing them away. The states before and after the string decays are related by a finite superrotation (see \cite{Barnich:2016lyg}), providing a physical interpretation of the role of these generators. 

This analysis, along with its AdS counterpart, offers insights into the dual structure of these spacetimes. Flat-space radiation helps uncover properties of the Carrollian fluid, such as dissipation and entropy, while the AdS fluid/gravity correspondence improves our grasp of gravitational radiation in AdS. Though the dual CFT state remains elusive, our results lay the foundation for further investigation.

\section{Conclusions}\label{Sec:Conclusions}

In this paper, we established a link between gravitational radiation in the bulk and dual boundary fluids, within the framework of the fluid/gravity correspondence. We did so in both asymptotically locally AdS and flat spacetimes. The connection between radiation and boundary fluid dissipation unearthed various threads that together formed the tapestry of our explorations. 

The first thread concerned AdS spacetimes, in which the conformal relativistic fluid living on the boundary is well-understood and under control. There, we linked its properties to the recently proposed notion of gravitational radiation, thereby importing this important GR tool into holography. We reviewed the salient features of AdS gravitational radiation in \autoref{Sec:Radiation}, discussing how the non-vanishing boundary value of the radiative vector gives a necessary condition for the presence of gravitational radiation in AdS. We provided its expression in terms of the boundary Cotton tensor and stress tensor, deriving that radiation reaching the boundary necessarily leads to a non-conformally flat boundary geometry. Then, in \autoref{Sec:Holo_fluid}, we evaluated this radiative vector as a function of the boundary fluid data, finding that it is indeed sourcing the hydrodynamic dissipative phenomena. Both the boundary heat current and viscous stress tensor appear in the formula for the radiative vector, corroborating the intuitive expectation that radiative fluxes reaching the boundary are dual to dissipation, and thus to a non-perfect hydrodynamic stress tensor. We then studied the sensitivity of the entropy current and its divergence to radiation. For this, we used the first order entropy current and found that bulk spacetimes with non-trivial radiation can still have a dual fluid with this notion of entropy current conserved. 

A second thread focused on the confirmation of our results in exact Einstein solutions. We thus tested our AdS results in the explicit example of accelerating black holes, confirming that they radiate due to the black hole acceleration parameter. Furthermore, their radiation is creating a non-trivial heat current and viscous stress tensor in the boundary. 

A third thread traced the connection between radiation and dual fluids in asymptotically flat spacetimes. Here, the state of the art is the exact opposite of AdS: bulk gravitational radiation has been studied and understood from many tackling angles, while the boundary fluid is much less understood and under control, mostly due to its elusive Carrollian nature. In \autoref{Sec:Flat_lim}, we utilized the full power of the flat limit procedures outlined in \cite{Ciambelli:2018wre, Campoleoni:2023fug} to rewrite the radiative vector as a function of the boundary fluid data, leading to the definition of the Carroll-radiative scalar and vector. The main result is that these boundary data controlling radiation are written in terms of the Carrollian heat current $Q_A$ and Carrollian viscous stress tensor $\Sigma_{AB}$, confirming their roles as dissipative tensors in the boundary fluid dynamics, and simultaneously questioning the role of the subleading heat and viscous tensors as drivers of dissipation --- we will touch on this point again presently. In a simplified setup, we identified the Carrollian viscous stress tensor and heat current with the time and space derivative of the news tensor, respectively. 

We then engaged again with the second thread and tested our asymptotically flat findings in exact bulk solutions. We first focused on Kerr-Taub-NUT spaces, confirming that they do not radiate. Interestingly, while the heat current $Q_A$ and viscous stress $\Sigma_{AB}$ are indeed zero, the subleading heat current $\pi_A$ does not vanish, questioning its very nature as a heat current. Indeed, one might expect that these spacetimes are non-radiative and dual to non-dissipative fluids. While we confirmed the former, the latter expectation would imply that the subleading heat current should not be interpreted as dissipation, but rather as a geometric tensor. The second example we presented was the Robinson-Trautman family of bulk solutions, in which we found that the bulk radiates, and this is captured in the boundary dual fluid by both heat and viscous dissipation.

The final thread addressed the flat limit in the framework of \cite{Poole:2018koa, Compere:2019bua} and the relationship to celestial holography. There, we demonstrated how the flat limit of the stress tensor and Cotton tensor carry information about the time derivative of the news tensor, and we reviewed how to construct the energy operator. While these results were known \cite{Ciambelli:2024kre}, their application to accelerating black holes unveiled some interesting new features, including the explicit form of the Bondi quantities both in AdS and in the flat limit. In AdS, this could help probe the microscopic nature of the dual CFT state, while in flat space, this can have interesting repercussions in the understanding of the Carrollian entropy current.

Our analysis has revealed several promising avenues for further investigation.
\begin{itemize}
    \item We have found radiative solutions in AdS whose dual fluids have vanishing first order entropy production. We plan to study higher order corrections to the entropy current, as detailed in \cite{Rezzolla:2013dea}, to see if there is a more compelling notion of fluid entropy that we can construct. Our expectation is that higher order terms in the diffusion tensors may help in clarifying the connection between bulk radiation and dual entropy production. A further reward of this analysis is the missing link between the results in \cite{Skenderis:2017dnh} and \cite{Ciambelli:2017wou} about the specific case of RT solutions. As a more far-reaching goal, a complete understanding of entropy production versus gravitational radiation, especially in flat spaces, will be useful in constructing a Carrollian entropy current for fluids dual to asymptotically flat spaces, for gaining insight into Carrollian dissipation. This is a timely and interesting direction to pursue, as entropic discussions in flat space holography are presently missing, and have been extremely informative in AdS/CFT. 
    \item The connection between radiative data and Carrollian fluids dissipation established in this paper can have important repercussions on the recent analysis performed in \cite{Ciambelli:2025mex}. Indeed, in that manuscript, the properties of Carrollian geometry and gravitational phase space on null hypersurfaces at finite distance have been related to those at asymptotic infinity. Building on our results, one can now transfer the newly understood dissipative phenomena back to finite-distance null hypersurfaces. This opens the intriguing possibility of reformulating the black hole membrane paradigm in a language akin to that of the boundary Carrollian fluid, thereby establishing a direct link between the gravitational dynamics induced on these two null hypersurfaces.
    \item Another interesting direction to explore is to study the memory effects associated with the C-metric. In \cite{Adami:2024mtu}, it is shown that the surface charge algebra associated with a generic RT spacetime corresponds to a Heisenberg algebra with its Casimir given by the ADM mass of the black hole. The canonical symplectic potential does not depend on the details of the geometry of the wavefront associated with gravitational radiation, so an asymptotic observer cannot distinguish the shape of the incoming radiation but only its total magnitude. Nonetheless, if the wavefront is non-smooth, one must add terms of higher codimension to the action principle, modifying the symplectic form. This modification might render it sensitive to the RT mode and thus to these associated memory effects.  The C-metric is an example of such a configuration, since the cosmic string can pierce through the wavefront, generating such punctures. This, in turn, produces IR divergences at the level of the on-shell action. For three-dimensional black holes \cite{Arenas-Henriquez:2022www, Arenas-Henriquez:2023hur, Arenas-Henriquez:2023vqp, Arenas-Henriquez:2024ypo,Cisterna:2023qhh} it is clearer how the presence of the cosmic string modifies the on-shell value of the action. We wish to study the action modification due to the cosmic string and how to incorporate corner terms in the symplectic potential, such that gravitational memory effects get modified due to defects. 
    \item One of the key consequences of our investigation is a unified framework to study radiation and the fluid/gravity correspondence in spacetimes with negative or vanishing cosmological constant. A natural future direction is to extend our results to positive cosmological constants. While the notion of gravitational radiation has already been established in asymptotically de Sitter spaces \cite{Fernandez-Alvarez:2021yog}, the fluid/gravity dictionary is still lacking. The reason is quite simple: the dual fluid would be hosted at future infinity, which is now a spacelike hypersurface. Therefore, one would have to formulate a hydrodynamic theory on an Euclidean background, which is a daunting task. Nonetheless, there has been recent progress in understanding de Sitter asymptotics, such as the flux-balance laws  \cite{Compere:2023ktn,Compere:2024ekl}, and the construction of pseudo-entropy \cite{Doi:2023zaf, Doi:2024nty}. Albeit exotic, perhaps a fluid/gravity duality for dS spacetimes is possible to be formulated, and here again, as for flat spacetimes, the characterization of gravitational radiation can be a guiding principle for this boundary macroscopic theory.
    \item While the role of the Cotton tensor in the $4d$ fluid/gravity correspondence has been appreciated,\footnote{Albeit Kubo formulas are subtle for the C-metric, due to cosmic string contributions to hydrodynamics.} its part in the microscopic AdS/CFT correspondence is still unknown. In fact, radiative solutions have fewer asymptotic conformal isometries, and their boundary CFT dual states have not yet been constructed. Our analysis of radiation and the explicit example of the C-metric can lead the way to a novel impulse in understanding how the Cotton tensor affects the Witten dictionary, and its CFT origin.
    \item Eventually, it would be interesting to reformulate the flat space dual Carrollian theory and the flat limit procedure in the abstract-index notation employed and developed, e.g., in  \cite{Freidel:2022bai, Freidel:2022vjq, Ciambelli:2023mir}. While this seems an academic exercise at first, a covariant reformulation of the boundary theory has important repercussions. First, it allows for the inclusion of all boundary diffeomorphisms, leading to a fully covariant --- as opposed to Carroll covariant --- ensemble of equations, which naturally connects to the asymptotic limit from the bulk \cite{Ciambelli:2025mex}. Secondly, it leads to a formulation of hydrodynamics which is more akin to its relativistic counterpart. In particular, this could lead to a simpler setup to explore Carrollian entropy currents and their dissipative phenomena. Lastly, we expect the covariant reformulation of the flat limit to lead to far-reaching consequences for flat holography, yet to be unveiled.
\end{itemize}



\paragraph{Acknowledgements} We thank Jos\'e Barrientos, Hern\'an Gonz\'alez, Sruthi Naranayan, Sabrina Pasterski, and Kostas Siampos for comments and discussions on related topics. We are especially thankful to José Senovilla, for comments on the drafts and discussions. WJ thanks Yingfei Gu for his hospitality and support at IAS, Tsinghua University, where this project was initiated. The work of WJ is also supported by a grant from the Research Grants Council of the Hong Kong Special Administrative Region, China (Project No.~AoE/P-404/18). The work of GAH is supported by the Beijing Natural Science Foundation (IS24016) and the Shuimu Scholars program of Tsinghua University. This research was supported by FONDECYT grant 1230853 and the Simons Collaboration on Celestial Holography. Research at Perimeter Institute is supported in part by the Government of Canada through the Department of Innovation, Science and Economic Development Canada and by the Province of Ontario through the Ministry of Colleges and Universities. 

\begin{appendix}
\section{AdS Boundary Weyl Geometry}\label{App:Weyl}
As presented in \autoref{Sec:Holo_fluid}, the 3-dimensional conformal boundary of AlAdS spacetimes enjoys covariance with respect to Weyl rescalings. Geometrically, this can be made manifest by introducing a metric-compatible Weyl-covariant derivative $\mathscr{D}_i$ and the corresponding Weyl connection. For the associated holographic relativistic fluid, we have introduced the Weyl connection \eqref{Weyl_conn} constructed from the kinematic quantities associated with the fluid velocity $u^i$, and the Weyl-covariant derivative acting on scalars and vectors is defined by \eqref{WD_scalar} and \eqref{WD_vec}, respectively. 

Similar to the Riemann curvature tensor of a covariant derivative, the \emph{Weyl-Riemann tensor} $\mathscr{R}^k_{\hphantom{k}lij}$ can be defined through the commutation relations of the Weyl-covariant derivatives. Suppose $V^k$ is a Weyl-covariant vector of weight $w$,\footnote{The Weyl weight corresponds to the power of $\mathscr{B}$ that appears in the Weyl transformation of the tensor under consideration. That is, a generic tensor is of weight $w$ if it transforms as $T^{i_1\dots i_n}_{\hphantom{i_1\dots i_n}j_1\dots j_n}\to\mathscr{B}^{w}T^{i_1\dots i_n}_{\hphantom{i_1\dots i_n}j_1\dots j_n}$ under Weyl rescaling.} we have
\begin{equation}\label{Weyl_rel_comm}
\left[\mathscr{D}_i,\mathscr{D}_j\right]V^k=\mathscr{R}^k_{\hphantom{k}lij}V^l+wF_{ij}V^k\,,
\end{equation}
where $F_{ij}=\partial_iA_j-\partial_jA_i$ is the curvature of the Weyl connection $A_i$. In terms of the Riemann tensor of the covariant derivative $\nabla^{(0)}$, the Weyl-Riemann tensor on a $d$-dimensional conformal boundary can be expressed as
\begin{align}\label{Weyl_Riemann}
\mathscr{R}^i{}_{jkl}={}&R^i{}_{jkl}+\nabla^{(0)}_l A_j\delta^i{}_k-\nabla^{(0)}_k A_j\delta^i{}_l
+(\nabla^{(0)}_l A_k-\nabla^{(0)}_k A_l)\delta^i{}_j
+\nabla^{(0)}_k A^i g^{(0)}_{jl}-\nabla^{(0)}_l A^i g^{(0)}_{jk}\nn\\
&+A_j(A_l\delta^i{}_k-A_k\delta^i{}_l)
+A^i(g^{(0)}_{jl}A_k-g^{(0)}_{jk}A_l)
+A^2(g^{(0)}_{jk}\delta^i{}_l-g^{(0)}_{jl}\delta^i{}_k)\,.
\end{align}
Then, the \textit{Weyl-Ricci tensor} and \textit{Weyl-Ricci scalar} are given by
\begin{align}
\label{Weyl_rel_Ricc} \mathscr{R}_{ij}&=R_{ij}+(d-2)(\nabla^{(0)}_{j}A_i+A_iA_j)+g^{(0)}_{ij}\left(\nabla^{(0)}_kA^k-(d-2)A_kA^k\right)-\frac{d}{2}F_{ij}\,,\\
\label{Weyl_rel_Rscalar}
\mathscr{R}&=R+2(d-1)\nabla^{(0)}_iA^i-(d-1)(d-2)A_iA^i\,.
\end{align}
The above curvature quantities transform covariantly under a Weyl transformation as
\begin{align}
\label{hatRtrans}
\mathscr{R}^i{}_{jkl}\to\mathscr{R}^i{}_{jkl}\,,\qquad\mathscr{R}_{ij}\to\mathscr{R}_{ij}\,,\qquad\mathscr{R}\to{\cal B}^2\mathscr{R}\,.
\end{align}
Furthermore, the Weyl-Schouten tensor $\mathscr{\hat S}_{ij}$ and Weyl-Cotton tensor $\mathscr{\hat C}_{ijk}$ can be expressed as
\begin{align}
\label{hatP}
\mathscr{\hat S}_{ij}&=\mathscr{S}_{ij}+\nabla^{(0)}_j A_{i}+A_{i}A_{j}-\frac{1}{2}A^2g^{(0)}_{ij}\,,\\
\mathscr{\hat C}_{ijk}&=\mathscr{C}_{ijk}-A_l W^l{}_{ikj}\,.
\end{align}
For a $3d$ conformal boundary, the Weyl tensor $W^l{}_{ikj}$ vanishes identically, and thus $\mathscr{\hat C}_{ijk}$ is equal to the Cotton tensor $\mathscr{C}_{ijk}$ defined in \eqref{Cott_Scho}, which makes clear that the Cotton tensor is conformally invariant in $3d$. For a more detailed review on the Weyl geometry and the Weyl covariant quantities, the reader may refer to \cite{Ciambelli:2019bzz} and Chapter 2 of \cite{Jia:2024ujz}.

\section{Carrollian Geometry in the Time/Space Splitting Formalism}\label{App:Carroll_boundary}

In the first half of \autoref{Sec:Flat_lim} we found that the conformal boundary of Ricci-flat spacetimes possesses a three-dimensional Carroll structure on $\mathcal{M}\equiv\mathbb{R}\times\mathcal{S}$ equipped with a degenerate metric \eqref{degen_metric}, where the time coordinate $u$ is aligned with the fibre of the structure \eqref{Carr_vec}. In this appendix, we discuss the Carroll covariance of the latter structure with emphasis on the definition of Carroll and Weyl-Carroll covariant derivatives used in this work and their associated geometric tensors.

The coordinate system of our Carroll structure is adapted to the time/space splitting, namely, it is preserved under the Carrollian diffeomorphism 
\begin{equation}\label{Carr_diffeos}
    u'=u'(u,\textbf{x})\,,\qquad \textbf{x}'=\textbf{x}'(\textbf{x})\,,
\end{equation}
 with Jacobian
\begin{equation}\label{Carr_Jac}
    J(u,\textbf{x})=\frac{\partial u'}{\partial u}\,,\qquad j_A(u,\textbf{x})=\frac{\partial u'}{\partial x^A}\,,\qquad J^{A}_{\hphantom{A}B}(\textbf{x})=\frac{\partial x'^{A}}{\partial x^B}\,.
\end{equation}

The above transformation acts on the building blocks of the Carroll geometry as 
\begin{equation}\label{geom_trans}
    \Omega'=\frac{1}{J}\Omega\,,\quad b'_A=\left(b_{B}+\frac{\Omega}{J}j_B\right)\left(J^{-1}\right)^B_{\hphantom{B}A}\,,\quad a'_{AB}=\left(J^{-1}\right)^C_{\hphantom{C}A}\left(J^{-1}\right)^D_{\hphantom{D}A}a_{CD}\,,
\end{equation}
while for the vector basis $\{\partial_u,\partial_A\}$ it acts as
\begin{equation}\label{basis_Trans1}
    \partial'_u=\frac{1}{J}\partial_u\,,\qquad \partial'_A=\left(J^{-1}\right)^B_{\hphantom{B}A}\left(\partial_B-\frac{j_B}{J}\partial_u\right)\,.
\end{equation}

There is a better choice of vector basis for the structure we are considering that allows for the separation of vector components in terms of Carroll scalars and Carroll vectors on the base space $\mathcal{S}$. For a Carroll vector $\textbf{X}$ in $\mathcal{M}\equiv\mathbb{R}\times\mathcal{S}$, the latter can be expressed as
\begin{equation}\label{Carr_vec_split}
    \textbf{X}=X^{u}\upupsilon+X^{A}\tilde{\partial}_A\,,
\end{equation}
with
\begin{equation}\label{Cpartial_A}
    \upupsilon=\frac{1}{\Omega}\partial_u\,, \qquad\tilde{\partial}_A=\partial_A+\frac{b_A}{\Omega}\partial_u\,.
\end{equation}

The vector basis $\{\upupsilon,\tilde{\partial}_A\}$ transforms covariantly with respect to \eqref{Carr_diffeos}, and thus the vertical component $X^{u}$ transforms as a Carrollian scalar and the transverse component $X^A$ transforms as a Carrollian vector of the base space $\mathcal{S}$. In this formalism, the indices of any Carroll vector and tensor $V^A$ and $V^{AB}$ are raised and lowered with the two-dimensional metric $a_{AB}$ and its inverse $a^{AB}$. In more general realizations of Carrollian structures, this basis can at best be attained in the frame bundle, and the whole theory is invariant under all diffeomorphisms, see, e.g., \cite{Ciambelli:2023mir}.

This adapted basis defines the \textit{Carrollian vorticity and acceleration} through its commutation relations as
\begin{equation}\label{commut_basis}
    \left[\upupsilon,\tilde{\partial}_A\right]=\varphi_A\upupsilon\,,\qquad \left[\tilde{\partial}_A,\tilde{\partial}_B\right]=2\varpi_{AB}\upupsilon\,,
\end{equation}
where 
\begin{equation}\label{vort_acc}
    \varphi_A=\frac{1}{\Omega}\left(\partial_u b_A+\partial_A\Omega\right)\,,\qquad \varpi_{AB}=\partial_{[A}b_{B]}+b_{[A}\varphi_{B]}\,.
\end{equation}

As we mentioned previously in \autoref{Sec:Flat_lim}, a strong Carroll structure is also equipped with a metric-compatible connection, which is ambiguous due to the degeneracy of the metric. Still, a choice can be made that corresponds to the one that naturally arises when considering the Carroll limit of a relativistic ascendant expressed in the Papapetrou-Randers gauge \eqref{PR_gauge}.\footnote{It also corresponds to the horizontal projection of the connection introduced in \cite{Freidel:2022bai, Freidel:2022vjq, Ciambelli:2023mir}, see also the appendix of \cite{Ciambelli:2023xqk}.} The latter defines a Carroll-Levi-Civita connection such that a spatial Carroll-covariant derivative acts on a Carroll scalar $\Phi$ and vector $V^A$ as
\begin{align}\label{Carr_covD_space}
    \tilde{\nabla}_A\Phi&=\tilde{\partial}_A\Phi\,,
    \\
    \tilde{\nabla}_AV^B&=\tilde{\partial}_AV^{B}+\tilde{\gamma}^{B}_{\hphantom{B}AC}V^{C}\,,
\end{align}
with
\begin{equation}\label{Carroll_LC}
    \tilde{\gamma}^{B}_{{\hphantom{B}}AC}=\frac{1}{2}a^{BD}\left(\tilde{\partial}_{A}a_{CD}+\tilde{\partial}_{C}a_{AD}-\tilde{\partial}_{D}a_{AC}\right)\,.
\end{equation}
The above obeys $\tilde{\gamma}^{B}_{\hphantom{B}[AC]}=0$ and $\tilde{\nabla}_{A}a_{BC}=0$. Note that the structure constants of the non-holonomic base do not intervene, as they are purely vertical in the fibre bundle.

We can also define a metric-compatible temporal Carroll connection. The latter use the extrinsic curvature of the spatial section $\mathcal{S}$ and is defined as 
\begin{equation}\label{Ext_Curv_S}
    \tilde{\gamma}_{AB}=\frac{1}{2\Omega}\partial_u a_{AB}\,.
\end{equation}
The latter is a symmetric Carroll tensor which can be decomposed into a pure trace part $\theta$, named \textit{Carrollian expansion},
\begin{align}
    \theta=a^{AB}\tilde{\gamma}_{AB}=\frac{1}{\Omega}\ln\sqrt{a}\,,\label{Carr_expan}
\end{align}
and a symmetric and traceless part, called \textit{Carrollian shear},
\begin{equation}\label{Carr_shear}
    \xi_{AB}=\tilde{\gamma}_{AB}-\frac{1}{2}a_{AB}\theta\,.
\end{equation}

With the above, we can define a time Carroll-covariant derivative, whose action on scalars and vectors is
\begin{align}\label{Carr_covD_time}
    \frac{1}{\Omega}\tilde{D}_u\Phi&=\frac{1}{\Omega}\partial_u\Phi\,,
\\
    \frac{1}{\Omega}\tilde{D}_uV^A&=\frac{1}{\Omega}\partial_uV^A+\tilde{\gamma}^{A}_{\hphantom{A}B}V^B\,.
\end{align}

The commutators of the temporal and spatial Carroll-covariant derivatives define the different curvature tensors of this geometry. From the commutator between two spatial Carroll covariant derivatives, we have
\begin{equation}\label{commu_CDspatial}
    \left[\tilde{\nabla}_{B},\tilde{\nabla}_C\right]V^A=\tilde{R}^{A}_{\hphantom{A}DBC}V^D+2\varpi_{BC}\frac{1}{\Omega}\partial_uV^A
\end{equation}
with the \textit{Riemann-Carroll tensor} defined as
\begin{equation}\label{Carr_Rieman}
    \tilde{R}^{A}_{\hphantom{A}DBC}=\tilde{\partial}_B\tilde{\gamma}^A_{\hphantom{A}CD}-\tilde{\partial}_C\tilde{\gamma}^A_{\hphantom{A}BD}+\tilde{\gamma}^A_{\hphantom{A}BE}\tilde{\gamma}^{E}_{\hphantom{E}CD}-\tilde{\gamma}^A_{\hphantom{A}CE}\tilde{\gamma}^{E}_{\hphantom{E}BD}\,.
\end{equation}

From the above, one can define the \textit{Ricci-Carroll tensor and Carrollian scalar curvature} by taking the traces of \eqref{Carr_Rieman}:
\begin{equation}
    \tilde{R}_{AB}=\tilde{R}^{C}_{\hphantom{C}ACB}\,,\qquad \tilde{R}=a^{AB}\tilde{R}_{AB}\,.
\end{equation}
Contrary to the relativistic case, the Ricci-Carroll tensor is not symmetric, namely, $\tilde{R}_{AB}\neq\tilde{R}_{BA}$.

In the same way, the commutator between the temporal and spatial Carroll-covariant derivatives gives
\begin{equation}\label{commu_CDtime_spa}
    \left[\frac{1}{\Omega}\tilde{D}_u,\tilde{\nabla}_A\right]V^{B}=\varphi_A\left(\left(\frac{1}{\Omega}+\theta\right)V^B-\tilde{\gamma}^B_{\hphantom{B}C}V^{C}\right)-\tilde{\gamma}^{C}_{\hphantom{C}A}\tilde{\nabla}_CV^{B}-2\tilde{R}^{B}_{\hphantom{B}AC}V^C\,,
\end{equation}
with
\begin{equation}\label{Carr_R3}
    \tilde{R}^B_{\hphantom{B}AC}=\frac{1}{2}\left(\theta\varphi_A\delta{^B}_{C}+\tilde{\nabla}_{A}\tilde{\gamma}^{B}_{\hphantom{B}C}-\frac{1}{\Omega}\partial_u\tilde{\gamma}^{B}_{\hphantom{B}AC}\right)
\end{equation}
a three-index Carroll curvature with sub-trace given by
\begin{equation}\label{Carr_R1}
    \tilde{R}^{B}_{\hphantom{B}AB}=\tilde{R}_A=\frac{1}{2}\left(\tilde{\nabla}_B\tilde{\gamma}^B_{\hphantom{B}A}-\tilde{\partial}_A\theta\right)\,.
\end{equation}

\subsubsection*{Weyl Covariance}

The Carrollian structure located at the null boundary of Ricci-flat spacetimes is conformal. Therefore, it is covariant with respect to an arbitrary rescaling $\mathscr{B}(u,\textbf{x})$ in the geometry. These are Weyl transformations that act on the Carrollian geometry as
\begin{equation}\label{Weyl_trans_geo}
    a_{AB}\to \frac{1}{\mathscr{B}^2}a_{AB}\,,\qquad b_{A}\to\frac{1}{\mathscr{B}}b_{A}\,,\qquad\Omega\to\frac{1}{\mathscr{B}}\Omega\,.
\end{equation}

One can define a Weyl-Carroll-covariant derivative that acts on Weyl-Carroll covariant tensors. To that end, one needs to construct the appropriate Weyl connection. From the Carrollian geometric quantities, we have that the Carrollian expansion and acceleration do transform as connections under Weyl transformation, i.e.,
\begin{equation}\label{theta_phi_Wtrans}
    \theta\to\mathscr{B}\theta-\frac{2}{\Omega}\partial_u\mathscr{B}\,,\qquad\varphi_A\to\varphi_A-\tilde{\partial}_A\ln\mathscr{B}\,.
\end{equation}
Therefore, Weyl-Carroll-covariant derivatives are defined with a Weyl connection built in terms of $\varphi_A$ and $\theta$, and thus are metric-compatible. 

For the transverse section $\mathcal{S}$ of the Carrollian manifold, the spatial Weyl-Carroll-covariant derivative is defined to act on scalars and vectors of weight $w$ as
\begin{align}\label{WC_covD_spatial}
    \tilde{\mathscr{D}}_A\Phi&=\tilde{\partial}_A\Phi+w\varphi_A\Phi\,,
    \\
    \tilde{\mathscr{D}}_AV^B&=\tilde{\nabla}_AV^B+\left(w-1\right)\varphi_AV^B+\varphi^BV_{A}-\delta^{B}_{A}\varphi_{C}V^C\,.
\end{align}
The latter is metric-compatible; that is, it satisfies $\tilde{\mathscr{D}}_Aa_{BC}=0$ and the weights of the objects it acts on are not altered.

In a similar way, the time Weyl-Carroll-covariant derivative is constructed such that its action on scalars and vectors is

\begin{align}\label{WC_covD_time}
    \frac{1}{\Omega}\tilde{\mathscr{D}}_u\Phi&=\frac{1}{\Omega}\tilde{D}_{u}\Phi+\frac{w}{2}\theta\Phi\,,
    \\
    \frac{1}{\Omega}\tilde{\mathscr{D}}_uV^A&=\frac{1}{\Omega}\tilde{D}_{u}V^A+\frac{w-1}{2}\theta V^A\,,
\end{align}
which is also metric-compatible, namely, $\frac{1}{\Omega}\tilde{\mathscr{D}}_ua_{AB}=0$ and increases the conformal weight of the object acted on by $w+1$.

The commutators of these Weyl-Carroll covariant derivatives also define different Weyl-Carroll-covariant curvature tensors. From the commutator between spatial derivatives, one gets
\begin{equation}\label{Weyl_Comm_spatial}
    \left[\tilde{\mathscr{D}}_B,\tilde{\mathscr{D}}_C\right]V^A=\left(\tilde{\mathscr{R}}^{A}_{\hphantom{A}DBC}-2\xi^{A}_{\hphantom{A}D}\varpi_{BC}\right)V^D+2\varpi_{BC}\frac{1}{\Omega}\tilde{\mathscr{D}}_uV^A+w\Omega_{BC}V^A\,,
\end{equation}
with 
\begin{equation}\label{Omega_AB}
    \Omega_{AB}=\tilde{\partial}_A\varphi_B-\tilde{\partial}_B\varphi_{A}-\varpi_{AB}\theta\,,
\end{equation}
antisymmetric and of weight-0, and the \textit{Weyl-Riemann-Carroll tensor} of weight-0 expressed as
\begin{align}\label{WeylCarrRiemann}
    \tilde{\mathscr{R}}^A_{\hphantom{A}DBC}&=\tilde{R}^{A}_{\hphantom{A}DBC}-\delta^{A}_D\left(\tilde{\partial}_{B}\varphi_{C}-\tilde{\partial}_C\varphi_{B}\right)-a_{DC}\tilde{\nabla}_B\varphi^{A}+\delta^{A}_{B}\tilde{\nabla}_{C}\varphi_{D}-\delta^{A}_{C}\tilde{\nabla}_B\varphi_{D}\nonumber
    \\
    &+\varphi^A\left(\varphi_{B}a_{DC}-\varphi_{C}a_{DB}\right)-\left(\delta^{A}_{B}a_{DC}-\delta^{A}_{C}a_{DB}\right)\varphi_{E}\varphi^{E}+\left(\delta^{A}_{B}\varphi_{C}-\delta^{A}_{C}\varphi_{B}\right)\varphi_{D}\,.
\end{align}

From the latter, the definition of the \textit{Weyl-Ricci-Carroll tensor and curvature scalar} follows from 
\begin{equation}\label{WeylRicc}
    \tilde{\mathscr{R}}_{AB}=\tilde{R}^{C}_{\hphantom{C}ACB}\,,\qquad\tilde{\mathscr{R}}=a^{AB}\tilde{\mathscr{R}}_{AB}\,,
\end{equation}
with $\tilde{\mathscr{R}}_{AB}\neq\tilde{\mathscr{R}}_{BA}$\,.

In the same way, one can obtain another curvature tensor from the commutator between the temporal and spatial Wey-Carroll covariant derivatives. From its action on a scalar, one gets
\begin{equation}\label{Weyl_Comm_spatial_time}
    \left[\frac{1}{\Omega}\tilde{\mathscr{D}}_u,\tilde{\mathscr{D}}_A\right]\Phi=w\tilde{\mathscr{R}}_A\Phi-\xi^{B}_{\hphantom{B}A}\tilde{\mathscr{D}}_B\Phi\,,
\end{equation}
where
\begin{equation}\label{WeylR1}
    \tilde{\mathscr{R}}_A=\frac{1}{\Omega}\partial_u\varphi_A-\frac12\left(\tilde{\partial}_A+\varphi_A\right)\theta\,.
\end{equation}

\subsubsection*{Carrollian Cotton Tensors}

In \autoref{Sec:Flat_lim} we saw that the components of the holographic Cotton tensor of a bulk asymptotically locally AdS spacetime can be decomposed in terms of a power expansion on the AdS radius, where at each power one finds different Carroll-Cotton descendants. In full generality, the expansion goes as follows
\begin{align}\label{Cotton_pieces_scalar}
    c&= \ell^4c_{(-2)}+\ell^2c_{(-1)}+c_{(0)}+\frac{1}{\ell^2}c_{(1)},
\\
c^A &= \ell^2z^{A}+\chi^{A}+\frac{1}{\ell^2}\psi^{A}\label{Cotton_pieces_vector}, 
\\
c^{AB}&= \ell^2Z^{AB}+X^{AB}+\frac{1}{\ell^2}\Psi^{AB}\,.\label{Cotton_pieces_tensor}
\end{align}
where for the expansion of the Cotton density we have
\begin{align}
    c_{(1)}&= 8\ast\!\varpi^3\,,\label{c1}
\\
c_{(0)}&=\left(\tilde{\mathscr{D}}_A\tilde{\mathscr{D}}^A+\tilde{\mathscr{R}}
\right)\ast\! \varpi\,,\label{c0}
\\
c_{(-1)}&=\tilde{\mathscr{D}}_A\tilde{\mathscr{D}}_B \ast\! \xi^{AB},\label{c_1}
\\
c_{(-2)} &= \ast \xi_{AB}\frac{1}{\Omega}\tilde{\mathscr{D}}_u \xi^{AB},\label{c_2}
\end{align}
From the Cotton current, we find
\begin{align}
    \psi^{A}&=3\tilde{\eta}^{BA}\tilde{\mathscr{D}}_B\ast\! \varpi^2\label{psi-Carroll}
\\
\chi^{A}&= \frac{1}{4}\tilde{\eta}^{BA}\tilde{\mathscr{D}}_B\hat{\mathscr{R}}+ \frac{1}{4}\tilde{\eta}^{BC}\tilde{\mathscr{D}}^A\tilde{\mathscr{R}}_{BC}-2\ast \!\varpi\left(\tilde{\mathscr{R}}^A + 2 \tilde{\mathscr{D}}_B  \xi^{AB}
\right)+3\tilde{\mathscr{D}}_B \left(\ast \varpi \xi^{AB}\right)\label{chi-Carroll},
\\
z^{A} &= \frac{1}{2}\tilde{\eta}^{AB}\tilde{\mathscr{D}}_B \xi^2 - \tilde{\mathscr{D}}_B\frac{1}{\Omega}\tilde{\mathscr{D}}_u \ast\! \xi^{AB} - \ast \xi^{A}_{\hphantom{A}B}\tilde{\mathscr{D}}_C \xi^{BC}
\end{align}
and from the Cotton stress tensor, we get
\begin{align}
    \Psi^{AB} &= -2 \ast \! \varpi^2  \ast \xi^{AB}
+\tilde{\mathscr{D}}^A \tilde{\mathscr{D}}^B\ast\! \varpi -\frac{1}{2}a^{AB} \tilde{\mathscr{D}}_C \tilde{\mathscr{D}}^C \ast\! \varpi -\tilde{\eta}^{AB}   \frac{1}{\Omega}\tilde{\mathscr{D}}_u\ast\! \varpi^2\label{Psi-Carroll}
\\
X^{AB}&=\frac{1}{2}\tilde{\eta}^{CA}\tilde{\mathscr{D}}_C
\left(\tilde{\mathscr{R}}^B+ \tilde{\mathscr{D}}_D  \xi^{BD}\right)+
\frac{1}{2} \tilde{\eta}^{CB}\tilde{\mathscr{D}}^A
\left(\tilde{\mathscr{R}}_C+ \tilde{\mathscr{D}}^D  \xi_{CD}\right)\nonumber
\\
&-\frac{3}{4}\tilde{\eta}^{CD}\tilde{\mathscr{R}}_{CD} \xi^{AB} -\frac12\tilde{\mathscr{R}}\ast \! \xi^{AB} +3\ast\! \varpi\frac{1}{\Omega} \tilde{\mathscr{D}}_u \xi^{AB}\label{X-Carroll}
\\
\label{ztens}
Z^{AB} &= 2 \ast\! \xi^{AB} \xi^2 - \frac{1}{\Omega}\tilde{\mathscr{D}}_u  \frac{1}{\Omega}\tilde{\mathscr{D}}_u \ast\!\xi^{AB}\,.
\end{align}

Here, $\tilde{\eta}_{AB}=\sqrt{a}\epsilon_{AB}$\footnote{$\epsilon_{AB}$ is the two-dimensional Levi-Civita symbol and $\tilde{\eta}_{AB}$ satisfies $\tilde{\eta}_{AC}\tilde{\eta}^{BC}=\delta{^B}_{A}$.} and defines a two-dimensional Hodge dual operator that acts on tensors as
\begin{equation}\label{2dHodge_dual}
    \ast V_A=\tilde{\eta}^B_{\hphantom{B}A}V_B\,,\qquad \ast V_{AB}=\tilde{\eta}^C_{\hphantom{C}A}V_{CB}\,.
\end{equation}

We have also defined
\begin{equation}
\label{xisquare}
 \xi^2 =\frac{1}{2}\xi^{AB}\xi_{AB}\quad
\Leftrightarrow\quad
\xi^{AC}\xi_{C}^{\hphantom{C}B}=
 \xi^2 a^{AB}\,.
\end{equation}
and
\begin{equation}\label{dual_Carr_vort}
    \ast\varpi=\frac{1}{2}\tilde{\eta}^{AB}\varpi_{AB}\,.
\end{equation}

These Carroll-Cotton tensors give a notion of conformal flatness for the Carroll structure at hand, which has been explored in great detail in \cite{Miskovic:2023zfz}. The latter also satisfies a number of geometrical identities associated with the Bianchi identity of its relativistic ascendant. These are
\begin{align}
\frac{1}{\Omega}\tilde{\mathscr{D}}_u c_{(1)}+\tilde{\mathscr{D}}_A \psi^{A}&=0\,,
\\
\frac{1}{\Omega}\tilde{\mathscr{D}}_u c_{(0)}+\tilde{\mathscr{D}}_A \chi^{A}-\Psi_{AB}\xi^{AB}
&=0,
\\
\frac{1}{\Omega}\tilde{\mathscr{D}}_u c_{(-1)}+\tilde{\mathscr{D}}_A z^{A}
-X_{AB}\xi^{AB}&=0\,,
\\
\frac{1}{\Omega}\tilde{\mathscr{D}}_u c_{(-2)}
-Z_{AB}\xi^{AB}&=0\,,
\\
\frac{1}{2}\tilde{\mathscr{D}}_A c_{(1)}
+2 \ast\! \varpi   \ast\!\psi_{A}&=0\,,
\\
\frac{1}{2}\tilde{\mathscr{D}}_A c_{(0)}- \tilde{\mathscr{D}}^B \Psi_{AB}
+2\ast\!\varpi \ast\!\chi_{A}
+ \frac{1}{\Omega}\tilde{\mathscr{D}}_u \psi_A
+\xi_{AB} \psi^B
&= 0,
\\
\frac{1}{2}\tilde{\mathscr{D}}_A c_{(-1)}- \tilde{\mathscr{D}}^B X_{AB}
+2 \ast\! \varpi   \ast\! z_{A}
+ \frac{1}{\Omega}\tilde{\mathscr{D}}_u \chi_A +\xi_{AB} \chi^B
&=0\,,
\\
\frac{1}{2}\tilde{\mathscr{D}}_A c_{(-2)}- \tilde{\mathscr{D}}^B Z_{AB}
+ \frac{1}{\Omega}\tilde{\mathscr{D}}_u z_A +\xi_{AB} z^B&=0\,.
\end{align}

When considering the flat limit of \eqref{PR_gauge}, Einstein's equations fix the Carrollian shear $\xi_{AB}$ to vanish. The latter implies that, for the null boundary of any Ricci-flat spacetime, $c_{(-2)}=c_{(-1)}=z_A=Z_{AB}=0$.

\section{AdS C-Metric and its boundary CFT}\label{App:WFG}

In this appendix, we describe the four-dimensional AdS C-metric using different coordinate charts. We show how to apply a coordinate transformation that allows us to describe the boundary CFT with ease and in a Weyl covariant way. 

\paragraph{Four-dimensional AdS C-metric.}  Using oblate coordinates $x^\mu = \{ \bar{t}, \bar{y},\bar{x},\bar{\phi}\}$\,, the AdS C-metric is described by the line element
\begin{align}\label{Coblate}
    {\rm d}s^2 = \frac{1}{A^2(\bar{x}+\bar{y})^2}\left[-\bar{F} \d\bar{t}^2 + \frac{\d\bar{y}^2}{\bar{F}}+\frac{\d \bar{x}^2}{\bar{G}} + \bar{G} \frac{\d\bar{\phi}^2}{{\cal{K}}^2} \right]\,,
\end{align}
with
\begin{align}
    \bar{F}= -\frac{\Lambda}{3 A^2}-1+\bar{y}^2-2Am\bar{y}^3\,,\qquad \bar{G} = 1-\bar{x}^2-2Am\bar{x}^3\,.
\end{align}
In these coordinates, the solution has a non-smooth zero-acceleration limit. A more natural description of these black holes is achieved in spherical coordinates which can be obtained through the following transformation $\bar{y} = (A \bar{r})^{-1}\,, \bar{t} = A \bt$\,, and $\bar{x} = \cos \bar{\theta}$. This renders \eqref{Coblate} to
\begin{align}\label{Cmetric_BL}
    {\rm d}s^2 = \frac{1}{\omega(\br,\bar{\theta})^2}\left[-\bar{f}(\br)\d\bt^2 + \frac{\d\br^2}{\bar{f}(\br)} + \br^2 \left(\frac{\d\bar{\theta}^2}{\Sigma(\bar{\theta})} + \Sigma(\bar{\theta})\sin^2\bar{\theta}\frac{\d\bar{\phi}^2}{{\cal K}^2}  \right)\right]\,,
\end{align}
with metric functions
\begin{align}
    \omega(\br,\bar\theta) \equiv {}& 1+A \br \cos\bar{\theta}\,,\\ \nonumber \Sigma(\bar\theta) \equiv {}& 1+2m A \cos\bar{\theta}\,,\\ \nonumber \bar{f}(\br) \equiv {}& 1 - \frac{2m}{\br} -\br^2\left(\frac{\Lambda}{3} + A^2\right) + 2m A^2\br \,.
\end{align}
This solution can be interpreted as a one-parameter extension of the AdS-Schwarzschild solution, but whose conformal boundary is now parametrized by
\begin{align}
    \frac{1}{\br_{\rm bdry}} = - {A\cos\bar\theta}\,. 
\end{align}
The parameter $m$ represents the black hole mass \cite{Anabalon:2018qfv}. In the weak-field limit $m=0$, the solution reduces to Rindler-AdS, describing an accelerated observer in anti-de Sitter with acceleration given by $A$.
The acceleration is induced by the presence of conical deficits at the north $\bar{\theta}_{+}=\pi$ and south pole $\bar{\theta}_{-}=0$. Near $\bar{\theta}=\bar{\theta}_{\pm}$, the induced metric is 
\begin{equation}
ds^{2}_{\bar{\theta}_\pm, \bar{\phi}} \approx \frac{1}{\Sigma_{\pm} \omega_{\pm}^2} \Bigg\{ \d \bar{\theta}^2 + \Sigma^2_{\pm} (\bar{\theta}  - \bar{\theta}_\pm)^2  \frac{\d\bar{\phi}^2}{{\cal K}^2} \Bigg\} + \mathcal{O} \left( \bar{\theta}^4 \right)   \,. 
\end{equation}
The corresponding deficit angle along $\theta=\theta_{\pm}$ is 
\begin{equation}
    \delta_{\pm}=2\pi \left(1-\frac{\Sigma_{\pm}}{{\cal K}} \right)\,,
\end{equation}
which can be interpreted as a cosmic string with tension 
\begin{equation}
   \mu_{\pm}= \frac{\delta_{\pm}}{8\pi}=\frac{1}{4}\left( 1 - \frac{1 \pm 2m A}{\cal K}\right) \,.
\end{equation}
From these expressions, it follows that it is not possible to choose $\cal K$ such that both tensions vanish simultaneously. This imbalance of tension between the north and south poles is what induces the acceleration in the system. More precisely, the difference of tensions between the south and north poles is
\begin{equation}
\mu_{-}-\mu_{+}= \frac{m A}{\cal K}   \,.  \label{tension}
\end{equation}
The causal structure of the black hole is determined by the roots of the function $f(\bar{r})$, which depend on the interplay between the mass parameter $m$, the acceleration parameter ${A}$, and the cosmological constant $\Lambda=-3/\ell^{2}$. Since $f(\bar{r})$ is a third-order polynomial, the geometry can admit up to three Killing horizons. When three horizons are present, two are interpreted as black hole horizons, while the other one corresponds to an accelerating (Rindler-type) horizon that causally disconnects both black holes. This occurs in the regime where ${A} \ell>1$, and hence is referred to as a \textit{rapidly accelerating phase}. As in the case of de Sitter space, the presence of an accelerating horizon disconnects the bulk from the conformal boundary, obscuring a standard holographic interpretation. In contrast, setting ${A} \ell \leqslant 1$, the cosmological constant counteracts the acceleration, resulting in a single-horizon black hole suspended away from the center of AdS, which is commonly referred to as being on a \textit{slowly accelerating phase}. The single horizon structure not only allows for a conventional treatment of black hole thermodynamics \cite{Appels:2016uha,Appels:2017xoe,Anabalon:2018ydc,Anabalon:2018qfv,Hale:2025veb}, but also facilitates the application of standard holographic techniques to explore their holographic dual \cite{Cassani:2021dwa}. 

In order to describe some holographic features of the solution, first, we use oblique coordinates
\begin{align}
    \bx = \cos\bar\theta\,,\qquad \by = {\bar{\phi}}{\cal K}^{-1}\,,
\end{align}
in which the metric reads
\begin{align}\label{cmetric}
    \text{d}s^2 = \frac{1}{\omega(\br,\bx)^2}\left[-\bar{f}(\br)\d\bt^2 + \frac{\d\br^2}{\bar{f}(\br)} + \br^2 \left(\frac{\d\bx^2}{\Sigma(\bx)} + \Sigma(\bx)\d\by^2  \right)\right]\,,
\end{align}
where now
\begin{align}
    \omega(\bar{r},\tilde{x}) =  1+A \bar{r} \tilde{x}\,,\qquad \Sigma(\tilde{x}) = (1-\tilde{x}^2)(1+2 A m \tilde{x})\,.
\end{align}

Finally, we now perform the coordinate transformation given in \cite{Arenas-Henriquez:2024ypo}
\begin{align}
    \frac{1}{\br} = \sum_{n\geqslant0} \zeta_{(n)}(\bx^i)\bz^n\,,
\end{align}
where ${\bx}^i = (\bt,\bx,\by)$ are the boundary coordinates, and
\begin{align}
    \zeta_{(0)} = -A\bx\,,\qquad \zeta_{(1)} = 1\,,\qquad \zeta_{(2)} = \frac{1}{2}\alpha(\bx^i)\,,
\end{align}
with $\alpha$ some arbitrary function and all higher coefficients $\zeta_{(n\geqslant3)}$ set to zero. Then, the metric becomes
\begin{align}\label{WFG}
    \d s^2 = \ell^2H^2\left(\frac{\d \bz}{\bz} -k_i \d \bx^i\right)^2 + h_{ij}\d\bx^i\d\bx^j\,,
\end{align}
where the conformal boundary is now located at $\bz=0$\,. The metric tensors are
\begin{align}
    H^{-2} = \frac{\ell^2\omega^2 \bar{f}}{(\bz+\alpha \bz^2)^2} \sum_{n\geqslant0}\zeta_{(n)}(\bx^i)\bz^n\,,\qquad k_i = \frac{1}{2\bz(1+\alpha \bz)}\left(2A\delta_i^{\bx}-\bz^2\partial_i\alpha \right)\,,
\end{align}
and
\begin{align}
    h_{ij}\d\bx^i\d\bx^j =\frac{1}{\omega^2}\left[-f \d\bt^2 + \br(z,\bx^i)^2\left(\frac{\d\bx^2}{\Sigma} + \Sigma^2 \d\by^2\right)\right]\,,
\end{align}
where now the metric functions depend on $\bz$ and $\bx^i$ as well.
An asymptotic expansion of the coefficients shows that $\alpha(\bx^i)$ corresponds to the boundary Weyl connection that allows for describing the dual theory Weyl covariantly and identifying the holographic Weyl anomaly in terms of geometric quantities \cite{Ciambelli:2019bzz, Jia:2021hgy, Jia:2023gmk, Ciambelli:2023ott, Arenas-Henriquez:2024ypo}. In principle, we could also use a more general $\zeta_{(1)}$ in terms of some arbitrary function of the boundary coordinates that generates a more complete boundary Weyl connection, but we will keep this minimal setup for the sake of simplicity. The solution is asymptotically locally AdS, i.e.,
\begin{align}
    \lim_{\bz\to 0}R^{\mu\nu}{}_{\alpha\beta} = \frac{\Lambda}{3}\left(\delta^\mu_\alpha\delta^\nu_\beta - \delta^\mu_\beta\delta^\nu_\alpha\right)\,.
\end{align}

\paragraph{Renormalized AdS action and thermodynamics:} 
As previously described, from a geometric standpoint, the existence of the conical deficit is associated with a cosmic string along the $\bz,\bt= \rm{const}$ surface. The variational principle in the presence of codimension-two defects has been previously investigated in the context of braneworld scenarios, see for instance \cite{Gregory:2001dn,Charmousis:2006pn}. Following this, the on-shell action can be obtained using the holographic renormalization method \cite{deHaro:2006wy} supplemented by the respective Nambu-Goto term associated with the cosmic string at the north and south poles. On-shell, the Nambu-Goto term generates an extra stress tensor localized at the cosmic string at the right-hand side of Einstein's equation, which, after projecting to the boundary, generates the Israel equations. Tracing over the Israel equations, one finds that, on-shell, the jump of the extrinsic curvature is proportional to minus the difference of tension between the south and north poles. Plugging back into the action, one finds that the Gibbons-Hawking term exactly cancels the contributions of the Nambu-Goto term. Then, the total action is solely given by 
\begin{align}
    I_{\rm ren} = \frac{1}{16\pi G}\int_{\cal M}\d^4\bx\sqrt{-g}\left(R+\frac{6}{\ell^2}\right) - \frac{1}{8\pi G}\int_{\partial\cal M}\d^3\bx\sqrt{-\gamma}\left(K-\frac{2}{\ell} + \frac{\ell}{2}{\cal R}(\gamma)\right)\,,
\end{align}
with $K=\gamma^{ij}K_{ij}$, the trace of the extrinsic curvature obtained by considering the covariant divergence of the outward-pointing normal to the hypersurface $\bz={\rm constant}$ with induced metric 
\begin{align}
    \gamma_{ij} = h_{ij}+\ell^2H^2k_i k_j\,.
\end{align}
For the metric \eqref{WFG}, we have
\begin{align}
    K_{ij} = -\frac{1}{2N}\left(\partial_{\bz} \gamma_{ij}- \bar\nabla_i N_j - \bar\nabla_j N_i\right)\,,
\end{align}
where $\bar\nabla$ is the covariant derivative with respect to $\gamma$ and
\begin{align}
    N^2 = \frac{\ell^2}{\bz^2}H^2\left(1-k_ik_j\gamma^{ij}\right)\,,\qquad N_i = -\frac{\ell^2}{\bz}H^2 k_i\,.
\end{align}
In the Euclidean regime, one finds
\begin{align}
    I_{\rm ren}^{\rm E} = \beta M - S\,,
\end{align}
with 
\begin{align}
    \beta =\frac{4\pi}{|f'(\br_+)|} = \frac{2 \pi  \ell^2 \br_+^2}{\ell^2 \left(A^2 m \br_+^2-A^2 \br_+^3+m\right)+\br_+^3}\,
\end{align}
the inverse of the Hawking temperature is given in terms of the horizon radius $\br_+$ (in spherical coordinates).

Moreover, the Bekenstein-Hawking black hole entropy is
\begin{align}
    S = \frac{{\rm Area}}{4G} = \frac{\pi \br_+^2}{{\cal K}(1-A^2\br_+^2)}\,,
\end{align}
while
\begin{align}
    M = \left(\frac{1-A^2\ell^2}{{\cal K}}\right)\frac{m}{G}
\end{align}
is the black hole energy \cite{Anabalon:2018qfv}. The black hole energy can also be extracted from the holographic stress tensor \cite{Myers:1999psa}, obtained as the asymptotic limit of the quasilocal stress tensor \cite{Balasubramanian:1999re}
\begin{align}
    t_{ij} := -\frac{2}{\sqrt{-\gamma}}\frac{\delta I_{\rm ren}}{\delta \gamma^{ij}} =-\frac{1}{8\pi G}\left( K_{ij}-K\gamma_{ij}+\frac{2}{\ell}\gamma_{ij} - \ell G_{ij}(\gamma)\right)\,,
\end{align}
with $G_{ij}(\gamma)$ the Einstein tensor of $\gamma_{ij}$. The holographic stress tensor is then
\begin{align}
    T_{ij} ={}& \lim_{\bz\to \infty}\frac{\ell}{\bz}t_{ij} = \frac{m\ell^4}{8\pi G}{\rm diag}(3A^2\ell^2\Sigma-2,1,1-3A^3\ell^2\Sigma)\,, 
\end{align}
which is traceless and covariantly conserved with respect to the boundary metric 
\begin{align}\label{g0WFG}
\gamma_{ij}^{(0)}\d x^i \d x^j = -(1-A^2\ell^2\Sigma)\d\bt^2 + \frac{\d\bx^2}{\Sigma(1-A^2\ell^2 \Sigma)} + \Sigma \d\by^2\,,
\end{align}
corresponding to the leading order coefficient in the asymptotic expansion of the induced metric $\gamma_{ij}$\,.

The boundary metric has a Killing vector field ${\cal L}_\xi \gamma^{(0)} = 0$. Therefore, the holographic stress tensor defines a conserved Noether current whose integration over the two-dimensional spacelike hypersurface $\Sigma_\infty$ yields the conserved charge
\begin{align}
    Q(\xi^i) = \oint_{\Sigma_\infty}d^2\bx \sqrt{-\sigma}\tilde{n}_j T^j{}_i \xi^i\,, 
\end{align}
with $\tilde{n}$ the outward-pointing normal to $\Sigma_\infty$\,. We find
\begin{align}
    Q(\partial_t) = M\,,
\end{align}
in agreement with \cite{Anabalon:2018ydc,Cassani:2021dwa}.

The stress tensor can also be written as that of a non-perfect fluid
\begin{align}
  T_{ij} & = T^{(0)}_{ij}+\Pi_{ij}\,,
\end{align}
where $T^{(0)}_{ij}$ is the perfect part, $u^i = (\gamma_{\bt\bt}^{(0)})^{-\frac12}(\partial_{\bt})^i$ is the fluid velocity, and 
\begin{align}
    \Pi_{ij} = \frac{3mA^2\ell^4}{16\pi G}{\rm diag}\left(0,\frac{1}{1-A^2\ell^2\Sigma},-\Sigma^2\right)\,.
\end{align}

The energy density is
\begin{eqnarray}
    \varepsilon = T_{ij} u^iu^j = \frac{\ell^4m}{8\pi G}\left(2-3A^2\ell^2\Sigma\right)\,,
\end{eqnarray}
and the heat current is trivial for this background. Then, the holographic fluid is naturally described in the Landau frame. The latter also implies that $\Pi_{ij}=\tau_{ij}$. For the boundary geometry \eqref{g0WFG}, the only non-trivial component of the Cotton-York tensor is
\begin{align}
    \mathscr{C}_{\bt\by} = 3Am\Sigma(1-A^2\ell^2\Sigma)\,,
\end{align}
indicating that the boundary is non-conformally flat, and its decomposition along the congruence $u_i$ gives $c=0$\,, $c_{ij}=0$\,, and 
\begin{equation}
    c_i=-3\ell^2Am\Sigma\sqrt{\left(1-A^2\ell^2\Sigma\right)}\delta^{\by}_i\,.
\end{equation}

In principle, we could describe all boundary quantities in terms of Weyl covariant tensors using the boundary Weyl connection, but this would not give much insight into the holographic fluid description and could generate confusion with the fluid Weyl connection, so we would rather avoid the Weyl covariant description here. 

The radiative bulk degrees of freedom are captured by the  radiative vector  \eqref{Super_P1} that can be read from the boundary Cotton and stress tensors as \cite{Ciambelli:2024kre}
\begin{align}
    \hat{\mathscr{P}}^i = {18Am^2\Sigma}\left[\frac{1}{\ell^2}-A^2\Sigma\left(3-2A^2\ell^2\Sigma\right)\right]\delta^i_x \,,
\end{align}
which is non-trivial for this solution, indicating that the AdS C-metric radiates. Notice that the radiation vanishes in the zero acceleration limit, and that it has a smooth flat limit as shown in \eqref{Carr_exp_Pu} and \eqref{Carr_exp_PA}. Notice that although the form of the radiation in the WFG gauge differs from the one obtained in \eqref{P_C_metric} for the NU gauge, both expressions are nontrivial and vanish in the zero acceleration limit, confirming that the physical properties that are encoded in the radiative vector are gauge invariant.

The non-vanishing asymptotic radiative vector for RT spacetimes is dictated by a differential equation for the metric function describing the solution \cite{Ciambelli:2024kre}.
The accelerating black holes actually correspond to a special case of the RT family.\footnote{In \cite{Hoenselaers:1992bm}, it is shown that the RT family of solutions, which is characterized by a spacelike Killing vector and a specific family of initial conditions, asymptotically approaches the C-metric.} This can be seen by using the NU coordinates obtained by applying the coordinate transformation \cite{Podolsky:2003gm}
\begin{align}\label{RTcoordTransf}
    {\breve{r}} &= \left[{A(x+y)}\right]^{-1}\,, \qquad A \d u = \d \bar{t}+ \frac{\d \bar{y}}{\bar{F}}\,, \\ \nonumber \frac{1}{\sqrt{2}}\left(\d \zeta + \d \bar\zeta\right) &= \d\bar{t} + \frac{\d\bar{y}}{\bar{F}} - \frac{\d\bar{x}}{\bar{G}}\,, \qquad \frac{i}{\sqrt{2}}\left(\d \zeta - \d \bar\zeta\right) = \d\bar{\phi}\,,
\end{align}
with $\bar\zeta$ the complex conjugate of $\zeta$, to the line element \eqref{Coblate} obtaining
\begin{align}\label{RobinsonTrautman}
    {\rm d}s^2 = -\breve{H}\d u^2 -2\d u \d {\breve{r}} +2\breve{r}^2\frac{\d\zeta\d\bar\zeta}{P^2}\,,
\end{align}
where $P^{2} = \bar G^{-1}$ and 
\begin{align}
    \breve{H} = -\frac{\Lambda}{3}\breve{r}^2 - 2\breve{r} \partial_u \log P + 2P^2\partial_\zeta \partial_{\bar\zeta} \log P - \frac{2m}{\breve{r}}\,.
\end{align}

\end{appendix}

\bibliographystyle{JHEP}
\bibliography{bib}
\end{document}